\documentclass[prx,aps,floatfix,twocolumn,superscriptaddress,amsmath,amssymb]{revtex4-2}
\usepackage{graphicx}
\usepackage{xcolor}
\usepackage{bm}
\usepackage{hyperref}
\hypersetup{colorlinks=true,citecolor=blue,linkcolor=blue,urlcolor=blue}
\usepackage[T1]{fontenc}
\usepackage{newtxtext,newtxmath}
\usepackage[normalem]{ulem}

\allowdisplaybreaks

\begin{document}

\title{%
Forecasting long-time dynamics in quantum many-body systems
by dynamic mode decomposition
}

\author{Ryui Kaneko}
\email{ryuikaneko@sophia.ac.jp}
\affiliation{%
Research Institute for Science and Engineering, Waseda University, Shinjuku, Tokyo 169-8555, Japan}
\affiliation{%
Physics Division, Sophia University, Chiyoda, Tokyo 102-8554, Japan}

\author{Masatoshi Imada}
\email{imada@g.ecc.u-tokyo.ac.jp}
\affiliation{%
Research Institute for Science and Engineering, Waseda University, Shinjuku, Tokyo 169-8555, Japan}
\affiliation{%
Physics Division, Sophia University, Chiyoda, Tokyo 102-8554, Japan}
\affiliation{%
Faculty of Engineering, The University of Tokyo, 7-3-1 Hongo, Bunkyo-ku, Tokyo 113-8656, Japan}

\author{Yoshiyuki Kabashima}
\affiliation{%
The Institute for Physics of Intelligence, The University of Tokyo, 7-3-1 Hongo, Bunkyo-ku, Tokyo 113-0033, Japan}
\affiliation{%
Trans-Scale Quantum Science Institute, The University of Tokyo, 
7-3-1 Hongo, Bunkyo-ku, Tokyo 113-0033, Japan
}

\author{Tomi Ohtsuki}
\affiliation{%
Physics Division, Sophia University, Chiyoda, Tokyo 102-8554, Japan}

\date{\today}

\begin{abstract}
Reliable numerical computation of quantum dynamics is a fundamental
challenge when the long-ranged quantum entanglement plays essential
roles as in the cases governed by quantum criticality in strongly
correlated systems. Here we apply a method that utilizes reliable
short-time data of physical quantities to accurately forecast long-time
behavior of the strongly entangled systems.
We straightforwardly employ the simple dynamic mode decomposition (DMD),
which is commonly used in fluid dynamics.
Despite the simplicity of the method, the effectiveness and applicability of
the DMD in quantum many-body systems such as the Ising model in the
transverse field at the critical point are
demonstrated, 
even when the time evolution at long time exhibits complicated
features such as a volume-law entanglement entropy and consequential
power-law decays of correlations characteristic of systems
with long-ranged quantum entanglements unlike fluid dynamics.
The present method, though simple, enables accurate forecasts
amazingly at time as long as nearly an order of magnitude longer than that
of the short-time training data.
Effects of noise on the accuracy of the forecast are also investigated, because they are important especially when
dealing with the experimental data.
We find that a few percentages of
noise do not affect the prediction accuracy destructively.
\end{abstract}

\maketitle

\section{Introduction}
\label{sec:intro}

Recent experimental developments have led to the evolution of diverse measurement techniques,
facilitating the detection of novel quantum states and quantum phase transitions
in the nonequilibrium dynamics of quantum many-body systems
including quantum critical fluctuations characteristic of long-ranged quantum-mechanically
entangled systems with strong correlation effects.
For instance, in cuprate superconductors,
time-dependent optical properties have been investigated by pump-probe measurements
employing short-pulse lasers and reported enhanced coherent transport reminiscent of the superconductivity far above the superconducting transition temperature in the equilibrium~\cite{hu2014,kaiser2014}.
In ultracold atomic gases confined in optical lattices
that serve as analog quantum simulators on various lattice systems in arbitrary spatial dimensions,
quantum gas microscopes~\cite{cheneau2012} or time-of-flight measurements~\cite{takasu2020}
allow for the observation of the time evolution of equal-time correlation functions
in quantum many-body systems.

The validation and prediction of experimental results
rely on comparing them with outcomes obtained from numerical simulations.
Despite the experimental progress in observing time-dependent physical quantities,
numerical simulations of the dynamics in quantum many-body systems are still challenging.
For example,
although obtaining the time-evolved states for systems in one spatial dimension (1D)
is fairly efficient using the tensor-network methods~\cite{schollwock2011},
calculating their dynamics by the tensor-network methods
are much harder for systems in two or more spatial dimensions
because of highly entangled structures
of quantum states after long-time
evolution~\cite{kshetrimayum2017,czarnik2019,hubig2019,hubig2020,schmitt2022,kaneko2022,kaneko2023}.
On the other hand,
the time-dependent variational Monte Carlo method~\cite{carleo2012,carleo2014,ido2015}
with neural-network quantum states~\cite{carleo2017,nomura2017}
is a promising candidate for calculating the dynamics
and would be able to simulate the dynamics for a much longer time~\cite{ido2017,schmitt2020}.
However, it often suffers from a high computational cost
because of a large number of variational parameters
and the requirement for sophisticated optimization of
wave functions at each time step.
In practice, most of the tensor-network and time-dependent variational Monte Carlo methods
can simulate the dynamics typically up to the time
$\sim \text{const}~\hbar/{\text{(characteristic energy scale)}}$
with an $\mathcal{O}(1)$ constant.

It would be advantageous to have a method
that leverages short-time data of physical quantities to reliably predict long-time behavior.
As for forecasting physical quantities after time evolution,
it is desired to develop a method that can predict long-time behavior from precise short-time data obtained by experiments or numerical simulations without knowing the time-evolved quantum state itself.
The lack of information about the wave function is not a drawback at all.
Instead, it allows for wide applications without detailed
knowledge of the wave function, which accelerates the application of
such a method to practical experimental data of quantum dynamics.
There has been a growing interest
in predicting the dynamics of observables
in quantum many-body systems
using machine learning and related
techniques~\cite{barthel2009,goto2016,huang2023,mohseni2023_arxiv,cemin2023_arxiv,wang2024_arxiv}.

To this end, we focus on the dynamic mode decomposition (DMD),
which is commonly used in the field of
fluid dynamics~\cite{schmid2008,rowley2009,schmid2010,tu2014,kutz2016,brunton2019,schmid2022}.
This method is essentially equivalent to
the matrix pencil method~\cite{pogorelyuk2018_arxiv},
which was introduced much earlier than the DMD
and
has been used for estimating frequencies and damping factors of
sinusoidal signals from noisy time-series
data~\cite{sarkar1980,hua1990,sarkar1995}.
The DMD
offers advantages in its simplicity of execution
and its minimal dependency on assumptions for the system.
The computational cost when one uses this algorithm
is primarily determined by the expense of performing a singular value decomposition (SVD)
of a certain matrix constructed from the short-time data,
making it a cost-effective method.
The DMD applies to systems in any spatial dimensions, crystal (or even noncrystal) structures and with any interactions;
it is versatile and applicable to theoretical simulations as well as to experimental data.
The DMD has been applied to the dynamics of quantum
systems~\cite{goldschmidt2021,sakata2021,kawashima2021,luchnikov2022,yin2022,yin2023,baddoo2023,
reeves2023,gu2024,hunstig2023_arxiv,maliyov2024}
and the quantum algorithm for dynamic
simulations~\cite{steffens2017,shen2023,gomes2023_arxiv,mizuno2023_arxiv,szasz2024_arxiv}
recently.
In Refs.~\cite{yin2022,yin2023,reeves2023,maliyov2024},
the DMD was successful in obtaining nonequilibrium long-time evolution
containing many-body effects within the perturbative self-energy.
However, whether the DMD offers a useful and efficient tool of
an accurate quantum many-body solver for long-time predictions
in cases of strongly correlated systems and/or systems
with critical fluctuations accompanied by strong quantum entanglement
still remain a big challenge.
The DMD applicable to data with quantum entanglement,
beyond the mean-field level and the perturbative range,
will open up significant potential for various applications.

In this paper, we specifically focus on
one of the most challenging problems in the DMD, i.e.,
the prediction of the long-time dynamics
from the short-time reliable data
that characterizes the evolution of entanglement entropy
and exhibits a resultant power-law decay, which stem from
the quantum many-body effects.
Such behavior already appears in the textbook example of
the time-dependent correlation functions
in the low-dimensional transverse-field Ising
models~\cite{tommet1975,muller1984,rossini2009,rossini2010}.
We demonstrate that the
conventional
DMD
that has often been used in classical systems
is already advantageous in predicting the long-time dynamics
and
makes it possible to predict the behavior at time as long as nearly one order of magnitude longer than the range of utilized short-time training data
even when the input data exhibits such complicated features
by taking the transverse-field Ising models
as model systems.
It is highly nontrivial that the simple procedure of the DMD works well
even in the presence of strong quantum entanglement and
quantum critical fluctuations.
Our findings suggest that the DMD is a powerful tool even in cases with competing orders and near continuous phase transitions, around which quantum criticality emerges.
Because of the simplicity of the DMD,
it has a wide range of applications and is expected to become a
standard method that can also be used with experimental data.

This paper is organized as follows.
In Sec.~\ref{sec:dmd},
we introduce the DMD and describe the details of
how one can apply it to the dynamics of the quantum many-body systems.
In Sec.~\ref{sec:application},
we demonstrate the effectiveness of the DMD
in predicting the time-dependent correlations
with oscillatory components.
We also examine to what extent the DMD deals with the challenging critical phenomena
that exhibits a power-law decay as a function of time
and converges to a nonzero value in the infinite-time limit.
To compare with exact time-series data
for arbitrarily long time,
the input data are prepared by the exact diagonalization method
or by the analytical calculation of an integrable system.
Moreover, to cope with more practical problems,
we show that the present DMD withstands
up to a moderate level of noise in the input data
and sustains the accuracy.
Last,
we estimate the systematic and statistical errors
of the DMD predictions.
In Sec.~\ref{sec:sum},
we draw our conclusions and look to further applications of the DMD
in quantum many-body systems.

\section{Dynamic mode decomposition}
\label{sec:dmd}

We briefly review how one can apply the DMD to a given time-series
data~\cite{kutz2016,brunton2019}.
We consider a time-dependent physical quantity
$f_n = f(n\cdot\Delta t) = f(t)$
that is measured or calculated at discrete time steps
$n = 0, 1, 2, \dots, N-1$.
Here we assume that the time interval between two consecutive time steps is
constant and denote the time interval by $\Delta t$.
We would like to predict the time-dependent physical quantity $f_n$
for $n = N, N+1, \dots, N_{\rm max}-1$,
where $N_{\rm max}-1$ is the maximum number of time steps to be predicted.

\begin{figure}[t!]
\centering
\includegraphics[width=1.00\columnwidth]{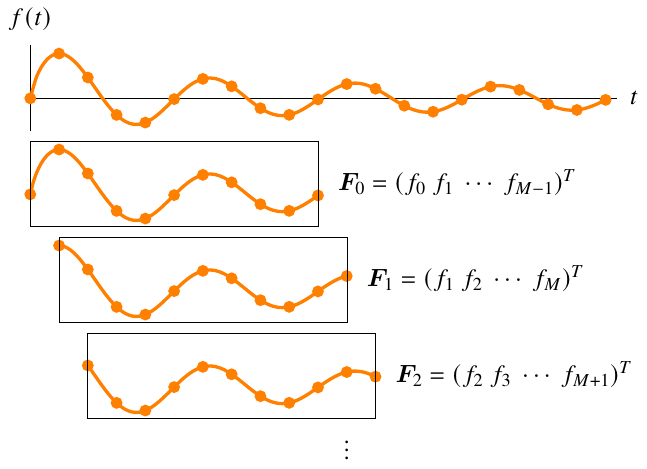}
\caption{%
Schematic picture of
time-series data of a physical quantity
$f_n = f(n\cdot\Delta t) = f(t)$
($n = 0, 1, 2, \dots, N-1$).
We separate the time-series data into several
short-interval sequences
and construct $N-M+1$ vectors.
Each column vector
contains $M$ elements.
}
\label{fig:time_series}
\end{figure}

We separate the time-series data into several
short-interval sequences
as shown in
Fig.~\ref{fig:time_series}.
We construct $N-M+1$ vectors
$\bm{F}_0$, $\bm{F}_1$, \dots, $\bm{F}_{N-M}$ from the
short-interval sequences.
Each column vector
$\bm{F}_n = (f_n~f_{n+1}~\cdots~f_{n+M-1})^{T}$
contains $M$ elements.
Hereafter,
we call each short-interval sequence (column vector) $\bm{F}_n$
a snapshot, following the convention~\cite{kutz2016,brunton2019},
although $\bm{F}_n$ contains the information of different time
from $n$ to $n+M-1$.
We then construct two $M \times (N-M)$ matrices $X_0$ and $X_1$ as
\begin{align}
 \label{eq:dmd_x0}
 X_0
 &=
 \begin{pmatrix}
 \vline & \vline & \vline & & \vline \\
 \bm{F}_0 & \bm{F}_1 & \bm{F}_2 & \cdots & \bm{F}_{N-M-1} \\
 \vline & \vline & \vline & & \vline \\
 \end{pmatrix}
\\
 &=
\begin{pmatrix}
 f_0 & f_1 & f_2 & \cdots & f_{N-M-1} \\
 f_1 & f_2 & f_3 & \cdots & f_{N-M} \\
 f_2 & f_3 & f_4 & \cdots & f_{N-M+1} \\
 \vdots & \vdots & \vdots & \ddots & \vdots \\
 f_{M-1} & f_{M} & f_{M+1} & \cdots & f_{N-2} \\
\end{pmatrix},
\\
 X_1
 &=
 \begin{pmatrix}
 \vline & \vline & \vline & & \vline \\
 \bm{F}_1 & \bm{F}_2 & \bm{F}_3 & \cdots & \bm{F}_{N-M} \\
 \vline & \vline & \vline & & \vline \\
 \end{pmatrix}
\\ 
 &=
\begin{pmatrix}
 f_1 & f_2 & f_3 & \cdots & f_{N-M} \\
 f_2 & f_3 & f_4 & \cdots & f_{N-M+1} \\
 f_3 & f_4 & f_5 & \cdots & f_{N-M+2} \\
 \vdots & \vdots & \vdots & \ddots & \vdots \\
 f_{M} & f_{M+1} & f_{M+2} & \cdots & f_{N-1} \\
\end{pmatrix}.
\end{align}
The DMD computes the leading eigendecomposition
of the linear operator $A$
that satisfies $X_1 = A X_0$.
The $M \times M$ matrix $A$ is given by
\begin{align}
 \label{eq:dmd_a}
 A = X_1 X_0^{-1},
\end{align}
where $X_0^{-1}$ is the Moore-Penrose pseudoinverse of $X_0$.
Note that $X_0^{-1}=X_0^\dagger(X_0 X_0^\dagger)^{-1}$ when the
number of rows is smaller than that of columns,
whereas
$X_0^{-1}=(X_0^\dagger X_0)^{-1} X_0^\dagger$ otherwise.
We can predict the time-series data $\bm{F}_n$ for
$n = N, N+1, \dots, N_{\rm max}-1$
by sequentially multiplying the matrix $A$ to the vector $\bm{F}_0$,
i.e.,
$\bm{F}_n \approx A^{n} \bm{F}_0$.

When the dimension $M$ of the matrix $A$ is large,
the computation of the eigendecomposition of $A$ becomes expensive.
Moreover, the matrix $A$ may contain
eigenvectors that cause numerical instability.
In such cases, we can reduce the computational cost
and improve the numerical stability
by the truncated SVD.
Here we describe how to utilize the truncated SVD
to perform the DMD.
Hereafter, for simplicity,
we assume that the matrix $A$ and that in the reduced subspace
(the matrix $\tilde{A}$) are diagonalizable.
When the matrix is not diagonalizable,
one can modify the following arguments
using the Jordan normal form of each matrix~\cite{mizuno2023_arxiv}:
\begin{enumerate}
 \item
We compute the SVD of the matrix $X_0$ as
\begin{align}
 \label{eq:dmd_svd}
 X_0 = U \Sigma V^{\dagger},
\end{align}
where $U$ is an $M \times M$ unitary matrix,
$V^{\dagger}$ is an $(N-M) \times (N-M)$ unitary matrix,
and $\Sigma$ is an $M \times (N-M)$ matrix
with only the first
$\min(M, N-M)$ diagonal elements being nonzero.
The diagonal elements of $\Sigma$ are denoted by
$\sigma_0 \ge \sigma_1 \ge \cdots \ge
\sigma_{\min(M, N-M)-1} \ge 0$.
We keep only the first $R$ columns of $U$ and $V$ and
the first $R \times R$ submatrix of $\Sigma$.
Here $R$
is the rank of the reduced SVD approximation to $X_0$
and
is determined by the smallest integer that satisfies
$\sigma_R/\sigma_0 < \epsilon$
with $\epsilon$ being a cutoff.
As a result, we obtain
an $M \times R$ matrix $U_R$,
an $R \times R$ diagonal matrix $\Sigma_R$,
and an $R \times (N-M)$ matrix $V_R^{\dagger}$,
respectively.
They satisfy
\begin{align}
 \label{eq:dmd_svd_r}
 X_0 \approx U_R \Sigma_R V_R^{\dagger}.
\end{align}
 \item
We compute the matrix $\tilde{A}$,
which is the $R\times R$ projection of $A$,
as
\begin{align}
 \label{eq:dmd_a_r}
 \tilde{A}
 =
 U_R^{\dagger} X_1 V_R \Sigma_R^{-1}.
\end{align}
Here we use the fact that
$
 \tilde{A}
 \approx
 U_R^{\dagger} A U_R
$,
$
 A = X_1 X_0^{-1}
 \approx X_1 V_R \Sigma_R^{-1} U_R^{\dagger}
$,
and
$U_R^{\dagger} U_R \approx 1_R$,
where $1_R$ is an $R \times R$ identity matrix.
 \item
We diagonalize the matrix $\tilde{A}$ as
\begin{align}
 \label{eq:dmd_a_r_diag}
 \tilde{A} W = W \Lambda,
\end{align}
where 
$\Lambda = {\rm diag}
(\lambda_0, \lambda_1, \dots, \lambda_{R-1})$
is a diagonal matrix
and
$W$ is an $R \times R$ unitary matrix.
 \item
We reconstruct the eigendecomposition of $A$
from $W$ and $\Lambda$ approximately.
The approximate dominant $R$ eigenvalues of $A$ are given by $\Lambda$,
and the corresponding $R$ eigenvectors of $A$ are given by
columns of the $M \times R$ matrix $\Phi$,
which is defined by
\begin{align}
 \label{eq:dmd_phi}
 \Phi = X_1 V_R \Sigma_R^{-1} W.
\end{align}
Indeed,
from Eq.~\eqref{eq:dmd_a_r}
and
Eq.~\eqref{eq:dmd_a_r_diag},
one can see that
\begin{align}
 \qquad
 A\Phi
 &\approx U_R \tilde{A} U_R^{\dagger} \Phi
  = U_R \tilde{A} U_R^{\dagger} X_1 V_R \Sigma_R^{-1} W
\nonumber
\\
 &= U_R \tilde{A} \tilde{A} W 
  = U_R \tilde{A} W \Lambda
\nonumber
\\
 &= U_R U_R^{\dagger} X_1 V_R \Sigma_R^{-1} W \Lambda
  = U_R U_R^{\dagger} \Phi \Lambda
  \approx \Phi \Lambda
\end{align}
holds.
 \item
We predict the time-series data
$f_n = f(n\cdot\Delta t) = f(t)$
for $n = N, N+1, \dots, N_{\rm max}-1$
from $\Phi$ and $\Lambda$.
From
$
 \bm{F}_n \approx A^n \Phi \Phi^{-1} \bm{F}_0
 \approx \Phi \Lambda^n \Phi^{-1} \bm{F}_0
$
with $\Phi^{-1}$
being the pseudoinverse of $\Phi$,
the time evolution of the time-series data is
approximately given by
\begin{align}
 \label{eq:dmd_predict}
 \bm{F}_n
 \approx
 \Phi \Lambda^n \bm{b}
 =
 \sum_{k=0}^{R-1}
 \bm{\phi}_k
 (\lambda_k)^n
 b_k.
\end{align}
Here the $M$ component vector
$\bm{\phi}_k$
is the $k$th column of $\Phi$,
and
the vector
$\bm{b} = (b_0~b_1~\cdots~b_{R-1})^{T}$
is determined by
the initial vector $\bm{F}_0$
and the pseudoinverse of $\Phi$ as
\begin{align}
 \label{eq:dmd_b}
 \bm{b} = \Phi^{-1} \bm{F}_0.
\end{align}
\end{enumerate}
The computational cost of the DMD is dominated by
the computation of the truncated SVD of $X_0$.
When the upper bound $R^{\rm upper}$ for the rank
of the truncated SVD is given in advance,
i.e., $R \le R^{\rm upper}$ and $R^{\rm upper} \ll M, (N-M)$,
the computational cost of the DMD becomes
${\cal O}[M(N-M)R^{\rm upper}]$.
The cost can be further reduced
by applying the randomized SVD~\cite{halko2011}
and is given by
${\cal O}[M(N-M) \log R^{\rm upper}]$.
$R^{\rm upper}$ is practically bounded by a few hundred,
and thus, the primal cost is 
scaled only by
the size of input data.
This cost is much cheaper than
the computational cost
${\cal O}(M^3)$
of the direct eigendecomposition of $A$
when $M\sim (N-M)$.

\section{Application to the quantum dynamics}
\label{sec:application}

We apply the DMD to the dynamics of quantum many-body systems
and discuss the accuracy and applicability of the DMD.
We specifically consider the following cases:
(i)
correlation functions that exhibit multiple oscillatory modes
(caused by the evolution of entanglement entropy, which stem from the
quantum many-body effects, see Appendix~\ref{sec:ee_dynamics})
and
(ii)
correlation functions that exhibit oscillatory behavior
and have a power-law decay.
In case (i),
we choose the two-dimensional
(2D)
transverse-field Ising model
as a model system.
For a small finite system,
the time evolution of equal-time spin-spin correlation functions after a sudden quench
exhibit oscillatory behavior without damping.
In case (ii),
we choose the
(1D)
transverse-field Ising model
at the critical point as a model system, where a strong long-range quantum entanglement is expected providing us with a
challenge
in conventional numerical simulations~\cite{perales2008}.
For a sufficiently large system,
the unequal-time (time-displaced) spin-spin correlation functions exhibit oscillatory behavior with a power-law decay on top of the convergence
to a nonzero value in the infinite-time limit.
We show that the DMD can predict the time evolution of
the correlation functions in both cases
with high accuracy.
Moreover,
we discuss the error analysis of the DMD.

\subsection{Time-dependent correlation functions without damping}
\label{subsec:corr_no_damp}

We focus on the transverse-field Ising model
\begin{align}
 \label{eq:ham_ising}
 H = - J \sum_{\langle i,j \rangle} S^z_i S^z_j
 - \Gamma \sum_{i} S^x_i,
\end{align}
under the periodic boundary condition,
where $S_i^{\alpha}$ ($\alpha = x, y, z$) are the $S=1/2$ Pauli spins,
$J$ is the strength of the nearest-neighbor interaction,
and $\Gamma$ is the strength of the transverse magnetic field.
The symbol $\langle \cdots \rangle$ denotes nearest-neighbor site pairs.
Hereafter, we set the lattice constant $a$
and the reduced Planck constant $\hbar$ to unity.
We take the units of energy and time as $J$ and $J^{-1}$, respectively,
unless otherwise noted.

We obtain the time evolution of the equal-time longitudinal spin-spin correlation
functions in the following manner.
We consider the model in Eq.~\eqref{eq:ham_ising}
on a square lattice with a system size
$N_{\mathrm{s}} = L^2$ for $L = 4$.
We first prepare the initial state as
\begin{align}
 \label{eq:ising_initial_state}
 |\psi_0\rangle
 =
 \bigotimes_{i=1}^{N_{\mathrm{s}}}
 \frac{|\uparrow\rangle_i + |\downarrow\rangle_i}{\sqrt{2}},
\end{align}
which is the ground state of the transverse-field Ising model
at $\Gamma/J = \infty$.
We then perform a sudden quench
by changing the transverse magnetic field from $\Gamma/J = \infty$
to the critical point
$\Gamma/J = \Gamma_{\rm c}^{\rm 2D}/J \approx 1.522$~\cite{rieger1999,bloete2002,kaneko2021}.
The time evolution of equal-time longitudinal spin-spin correlation functions
at distance $\bm{r}$ and time $t$ are given by
\begin{align}
 \label{eq:ising_2d_corr}
 C^{zz}_{\rm eq}(\bm{r}, t)
 =
 \langle \psi_0 |
 S^z_{\bm{0}}(t) S^z_{\bm{r}}(t)
 |\psi_0\rangle\,,
\end{align}
with $S^z_{\bm{r}}(t) = e^{iHt} S^z_{\bm{r}} e^{-iHt}$.
As for a small finite system,
we can use the exact diagonalization method
to calculate the equal-time correlation functions
over arbitrarily long periods.
We calculate the correlation functions
for $t \in [0,1000)$ with a time step $\Delta t = 0.05$
and $N_{\rm max} = 20000$
using the \textsc{QuSpin}
library~\cite{weinberg2017,weinberg2019}.

\begin{figure}[t!]
\centering
\includegraphics[width=1.00\columnwidth]{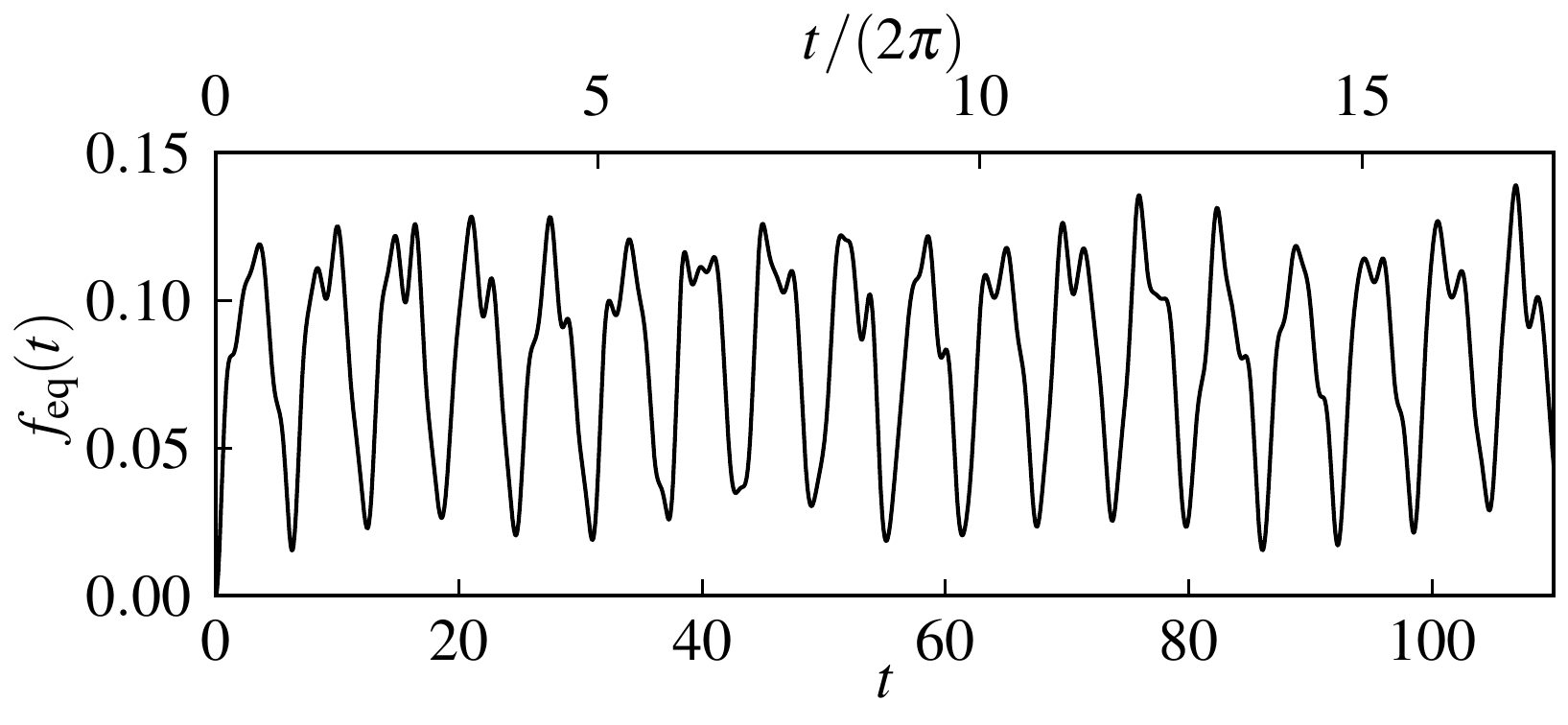}
\caption{%
Exact
time evolution of the equal-time longitudinal spin-spin correlation function
after a sudden quench in the
2D
transverse-field Ising model
on a finite-size square lattice
($4\times 4$ sites).
}
\label{fig:ising_2d_corr}
\end{figure}

\begin{figure}[t!]
\centering
\includegraphics[width=.525\columnwidth]{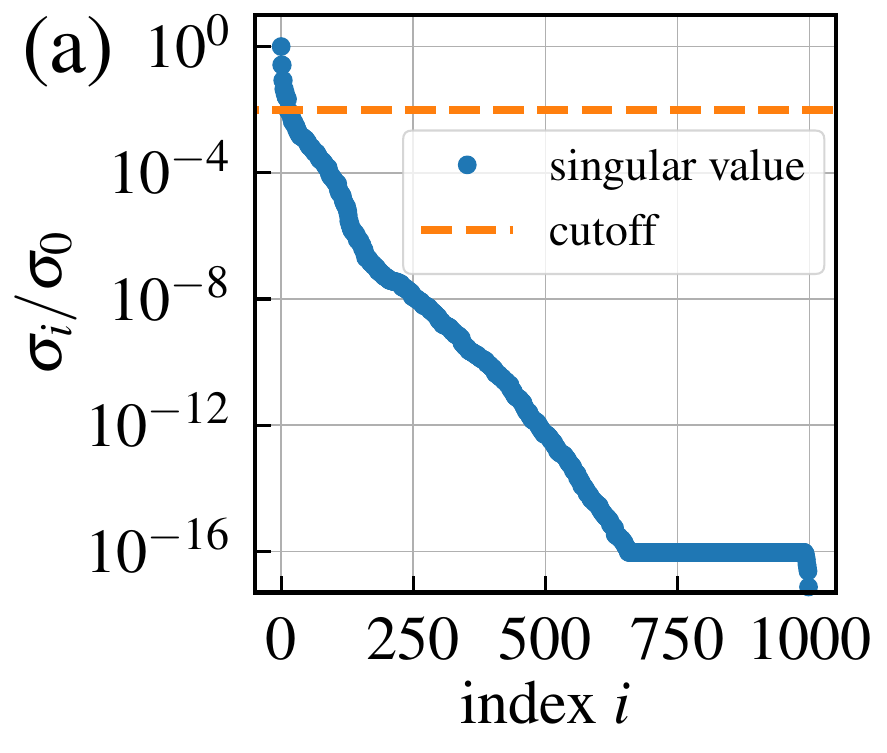}
\hfil
\includegraphics[width=.455\columnwidth]{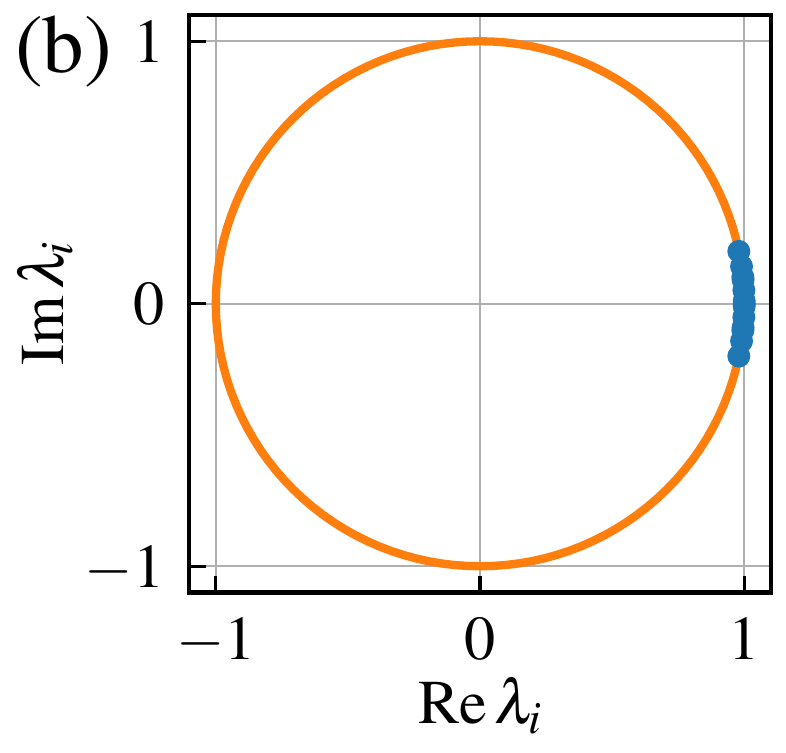}
\caption{%
DMD parameters in the case of
the equal-time longitudinal spin-spin correlation function
after a sudden quench in the
2D
transverse-field Ising model
on a square lattice.
(a) Singular values $\sigma_i$ of the truncated SVD of the matrix $X_0$.
(b) Eigenvalues $\lambda_i$ of the matrix $\tilde{A}$.
The orange unit circle is a guide to the eye.
}
\label{fig:ising_2d_corr_lmd_sgm}
\end{figure}

We show the exact time evolution of the correlation function
at distance $|\bm{r}| = 1$,
i.e.,
\begin{align}
\label{eq:def_ft_2d}
 f_{\rm eq}(t) = C^{zz}_{\rm eq}(|\bm{r}|=1, t)
\end{align}
in Fig.~\ref{fig:ising_2d_corr}.
Although the
2D
transverse-field Ising model
is nonintegrable, and thus the thermalization would occur
and the correlation function would converge to a nearly constant value
after a long time in the thermodynamic limit,
the correlation function does not exhibit damping
in a small finite system.
The correlation function exhibits oscillatory behavior,
and
the period of the dominant oscillation is
found to be $T \approx 2\pi/J$.

For the DMD,
we choose $M = 1000$ and $N = 2M$,
which correspond to the
time length of the short-interval sequence
$t_{\rm snap} = M\cdot\Delta t = 50$
and the whole time interval of input data
$t_{\rm input} = N\cdot\Delta t = 100$,
respectively.
The calculated singular values $\sigma_i$ 
decay exponentially as a function of the index $i$,
with a steep exponent at small index $i\leq 13$ and milder exponents at larger index $i$ as shown in Fig.~\ref{fig:ising_2d_corr_lmd_sgm}(a).
We should choose a sufficiently small cutoff $\epsilon$
to include relevant modes of the dynamics as many as possible;
at the same time,
we also need a reasonably not too-small cutoff $\epsilon$
to avoid the inclusion of irrelevant modes
that cause the divergence of time series for long $t$.
In the case of the present model,
we numerically find that
the DMD prediction becomes unstable
when the cutoff $\epsilon$ is smaller than $0.01$.
Therefore,
the cutoff is chosen to be $\epsilon = 0.01$,
and the rank $R$ of the truncated SVD becomes only $R = 13$.
The absolute values of the calculated eigenvalues
$|\lambda_i|$ of the matrix $\tilde{A}$
are smaller than or equal to unity
[see Fig.~\ref{fig:ising_2d_corr_lmd_sgm}(b)],
indicating that the dynamics obtained by the DMD is
stable.

\begin{figure}[t!]
\centering
\includegraphics[width=1.00\columnwidth]{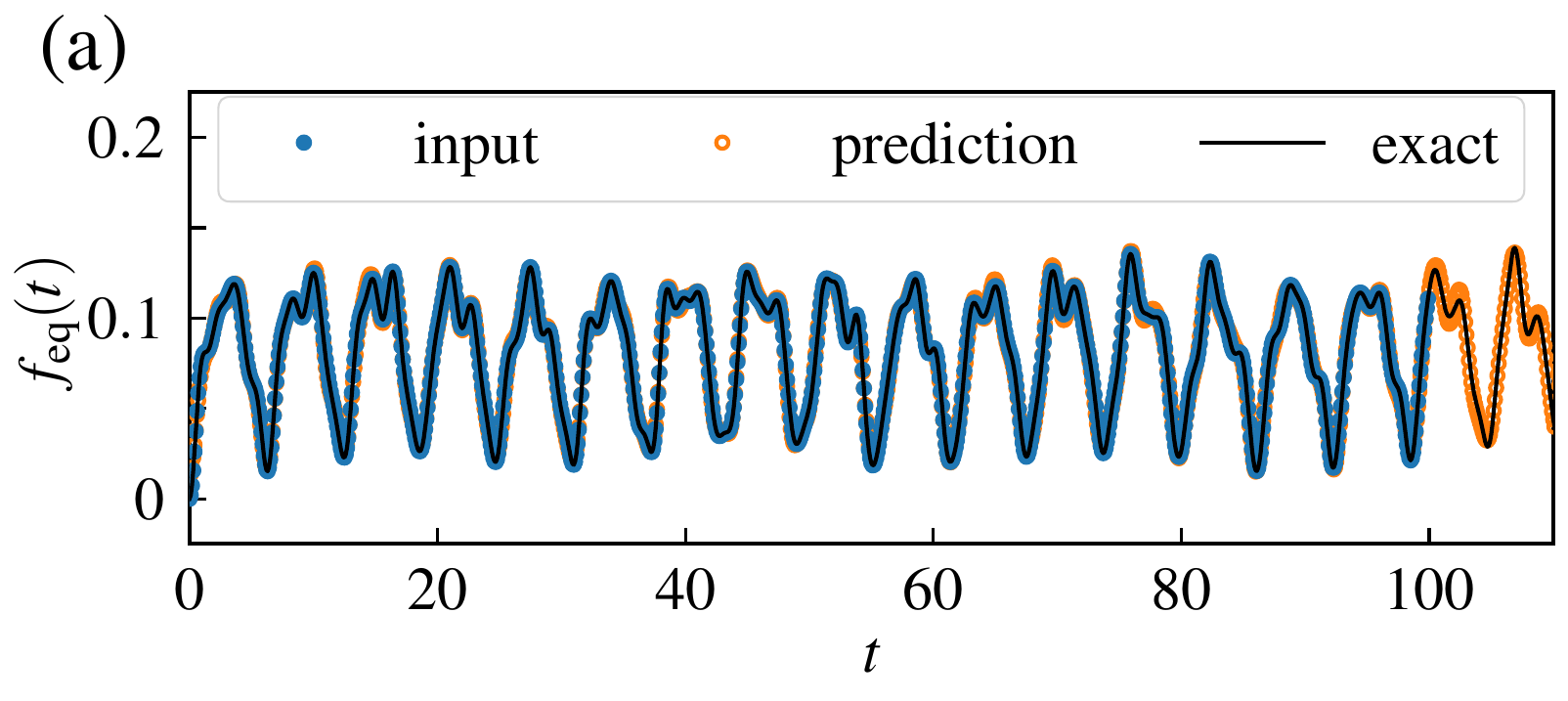}\\
\includegraphics[width=1.00\columnwidth]{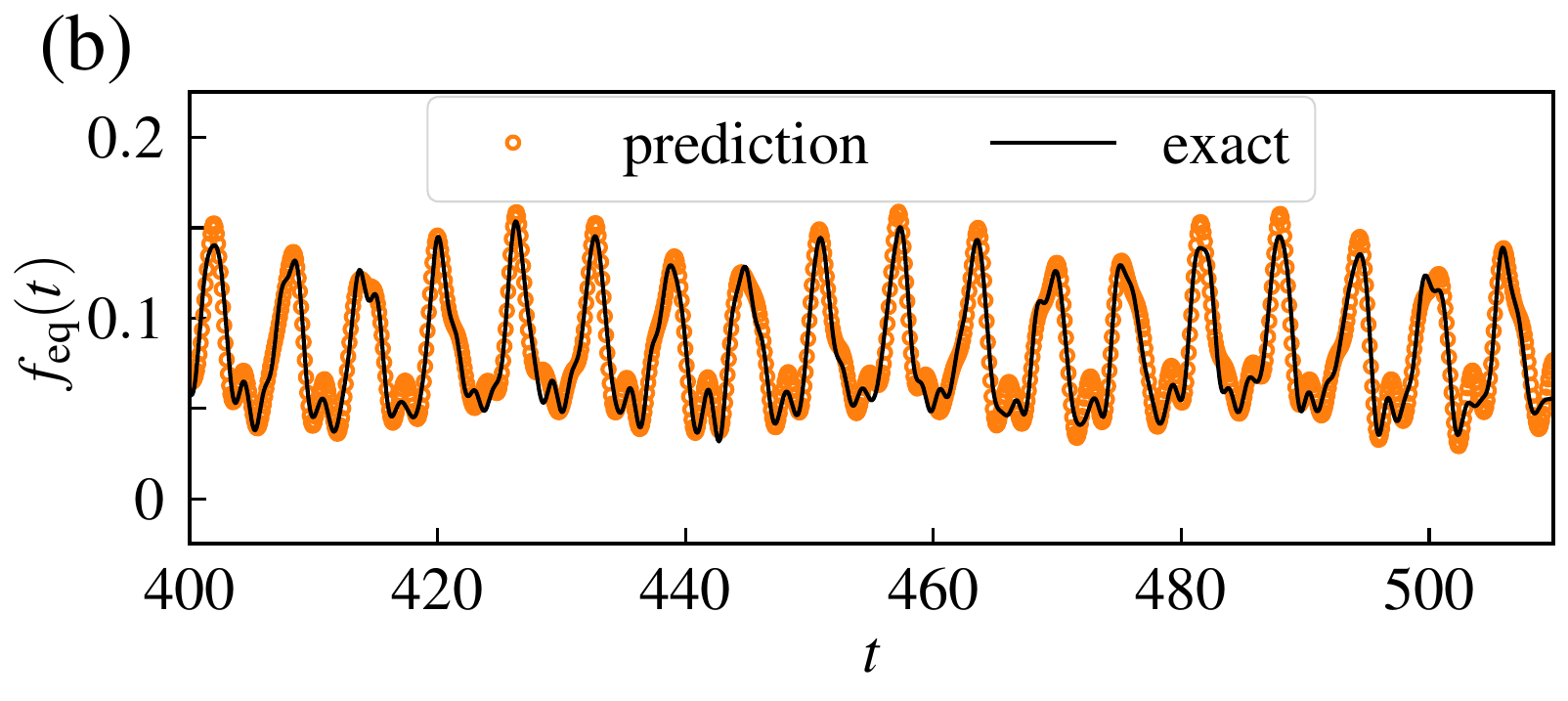}\\
\includegraphics[width=1.00\columnwidth]{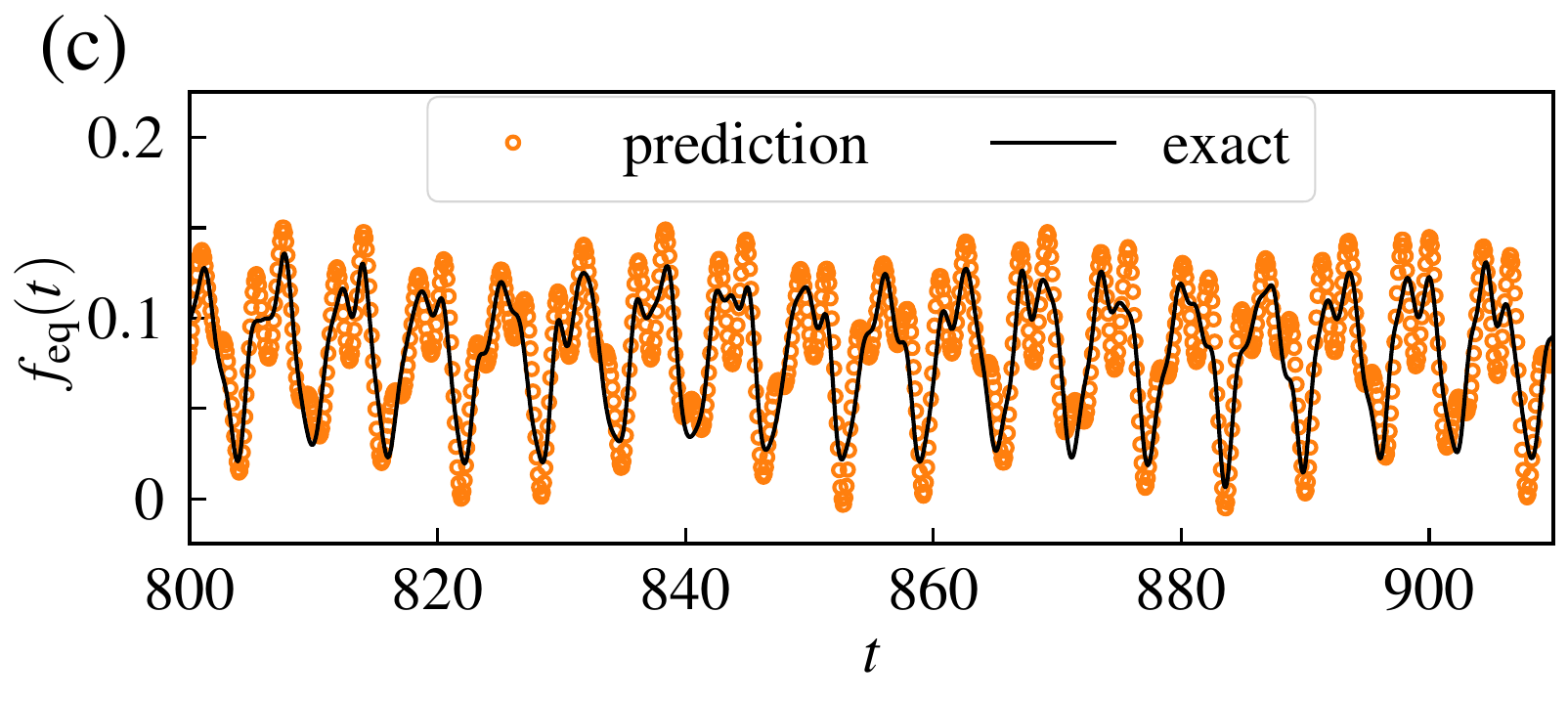}
\caption{%
DMD prediction of the equal-time longitudinal spin-spin correlation function
in the 2D transverse-field Ising model.
We show the time evolution of the correlation function for 
(a) $t \in [0,110]$,
(b) $t \in [400,510]$,
and
(c) $t \in [800,910]$.
The solid line is the exact result,
the filled blue circles are the input data,
and the open orange circles are the predicted data.
The data points $f_{{\rm eq},n}=f_{\rm eq}(n\cdot\Delta t)=f_{\rm eq}(t)$ are plotted
only when $n$ is an even number.
}
\label{fig:ising_2d_corr_dmd}
\end{figure}

\begin{figure}[t!]
\centering
\includegraphics[width=1.00\columnwidth]{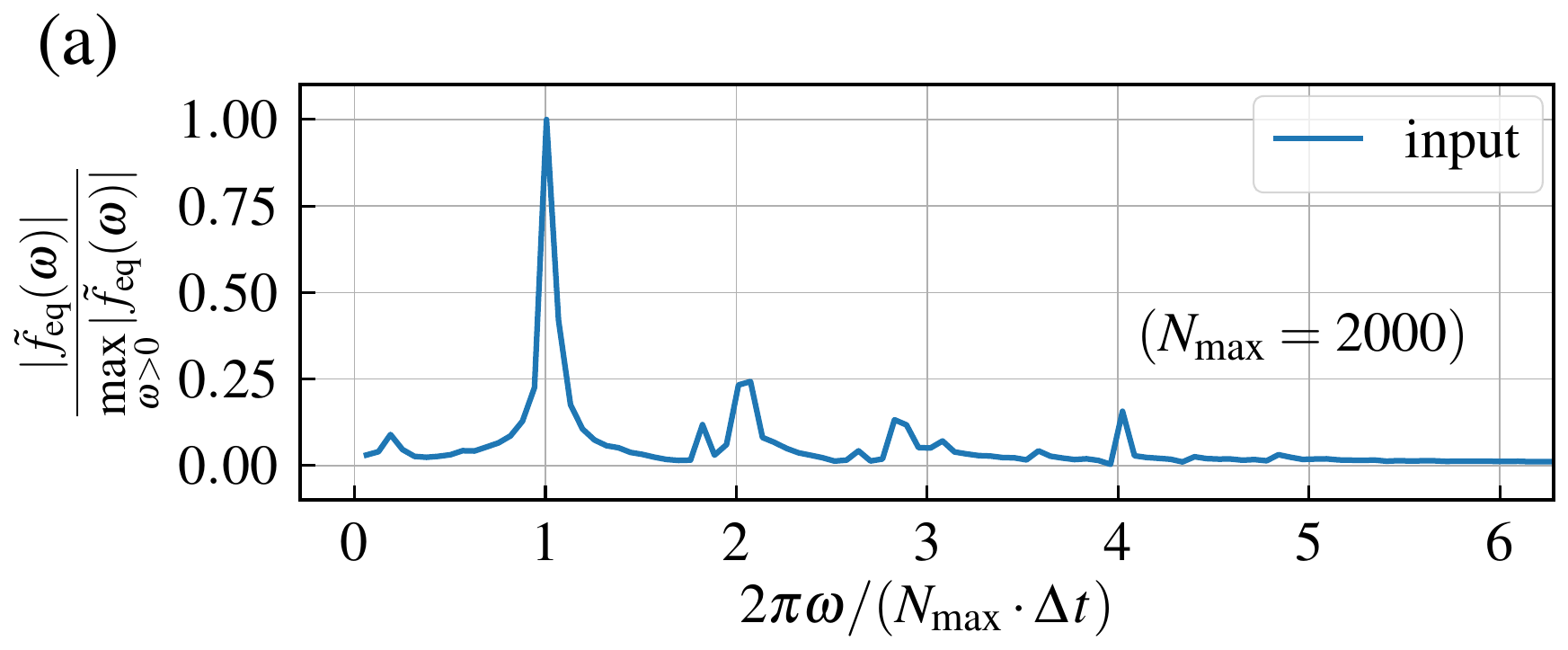}\\
\includegraphics[width=1.00\columnwidth]{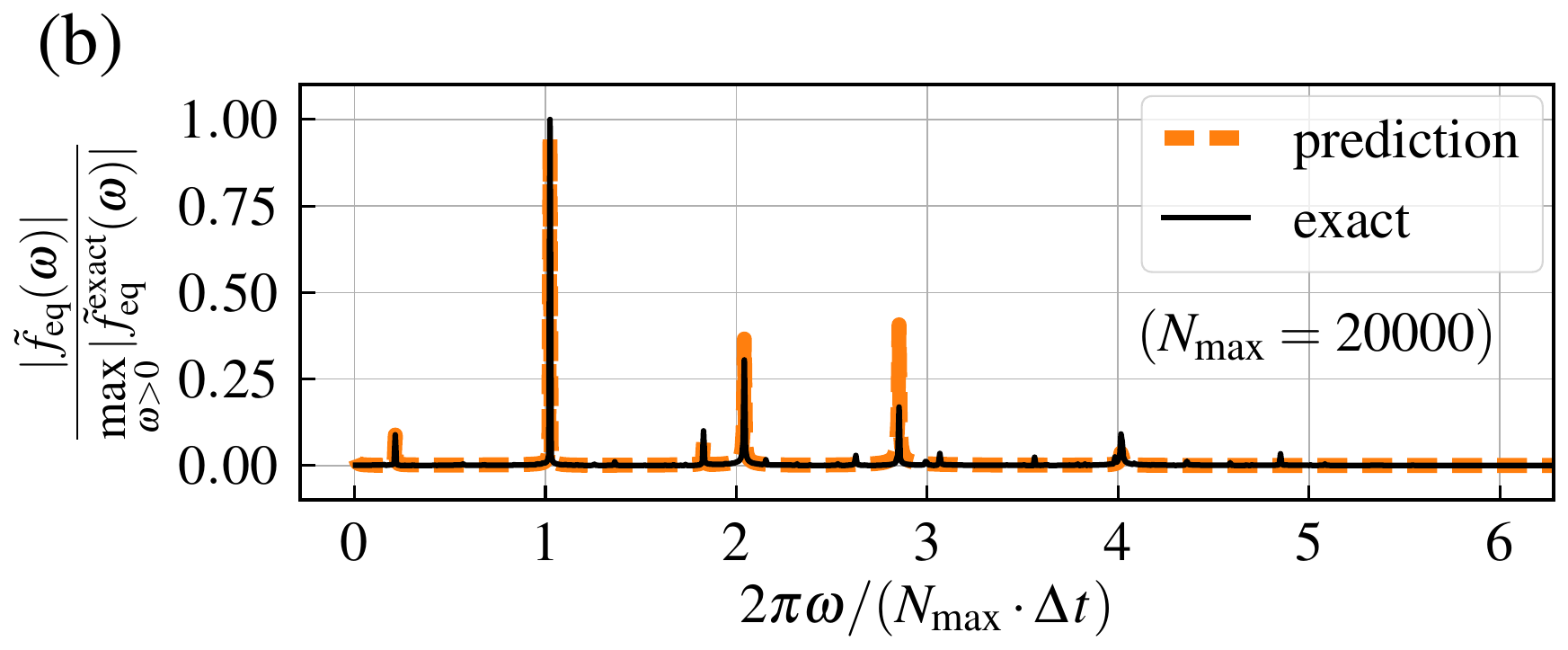}\\
\includegraphics[width=1.00\columnwidth]{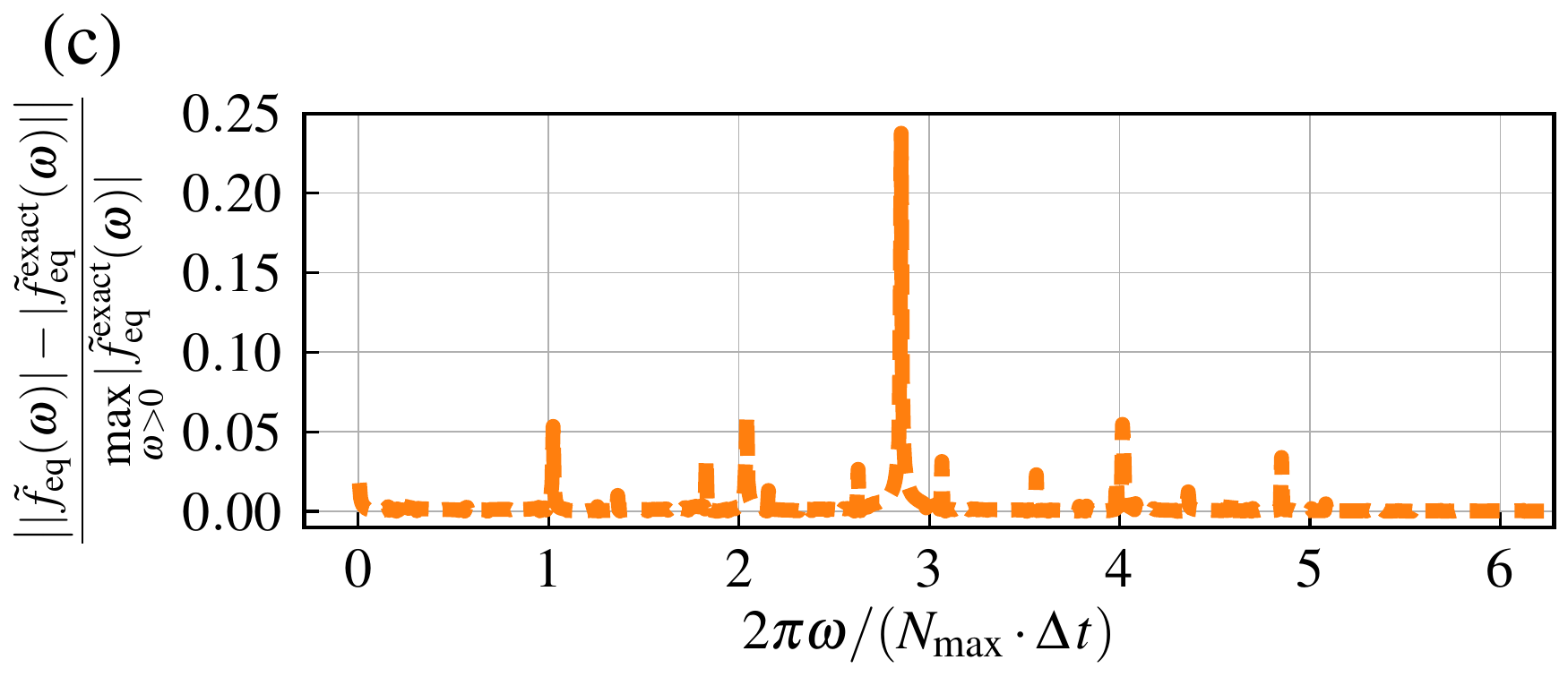}\\
\includegraphics[width=1.00\columnwidth]{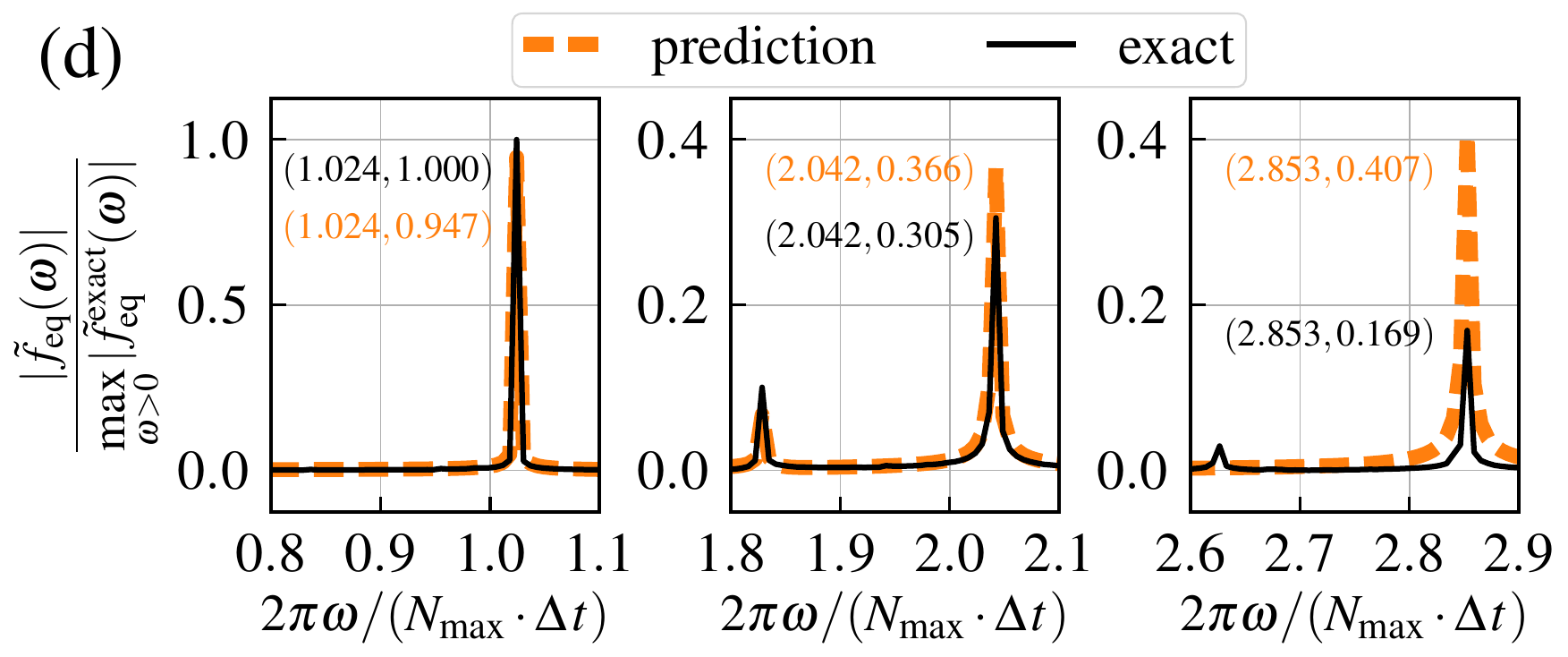}
\caption{%
Fourier transform of the equal-time longitudinal spin-spin correlation function
in the 2D transverse-field Ising model.
We remove a large value at $\omega = 0$.
(a) Exact result when $t \in [0,100)$ ($N_{\rm max} = 2000$).
(b) Comparison of the exact result (solid line)
and the DMD prediction (dashed line)
for the absolute value of the Fourier-transformed correlation function
when $t \in [0,1000)$ ($N_{\rm max} = 20000$).
(c) Relative difference
between the exact result and the DMD prediction.
(d) Magnified views of Fig.~\ref{fig:ising_2d_corr_fourier}(b).
The position and intensity of each peak
are also shown in the figure in the order (position, intensity).
All the positions of the peaks of the DMD prediction
coincide with those of the exact result.
}
\label{fig:ising_2d_corr_fourier}
\end{figure}

We show the selected time evolution of the
predicted correlation function
for $t \in [0,110]$,
$t \in [400,510]$,
and
$t \in [800,910]$
in Figs.~\ref{fig:ising_2d_corr_dmd}(a)--\ref{fig:ising_2d_corr_dmd}(c).
For $t \lesssim 500$,
the DMD prediction (open circles)
and the exact result (solid line)
are in good agreement.
In this sense,
the DMD can predict the time evolution of the correlation function
with high accuracy
up to approximately five times the duration of the input time.
As the time $t$ increases,
the DMD prediction deviates from the exact result.
Indeed, for $t \in [800,910]$,
we observe that the amplitude of the oscillation
in the DMD prediction is slightly larger than that in the exact result.
On the other hand,
the overall structure of the DMD prediction
remains similar to that of the exact result even at $t \ge 800$, namely at an order of magnitude longer time than $t_{\rm input}$.

We also show the Fourier transform of the
predicted correlation function,
which is defined by
\begin{align}
 \label{eq:ising_2d_corr_fourier}
 \tilde{f}(\omega)
 =
 \sum_{n=0}^{N_{\rm max}-1}
 f_n
 \exp\left(
   - \frac{2\pi i \omega n}{N_{\rm max}}
  \right)
\end{align}
for $\omega = 0, 1, \dots, N_{\rm max}-1$
in Fig.~\ref{fig:ising_2d_corr_fourier}.
We remove a relatively larger component at $\omega = 0$ hereafter.
As a reference,
we plot the exact result
when $t \in [0,100)$ ($N_{\rm max} = 2000$)
in Fig.~\ref{fig:ising_2d_corr_fourier}(a).
We will examine how the DMD prediction reproduces the exact result
for $t \in [0,1000)$ ($N_{\rm max} = 20000$)
when using this reference as input data.
Note that the prediction
based on the Gaussian process regression
(GPR) method~\cite{rasmussen2006,pedregosa2011},
which is a conventional machine learning method,
fails to reproduce the exact peak positions
as shown in Appendix~\ref{sec:gpr}.

As shown in Fig.~\ref{fig:ising_2d_corr_fourier}(b),
the DMD prediction (dashed bold line)
of the absolute value of the Fourier-transformed correlation function
is in good agreement with the exact result (thin solid line)
for all the frequencies $\omega$.
When the data are normalized by
$\max_{\omega>0} |\tilde{f}^{\rm exact}(\omega)|$
for all frequencies $\omega$,
the
relative
difference between the exact result and the DMD prediction
is at most $25\%$
and is nearly less than $5\%$ for most of the frequencies $\omega$
as shown in Fig.~\ref{fig:ising_2d_corr_fourier}(c).
Although the difference becomes larger for peaks with smaller exact intensities
as shown in Fig.~\ref{fig:ising_2d_corr_fourier}(d),
the peak positions are exactly reproduced
when the peak intensity is larger than $10\%$
of the maximum value of the exact $\tilde{f}(\omega)$
(see Appendix~\ref{sec:fourier_log} for more detailed comparisons).
In this sense,
the DMD is a more powerful method than the GPR method
at least in the prediction of the time evolution.

\subsection{Time-dependent correlation functions with a power-law decay}
\label{subsec:corr_damp}

Next, we focus on a more challenging case
where the correlation functions exhibit
a power-law decay arising from the long-distance long-time quantum entanglement at a critical point of phase
transition~\cite{perales2008}.
We consider the transverse-field Ising model
in Eq.~\eqref{eq:ham_ising}
on a chain with an infinite system size.
We prepare the initial state
$|\psi_0'\rangle$
as the ground state of the transverse-field Ising model
at the critical point
$\Gamma/J = \Gamma_{\rm c}^{\rm 1D}/J = 0.5$~\cite{pfeuty1970}.
The transverse unequal-time (time displaced) spin-spin correlation functions
at distance $r$ and time displacement $t$ are defined by
\begin{align}
 \label{eq:ising_1d_corr}
 C^{xx}_{\rm uneq}(r, t)
 =
 \langle \psi_0' |
 S^x_{0}(0) S^x_{r}(t)
 |\psi_0'\rangle
\end{align}
with $S^x_{r}(t) = e^{iHt} S^x_{r} e^{-iHt}$.
The
1D
transverse-field Ising model
is integrable.
The exact correlation functions can be calculated
after the Jordan-Wigner transformation
from spin operators to fermion operators~\cite{mccoy1971}
and, at $\Gamma = \Gamma_{\rm c}^{\rm 1D}$,
they are given by
\begin{align}
 \label{eq:ising_1d_corr_exact}
 C^{xx}_{\rm uneq}(r, t)
 &=
 \frac{1}{\pi^2}
 +
 \frac{1}{4}
 \left[
 J_{2r}(2\Gamma t) + iE_{2r}(2\Gamma t)
 \right]^2
\nonumber
\\
 &~\phantom{=}~
 - \frac{1}{4}
 \left[
 J_{2r-1}(2\Gamma t) + iE_{2r-1}(2\Gamma t)
 \right]
\nonumber
\\
 &~\phantom{==}~
 \times
 \left[
 J_{2r+1}(2\Gamma t) + iE_{2r+1}(2\Gamma t)
 \right],
\end{align}
where
\begin{align}
 \label{eq:ising_1d_corr_exact_j}
 J_r(x) = \frac{1}{\pi} \int_{0}^{\pi}
 \cos(r\theta - x\sin\theta)
 \,d\theta
\end{align}
is the Bessel function of $r$th order
and
\begin{align}
 \label{eq:ising_1d_corr_exact_e}
 E_r(x) = \frac{1}{\pi} \int_{0}^{\pi}
 \sin(r\theta - x\sin\theta)
 \,d\theta
\end{align}
is the related Anger-Weber or Lommel-Weber
function~\cite{tommet1975,muller1984}.

\begin{figure}[t!]
\centering
\includegraphics[width=1.00\columnwidth]{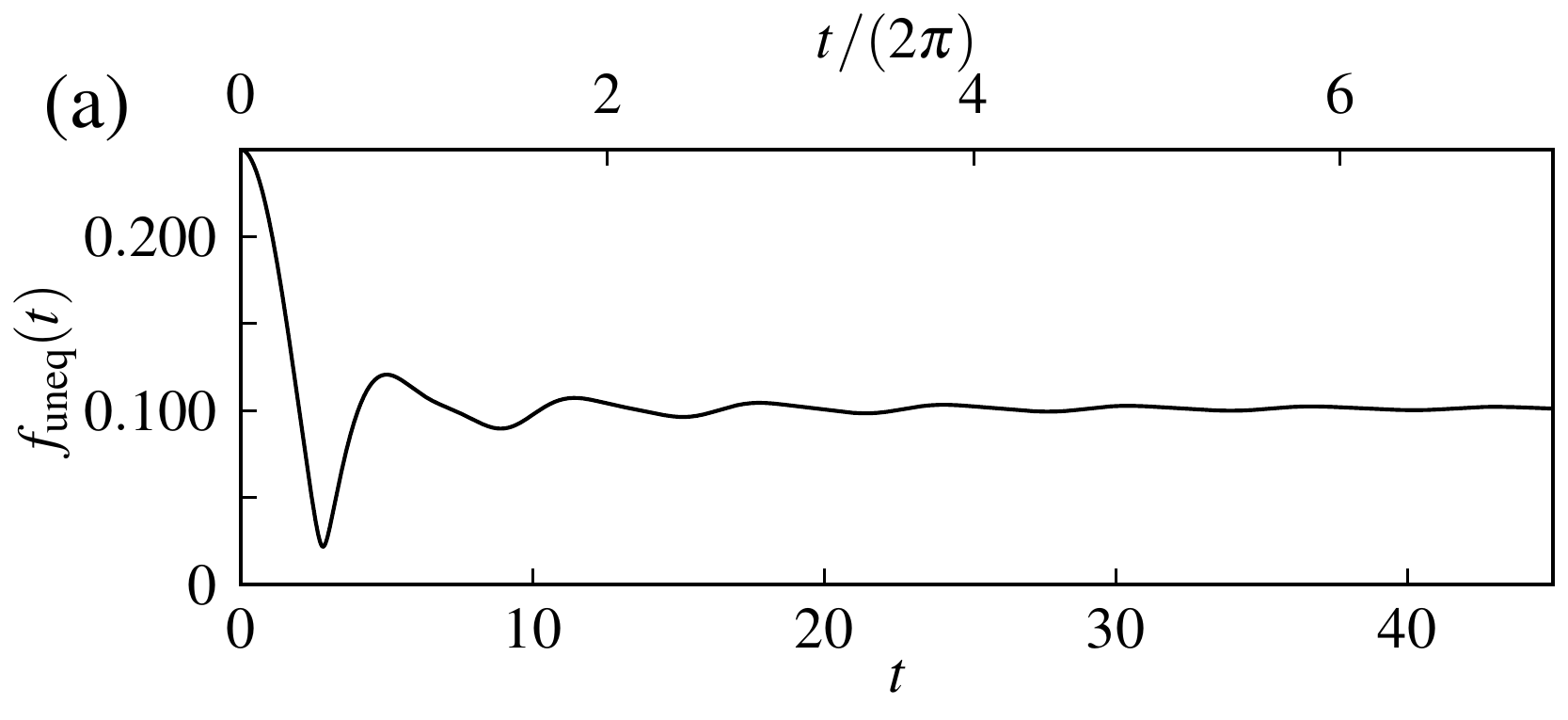}\\
\includegraphics[width=1.00\columnwidth]{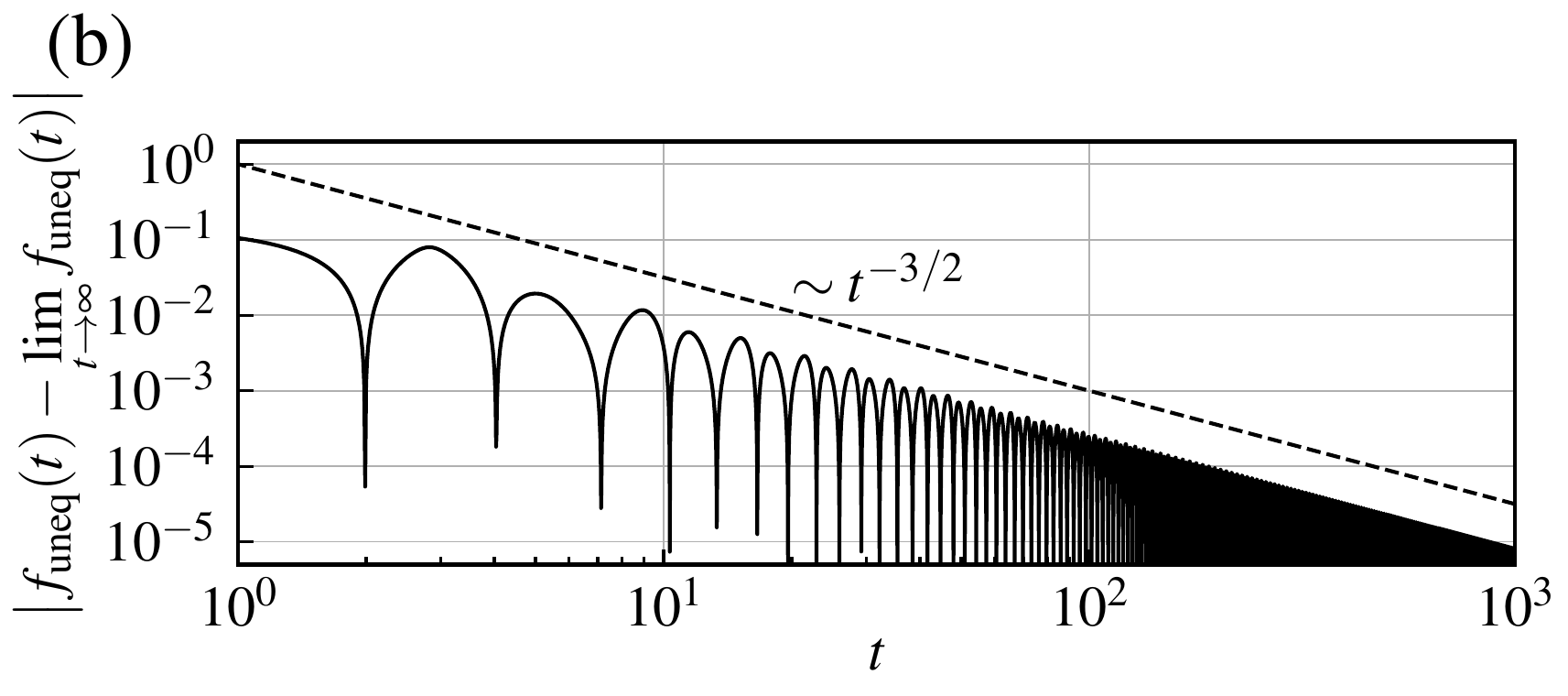}
\caption{%
Exact time evolution of the absolute value of the
unequal-time onsite transverse spin-spin correlation function
in the
1D
transverse-field Ising model
on an infinite chain.
}
\label{fig:ising_1d_corr}
\end{figure}

\begin{figure}[t!]
\centering
\includegraphics[width=.525\columnwidth]{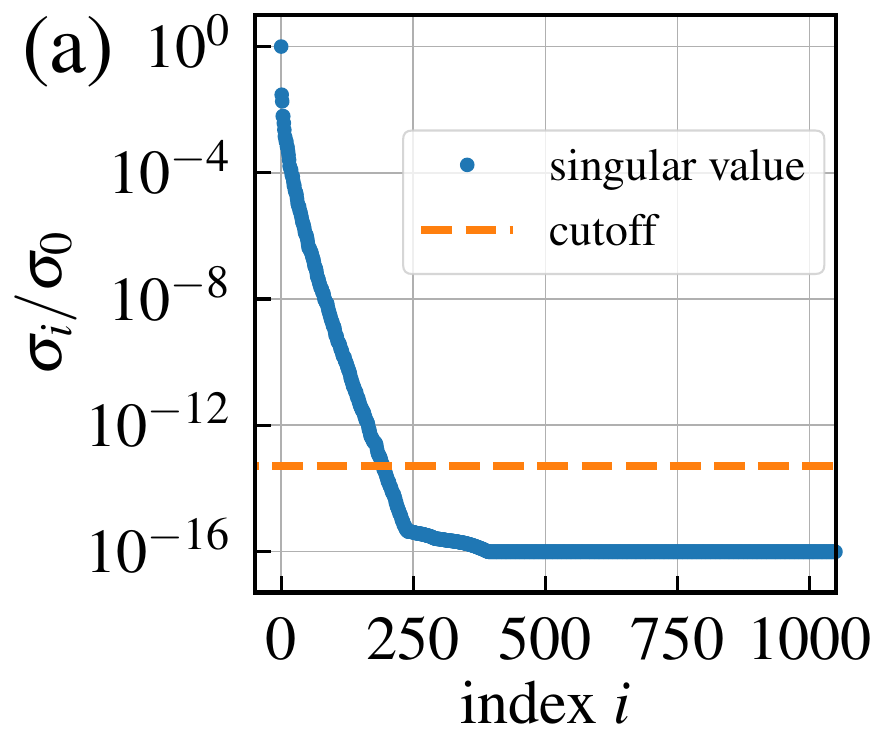}
\hfil
\includegraphics[width=.455\columnwidth]{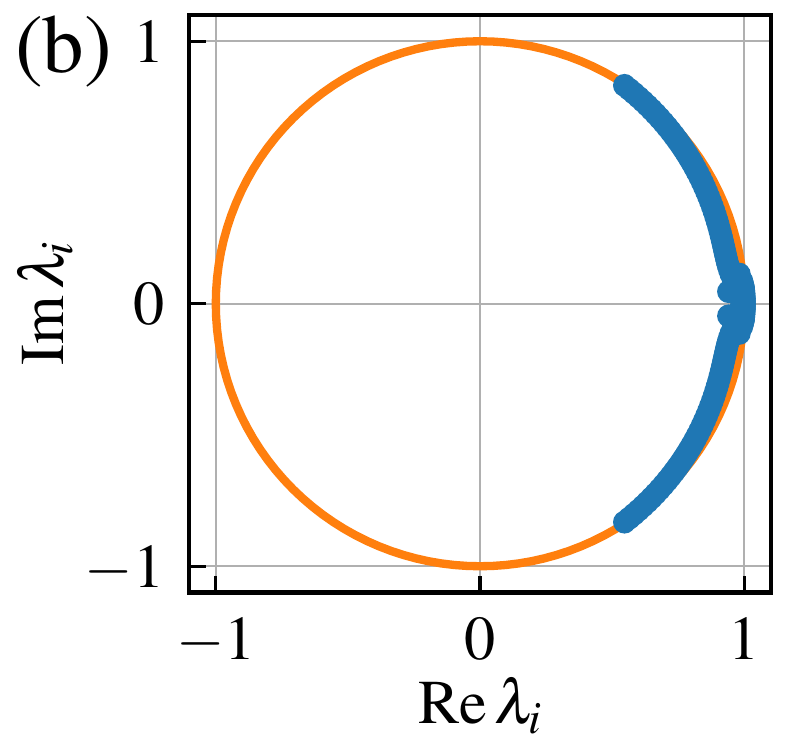}
\caption{%
DMD parameters in the case of
the unequal-time onsite transverse spin-spin correlation function
for the 1D transverse-field Ising model
on an infinite chain.
(a) Singular values $\sigma_i$ of the truncated SVD of the matrix $X_0$.
(b) Eigenvalues $\lambda_i$ of the matrix $\tilde{A}$.
}
\label{fig:ising_1d_corr_lmd_sgm}
\end{figure}

We show the exact time evolution of
the absolute value of the
unequal-time onsite ($r=0$) correlation function,
i.e.,
\begin{align}
\label{eq:def_ft_1d}
 f_{\rm uneq}(t) = |C^{xx}_{\rm uneq}(r=0, t)|
\end{align}
in Fig.~\ref{fig:ising_1d_corr}.
The correlation function exhibits oscillatory behavior
with a power-law decay~\cite{muller1984,rossini2009,rossini2010}
and converges to a nonzero value
corresponding to the squared transverse magnetization
$m_x^2 = 1/\pi^2$ at $\Gamma = \Gamma_{\rm c}^{\rm 1D}$
in the infinite-time limit~\cite{rossini2010}.
The long-time asymptotic envelope function is scaled as
\begin{align}
 \label{eq:ising_1d_corr_exact_asym}
 \left| \,
 \left| C^{xx}_{\rm uneq}(r=0, t) \right| - m_x^2
 \right|
 \sim
 t^{-3/2}
\end{align}
for $t \gg 1$~\cite{muller1984}.
The period of the dominant oscillation is
$T \approx \pi/\Gamma = 2\pi/J$.

\begin{figure}[t!]
\centering
\includegraphics[width=1.00\columnwidth]{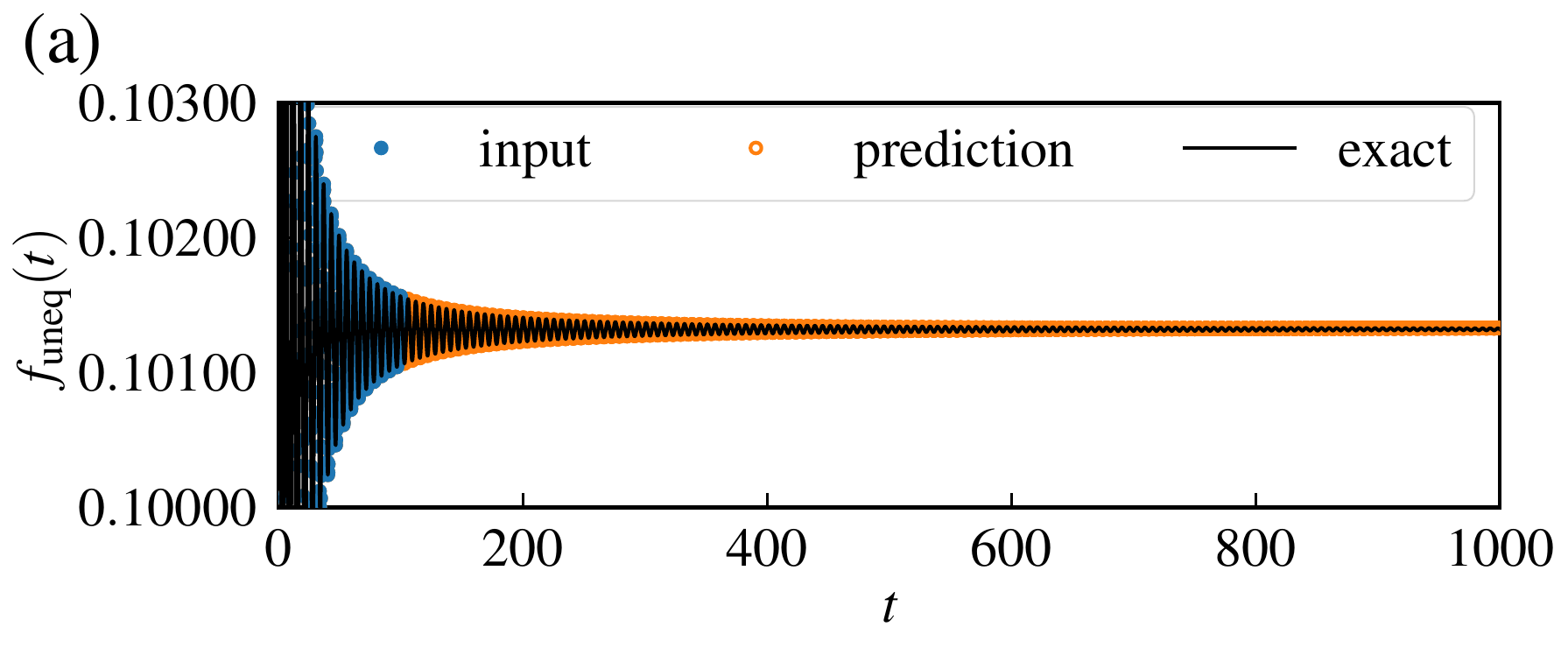}\\
\includegraphics[width=1.00\columnwidth]{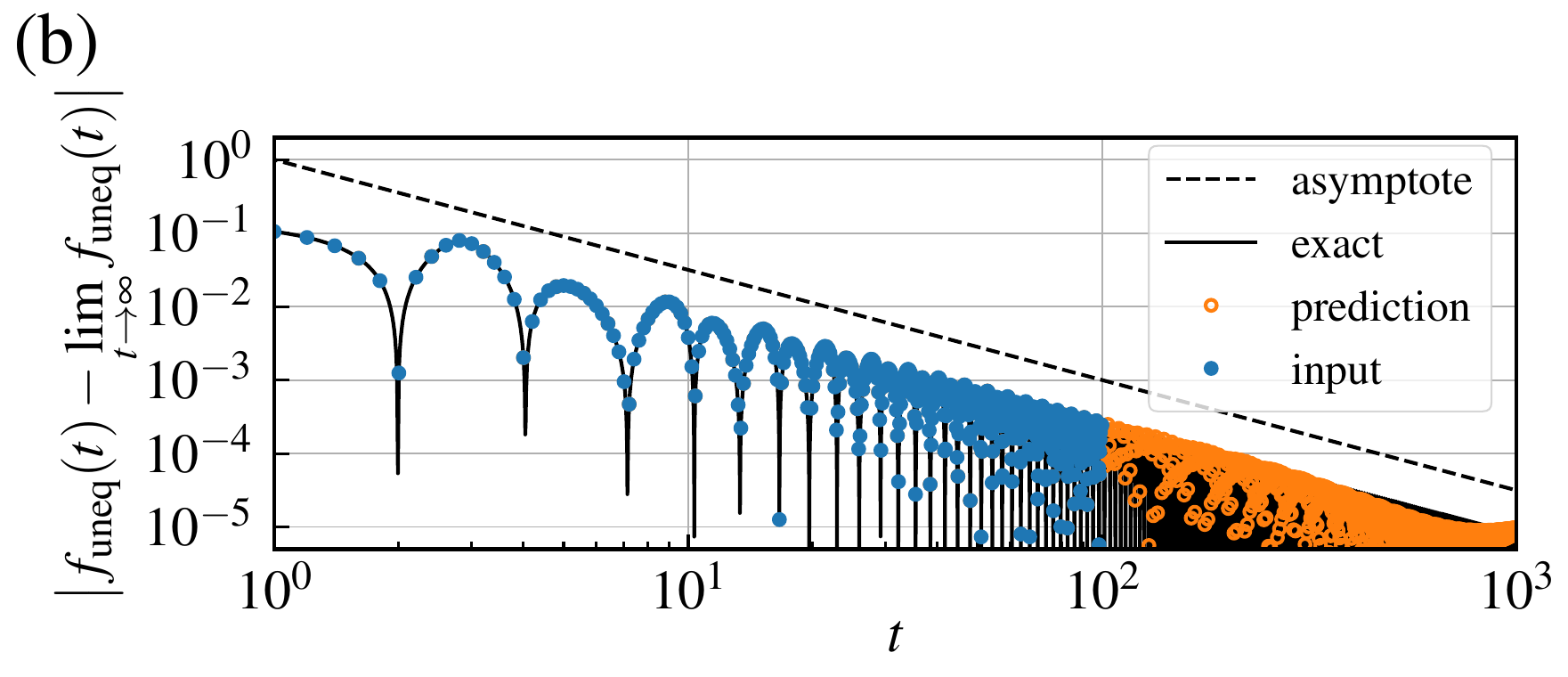}\\
\includegraphics[width=1.00\columnwidth]{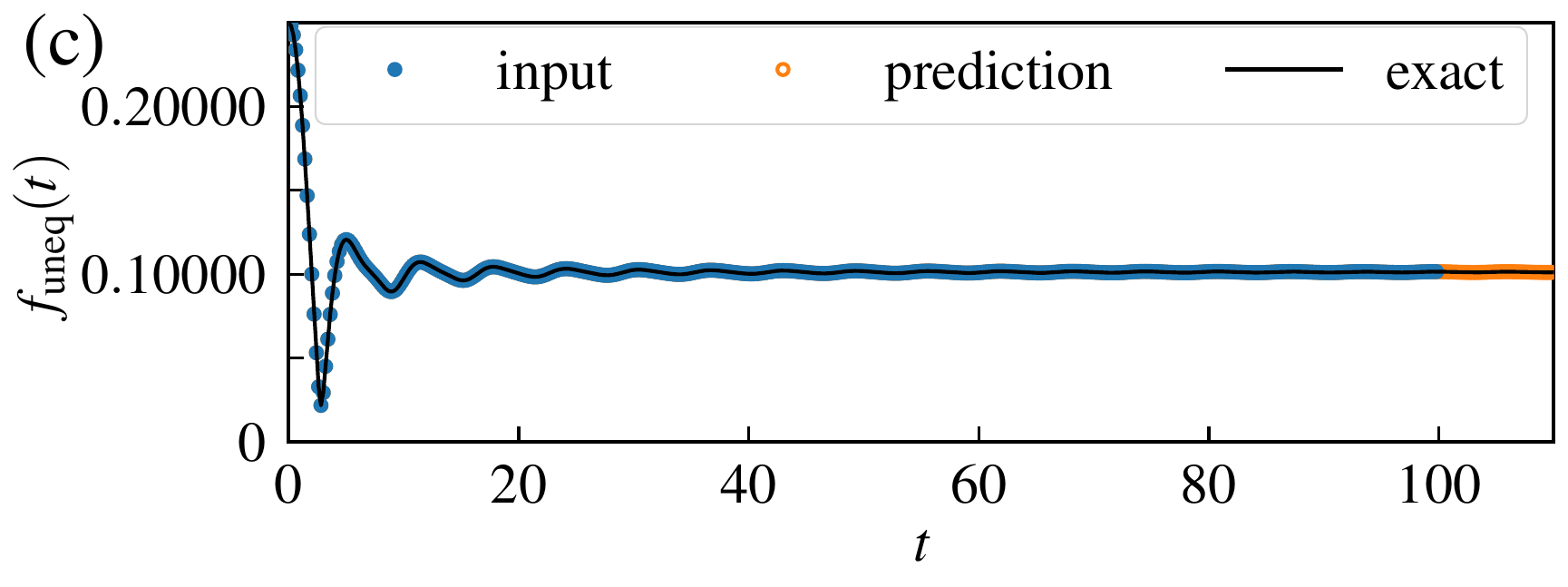}\\
\includegraphics[width=1.00\columnwidth]{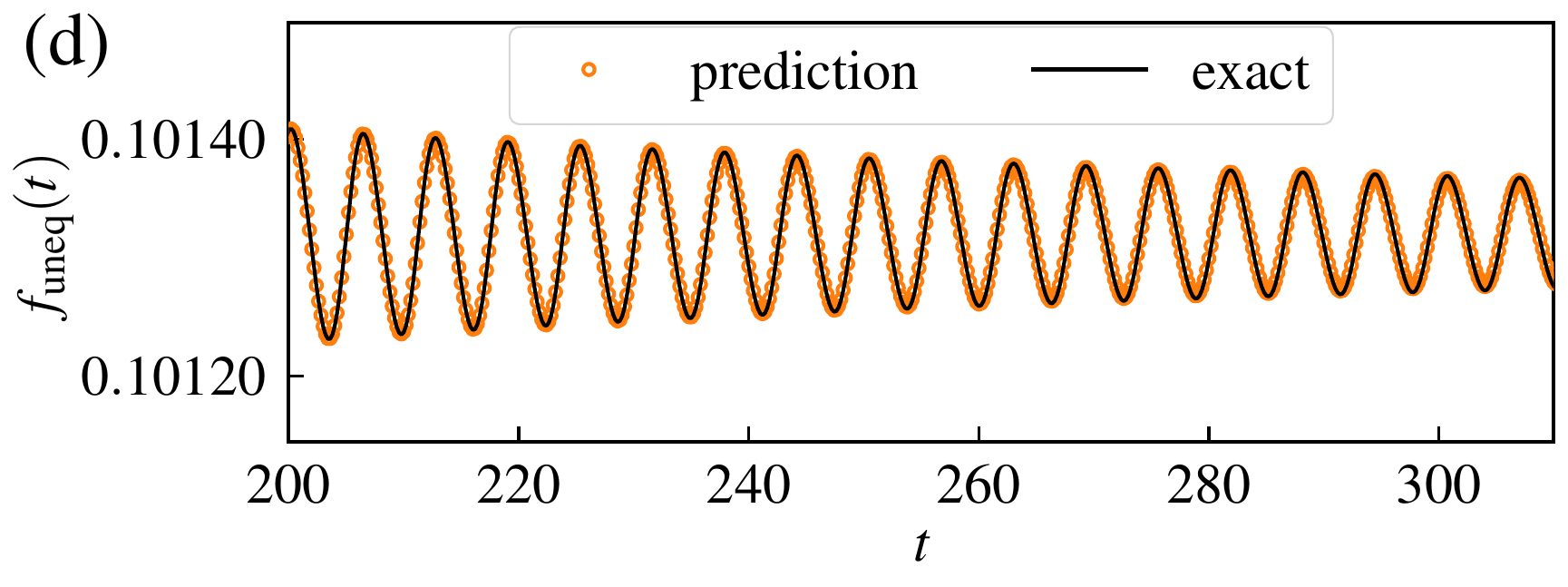}\\
\includegraphics[width=1.00\columnwidth]{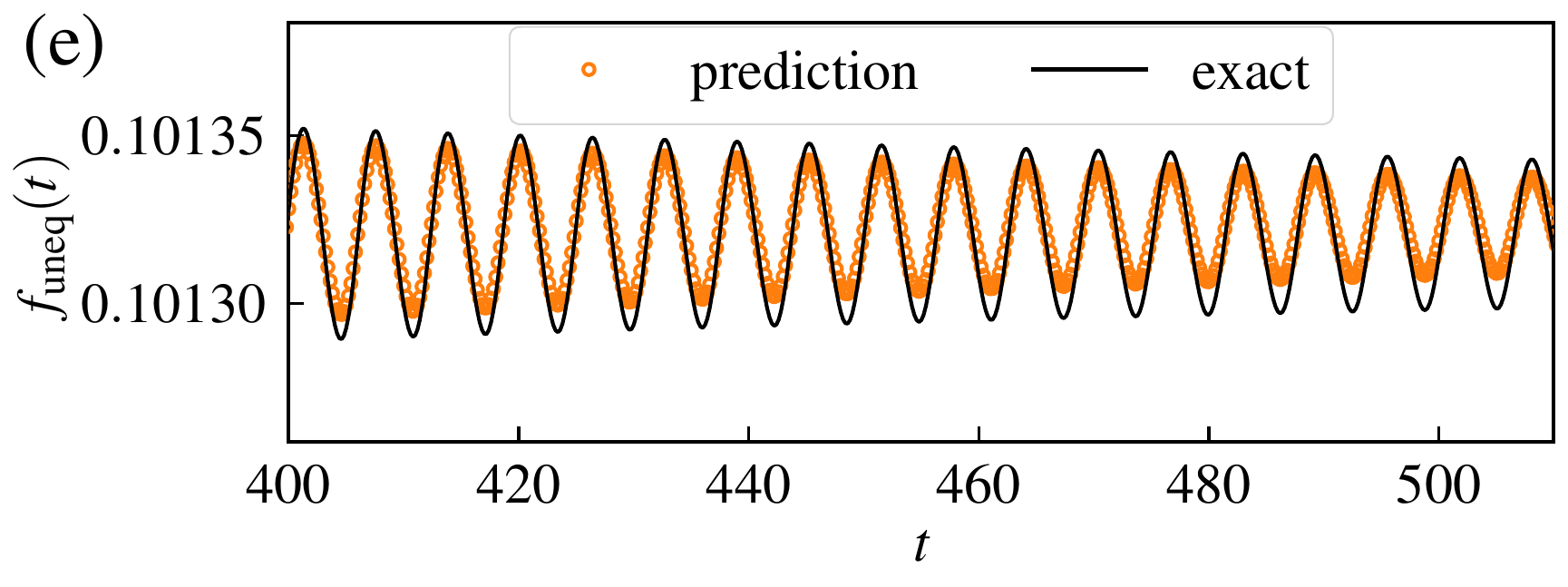}
\caption{%
DMD prediction of the unequal-time onsite transverse spin-spin correlation
function
in the 1D transverse-field Ising model.
We show the time evolution of
the absolute value of the correlation function for
(a) $t \in [0, 1000)$
and
(b) that in the logarithmic scale.
The magnified views of the time evolution for
(c) $t \in [0, 110]$,
(d) $t \in [200, 310]$,
and
(e) $t \in [400, 510]$
are also shown.
The solid line is the exact result,
the filled circles are the input data,
and the open circles are the predicted data.
The data points $f_{{\rm uneq},n}=f_{\rm uneq}(n\cdot\Delta t)=f_{\rm uneq}(t)$ are plotted
only when $n$ is a multiple of $20$.
}
\label{fig:ising_1d_corr_dmd}
\end{figure}

\begin{figure}[t!]
\centering
\includegraphics[width=1.00\columnwidth]{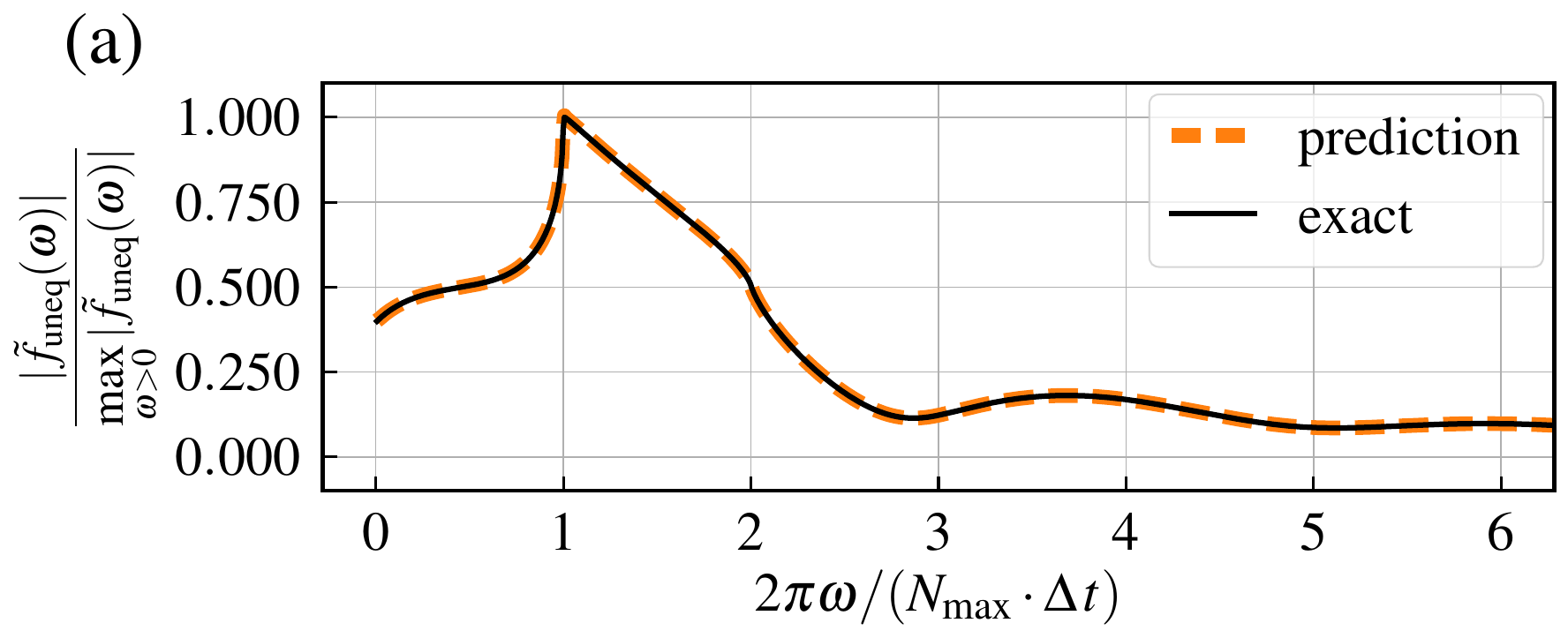}\\
\includegraphics[width=1.00\columnwidth]{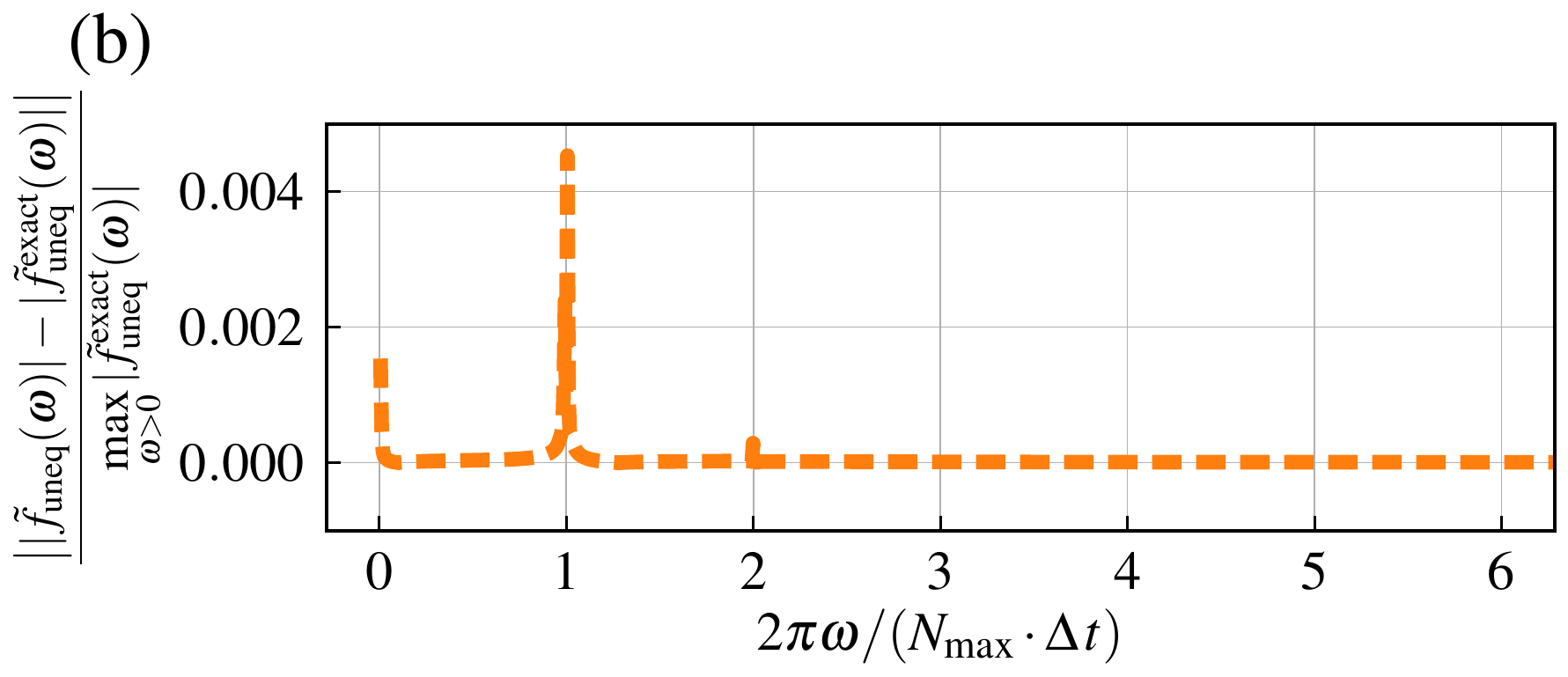}
\caption{%
Fourier transform of the unequal-time onsite transverse spin-spin correlation
function
in the 1D transverse-field Ising model.
We remove a divergently large value at $\omega = 0$.
(a) Comparison of the exact result (solid line)
and the DMD prediction (dashed line)
for the absolute value of the Fourier-transformed correlation function.
(b) Relative difference
between the exact result and the DMD prediction.
}
\label{fig:ising_1d_corr_fourier}
\end{figure}

For the DMD,
we choose $\Delta t = 0.01$,
$M = 5000$, $N = 2M$,
and $N_{\rm max} = 100000$.
The calculated singular values $\sigma_i$ decay exponentially
as a function of the index $i$,
as shown in Fig.~\ref{fig:ising_1d_corr_lmd_sgm}(a).
In the case of the
1D
transverse-field Ising model,
we numerically find that
the DMD prediction is rather stable
even when the cutoff $\epsilon$ is smaller than $0.01$,
which is the value for the
2D
transverse-field Ising case.
This observation suggests that
the number of relevant eigenmodes in the case with damping
is larger than that in the case without damping,
and many eigenmodes are responsible for the dynamics
exhibiting a power-law decay.
Therefore, the cutoff is chosen to be
a very small value $\epsilon = 5\times 10^{-14}$,
and the rank $R$ of the truncated SVD becomes $R = 193$.
Even with such a small cutoff,
the absolute values of the calculated eigenvalues
$|\lambda_i|$ of the matrix $\tilde{A}$
are smaller than unity
[see Fig.~\ref{fig:ising_1d_corr_lmd_sgm}(b)],
indicating that the dynamics obtained by the DMD is
stable.
Note that the application of the DMD
to an infinite-size system does not lead to a divergent increase
in the number of dominant eigenmodes required for the time-series prediction.

We show the time evolution of the
predicted correlation function
for $t \in [0, 1000)$ in Fig.~\ref{fig:ising_1d_corr_dmd}(a)
and that in the logarithmic scale in Fig.~\ref{fig:ising_1d_corr_dmd}(b).
The predicted correlation function (open circles)
appears to converge to a nonzero value in the very long-time limit,
which is consistent with the exact result (solid line).
We also show the magnified time evolution of the
predicted correlation function
for $t \in [0,110]$,
$t \in [200,310]$,
and $t \in [400,510]$
in Figs.~\ref{fig:ising_1d_corr_dmd}(c)--\ref{fig:ising_1d_corr_dmd}(e).
For $t \lesssim 400$,
the DMD prediction
and the exact result
agree very
well.
The period of the dominant oscillation
and the convergent value in the infinite-time limit
are well reproduced by the DMD prediction.
On the other hand,
for $t \gtrsim 400$,
the amplitude of the oscillation
in the DMD prediction is slightly smaller than that in the exact result.
The deviation gets slightly larger as the time $t$ increases,
although the period and center of the oscillation
are still well reproduced by the DMD prediction.
We also examine the asymptotic behavior of
the predicted correlation function
[see the logarithmic plot in Fig.~\ref{fig:ising_1d_corr_dmd}(b)].
For $t \lesssim 700$,
The DMD prediction decays as a power law
with the exponent close to $-3/2$,
which is the same as the exact result.
When the eigenmodes are sufficiently included in the DMD procedure,
the DMD prediction
nicely reproduces the exact result
for a very long time,
which is more than five times the duration of the input time $t_{\rm input}$.

Note that the DMD prediction gets slightly better
when the origin of the time series is shifted
to later times.
This is because we can neglect the initial transient behavior
that does not follow the power-law decay in a strict sense.
The effect of the shift of the origin of the time series
will be discussed in Appendix~\ref{sec:shift_origin}.

The comparison of the Fourier transform of the
predicted correlation function
and the exact result
is shown in Fig.~\ref{fig:ising_1d_corr_fourier}(a).
Because the $\omega=0$ component of the Fourier transform
is divergently large,
we remove the value at $\omega=0$.
The relative difference between the exact result and the DMD prediction
is less than $0.4\%$
over all frequencies
[see Fig.~\ref{fig:ising_1d_corr_fourier}(b)].

\subsection{Effects of noise}
\label{subsec:effect_noise}

\begin{figure}[t!]
\centering
\includegraphics[width=1.00\columnwidth]{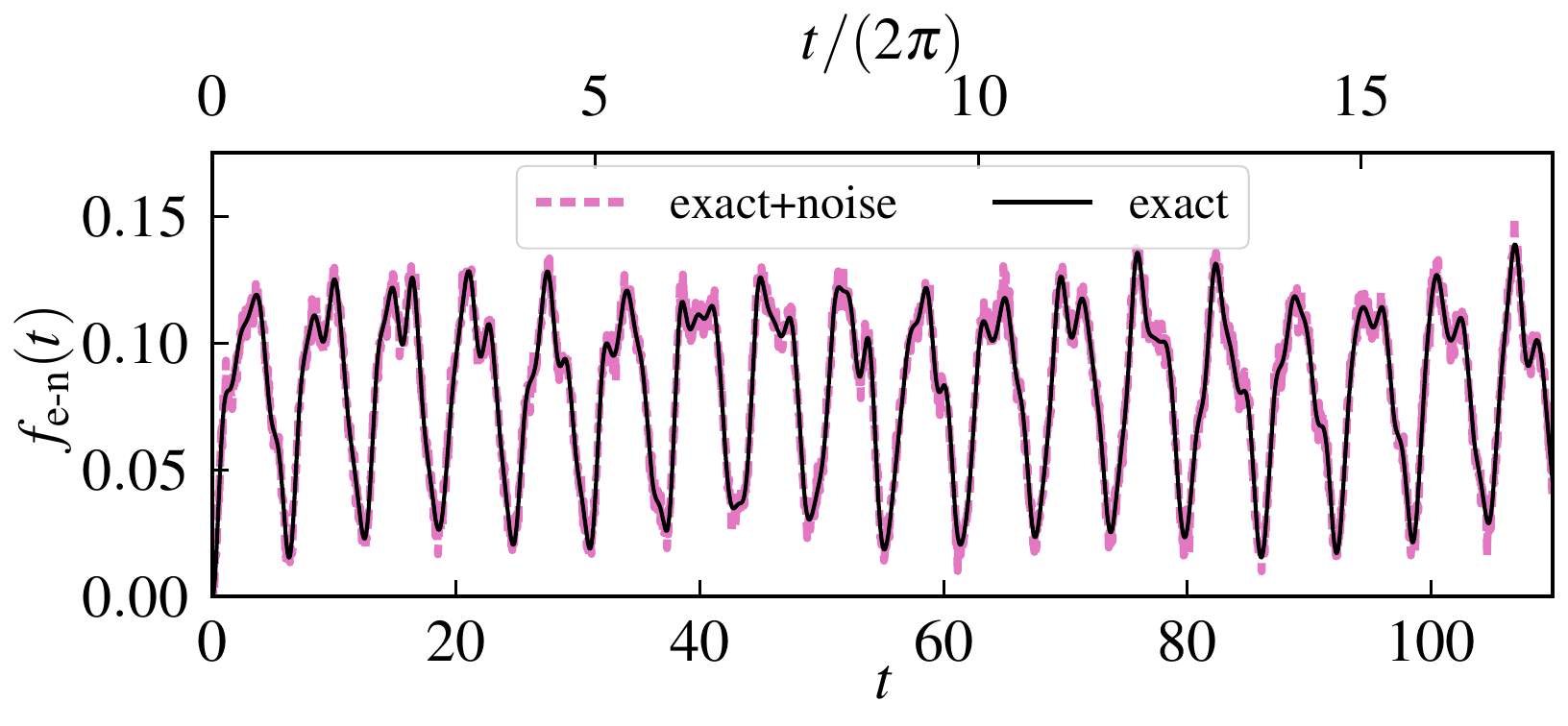}
\caption{%
Time evolution of the equal-time longitudinal spin-spin correlation function
after a sudden quench in the
2D   
transverse-field Ising model
on a finite-size square lattice
with additive white Gaussian noise
(dashed pink line).
As a reference, the time evolution without noise is also shown
(solid black line).
}
\label{fig:ising_2d_corr_noise}
\end{figure}

\begin{figure}[t!]
\centering
\includegraphics[width=.525\columnwidth]{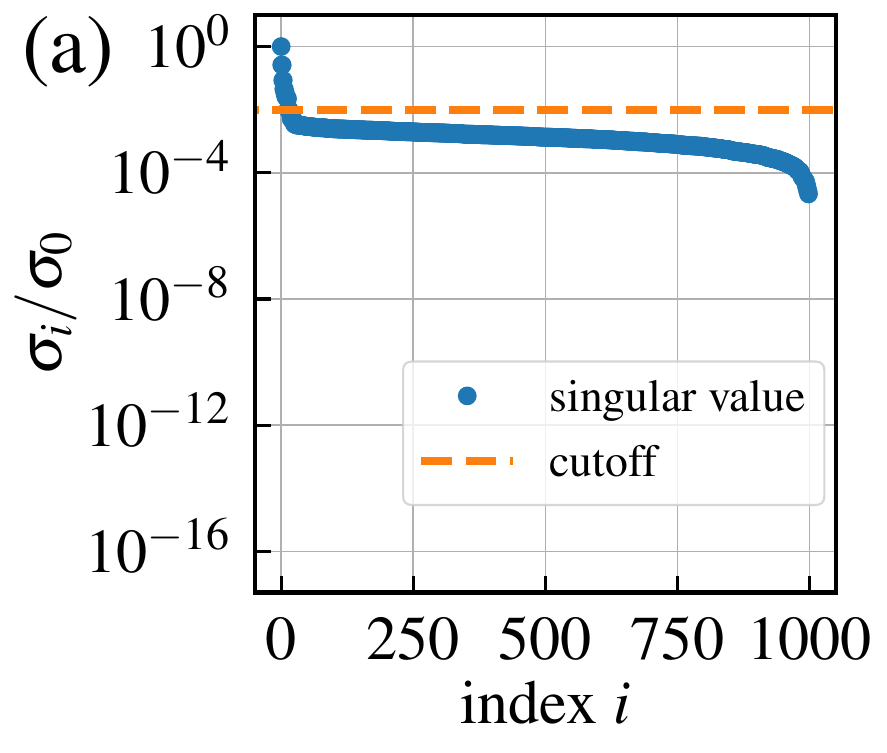}
\hfil
\includegraphics[width=.455\columnwidth]{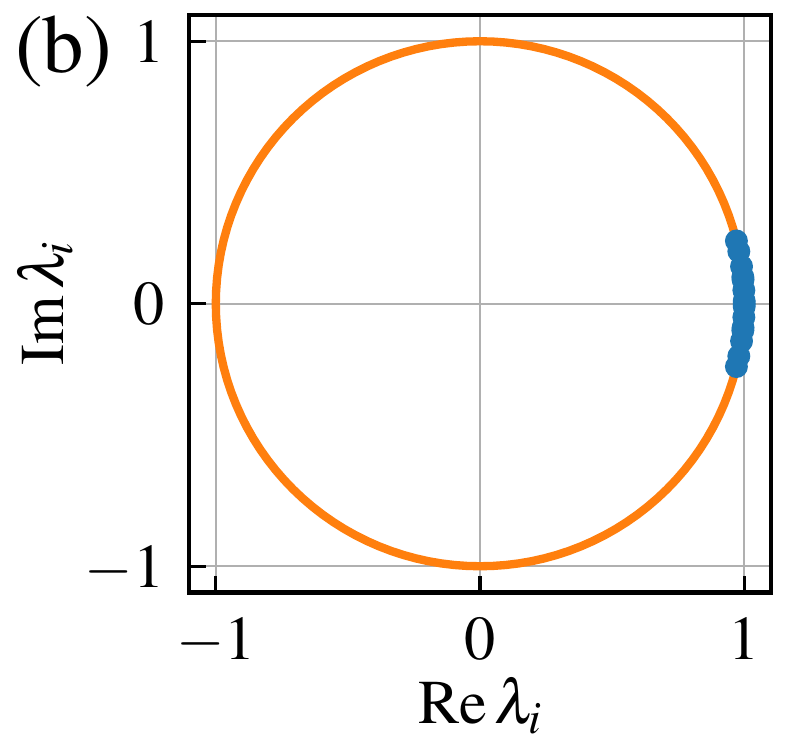}
\caption{%
DMD parameters in the case of
the equal-time longitudinal spin-spin correlation function
after a sudden quench in the
2D
transverse-field Ising model
on a square lattice
with additive white Gaussian noise.
(a) Singular values $\sigma_i$ of the truncated SVD of the matrix $X_0$.
Singular values with $\sigma_i/\sigma_0 \lesssim 10^{-3}$
as a function of index $i$ exhibit wide plateaulike behavior.
The cutoff is chosen to be larger than the value at the plateau.
(b) Eigenvalues $\lambda_i$ of the matrix $\tilde{A}$.
}
\label{fig:ising_2d_corr_lmd_sgm_noise}
\end{figure}

\begin{figure}[t!]
\centering
\includegraphics[width=1.00\columnwidth]{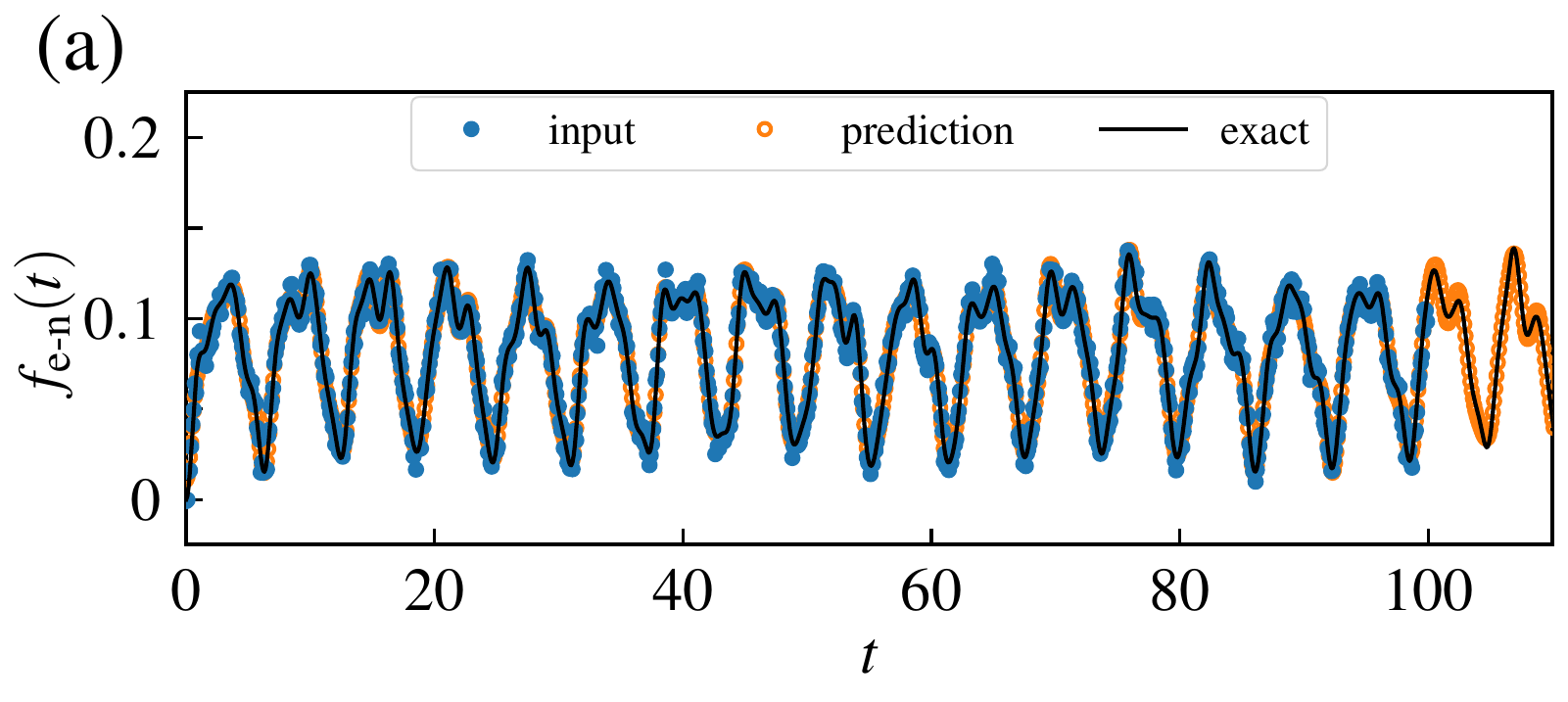}\\
\includegraphics[width=1.00\columnwidth]{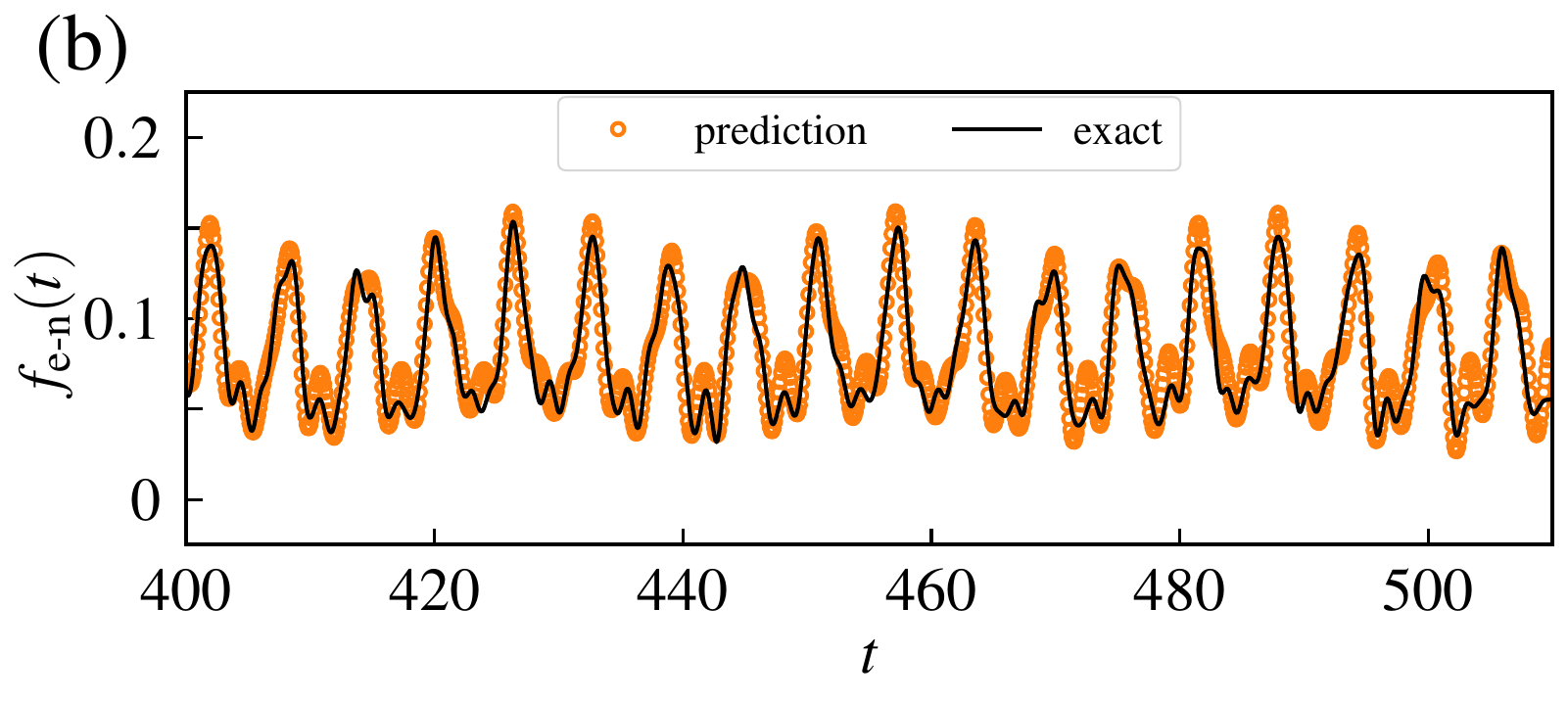}\\
\includegraphics[width=1.00\columnwidth]{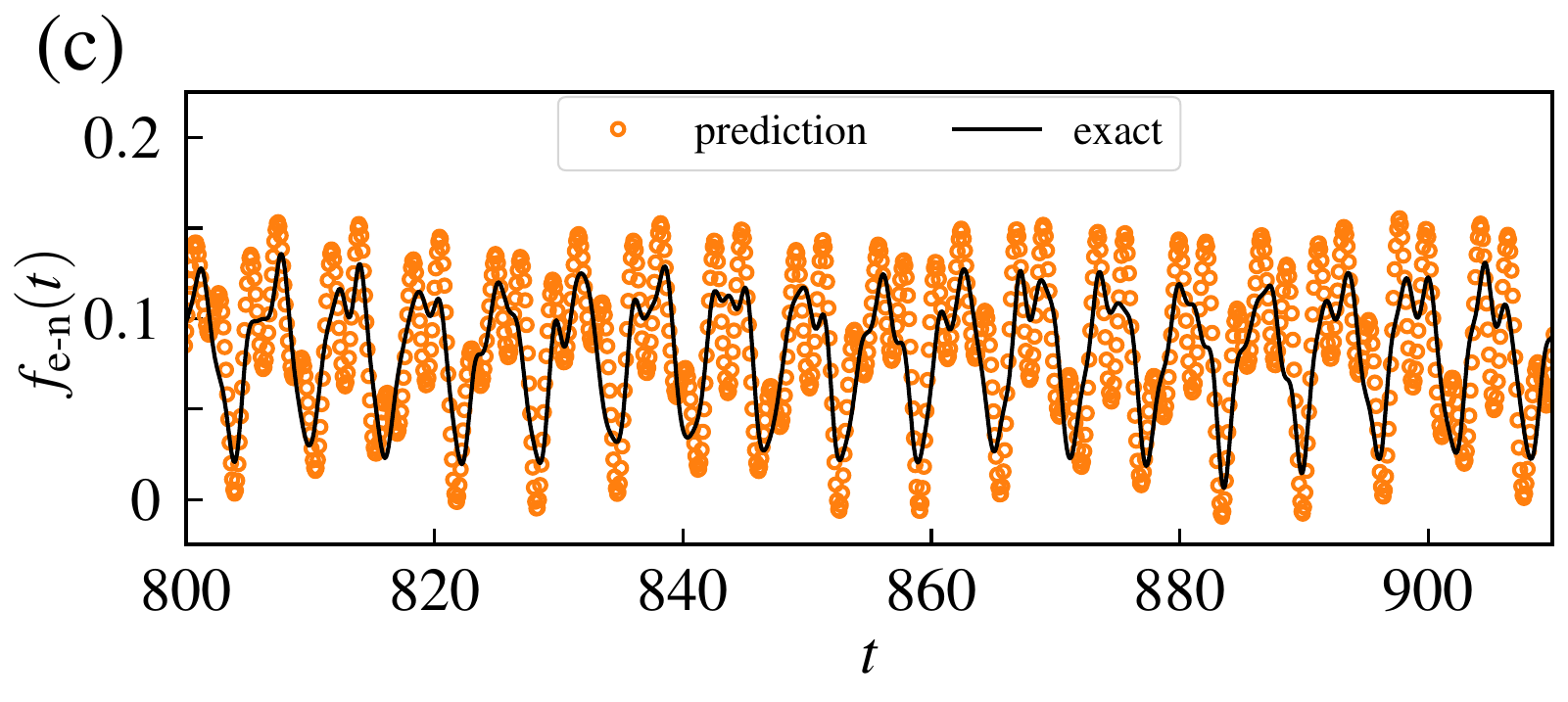}
\caption{%
DMD prediction of the equal-time longitudinal spin-spin correlation function
in the 2D transverse-field Ising model
with additive white Gaussian noise.
We show the time evolution of the correlation function for
(a) $t \in [0,110]$,
(b) $t \in [400,510]$,
and
(c) $t \in [800,910]$.
The solid line is the exact result,
the filled blue circles are the input data,
and the open orange circles are the predicted data.
The data points $f_{{\rm e\text{-}n},n}=f_{\rm e\text{-}n}(n\cdot\Delta t)=f_{\rm e\text{-}n}(t)$ are plotted
only when $n$ is an even number.
}
\label{fig:ising_2d_corr_dmd_noise}
\end{figure}

\begin{figure}[t!]
\centering
\includegraphics[width=1.00\columnwidth]{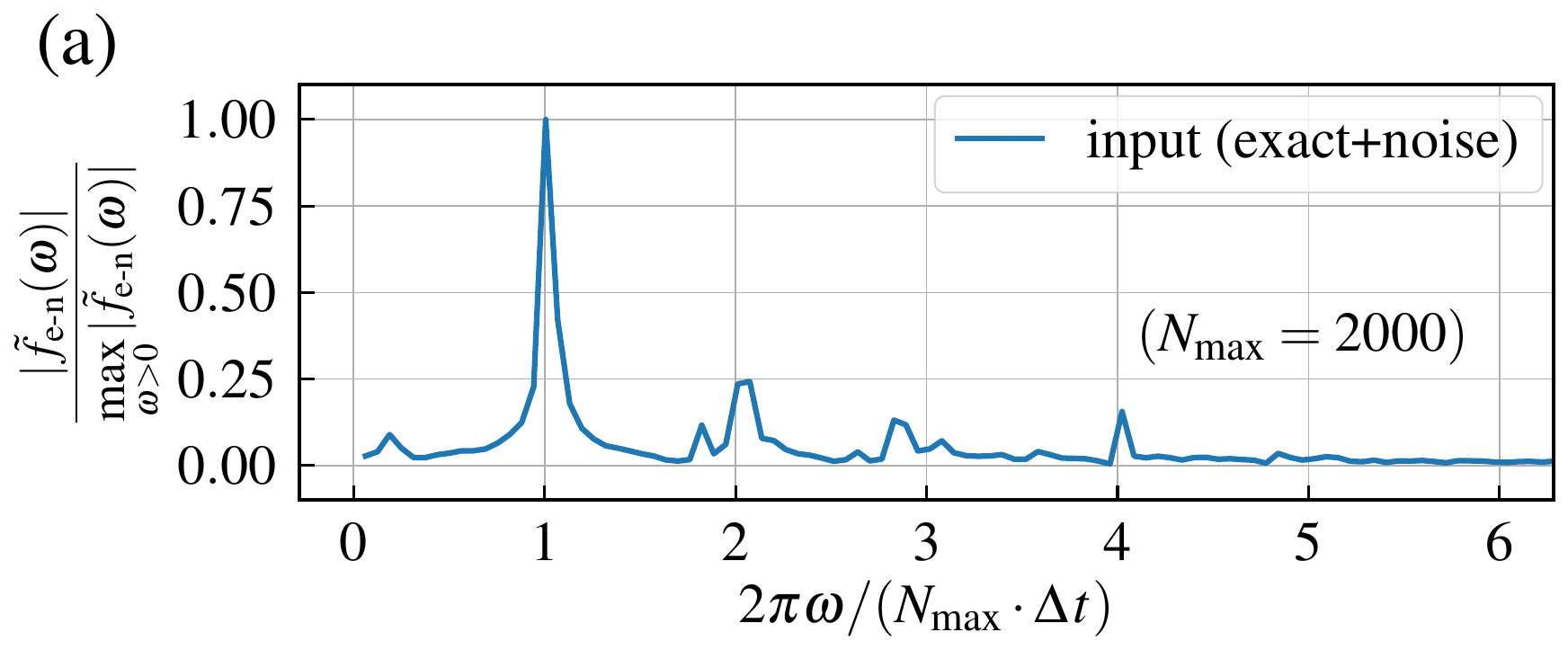}\\
\includegraphics[width=1.00\columnwidth]{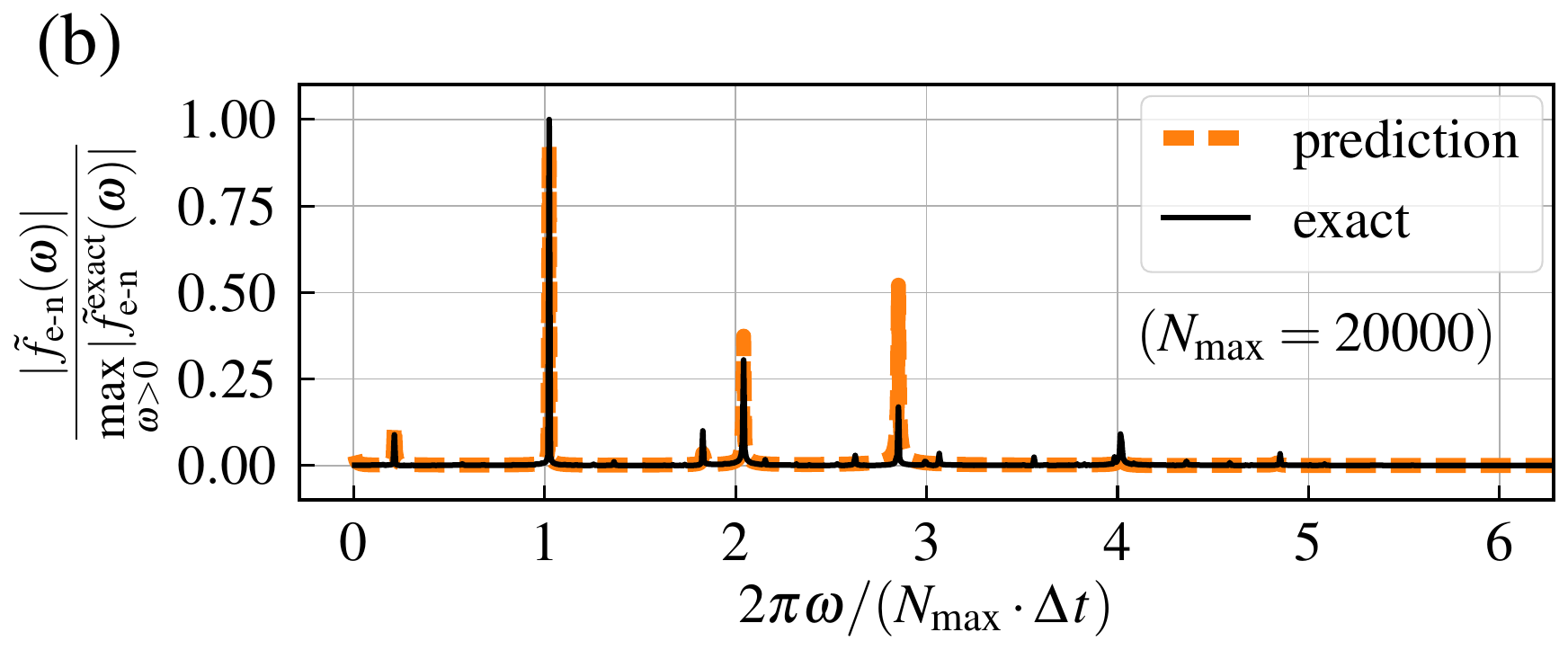}\\
\includegraphics[width=1.00\columnwidth]{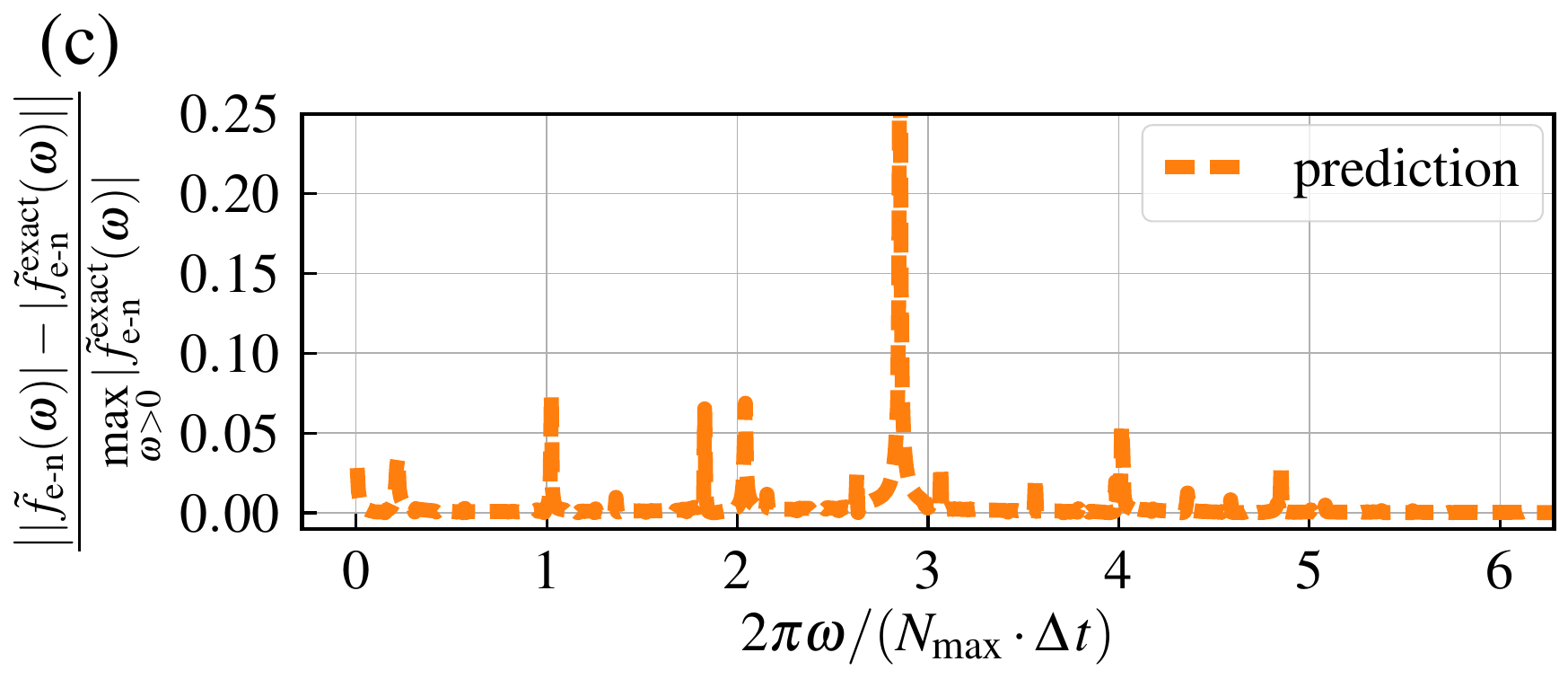}\\
\includegraphics[width=1.00\columnwidth]{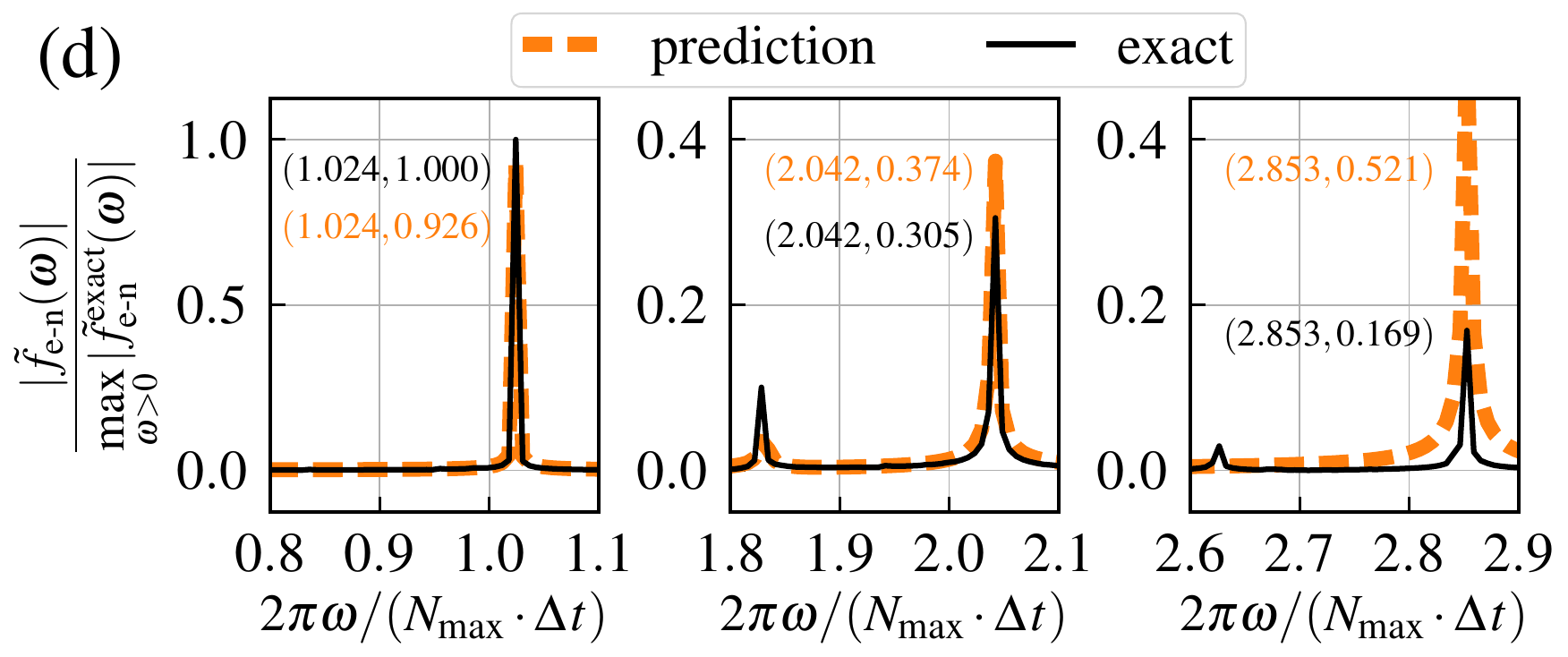}
\caption{%
Fourier transform of the equal-time longitudinal spin-spin correlation function
in the 2D transverse-field Ising model
with additive white Gaussian noise.
We remove a large value at $\omega = 0$.
(a) Exact result when $t \in [0,100)$ ($N_{\rm max} = 2000$).
(b) Comparison of the exact result (solid line)
and the DMD prediction (dashed line)
for the absolute value of the Fourier-transformed correlation function
when $t \in [0,1000)$ ($N_{\rm max} = 20000$).
(c) Relative difference
between the exact result and the DMD prediction.
(d) Magnified views of Fig.~\ref{fig:ising_2d_corr_fourier_noise}(b).
The position and intensity of each peak
are also shown in the figure in the order (position, intensity).
All the positions of the peaks of the DMD prediction
coincide with those of the exact result.
}
\label{fig:ising_2d_corr_fourier_noise}
\end{figure}

We investigate how noise in input data
affects the prediction by the DMD method.
In experiments, time-series data are always affected by noise.
In numerical simulations based on the Monte Carlo method,
we often need to average over many samples to obtain the time-series
data; consequently, the data inevitably have statistical errors.
Therefore, examination of the effects of noise is important
for practical applications of the DMD method.

In this section,
we consider time-series data $f_{\rm n}(t)$ affected by noise $\eta(t)$,
which is given by Gaussian white random variables, namely, Gaussian noise with zero mean and variance
$\sigma_{\rm noise}^2$ without correlations between different time steps.
We will demonstrate that the DMD method withstands noise
when the noise level is 
moderate and not too high by taking the time-dependent correlation function
in the transverse-field Ising model
as an example.
The standard deviation $\sigma_{\rm noise}$ is assumed to be
substantially smaller
than the amplitude of original $f_{\rm orig}(t)$ obtained by the evolution without noise
and is defined as
\begin{align}
 \sigma_{\rm noise}
 =
 \epsilon_{\rm noise}
 \max_{t\in [0,(N-1)\cdot\Delta t]} |f_{\rm orig}(t)|,
\end{align}
where $\epsilon_{\rm noise}$ is a small positive parameter
that characterizes the strength of noise.

Note, in general, that
the noise is not necessarily given by Gaussian white random variables.
For example,
if the short-time input data are generated by numerical simulations
that accumulate errors in each time step,
the noise would be correlated in time.
Such small systematic errors might be amplified over long timescales
and would lead to a deviation from the exact result
when the DMD prediction is performed.
To avoid such a situation,
we aim to apply a method of data generation that,
despite its high computational cost,
provides not only high precision but also systematically improvable
short-time data.
One can estimate the noise level at long timescales
by monitoring the effect of improved accuracy of short-time data
obtained by the direct simulation or measurement.
The DMD would apply to more practical data,
such as those obtained in the hardware simulations on IBM 127-qubit devices,
where the level of systematic error is controlled
to obtain multiple time-series data and then
extrapolated to the systematic-error-free limit for data prediction~\cite{kim2023}.
Hereafter, we focus on the case of Gaussian white noise
on top of the exact time-series data
obtainable by the exact diagonalization of the Hamiltonian
for a finite system
or by the analytical calculation
for an infinite-size integrable system.

For simplicity, as a
noiseless time-series data,
we choose the time-dependent correlation
function without damping
of the 2D transverse-field Ising model
in Sec.~\ref{subsec:corr_no_damp}.
(See also Appendix~\ref{sec:effect_noise_damping} for
the effects of noise on
the time-dependent correlation function with a power-law decay
in Sec.~\ref{subsec:corr_damp}.)
Then, the time-series data $f_{\rm e\text{-}n}(t)$ affected by noise $\eta(t)$ is given by
\begin{align}
 f_{\rm e\text{-}n}(t) = C^{zz}_{\rm eq}(|\bm{r}|=1, t) + \eta(t),
\end{align}
where $C^{zz}_{\rm eq}(|\bm{r}|=1, t)$ is defined in
Eq.~\eqref{eq:ising_2d_corr}.
As for this input data,
we have numerically confirmed that the prediction
up to $N_{\rm max} = 20000$ by the DMD method withstands
against the noise when $\epsilon_{\rm noise} \lesssim 0.03$.
Hereafter, we set $\epsilon_{\rm noise} = 0.03$
and compare the differences of the DMD predictions
with and without noise.
The corresponding input data $f_{\rm e\text{-}n}(t)$ with noise is shown in
Fig.~\ref{fig:ising_2d_corr_noise}.

For the DMD,
we choose the same parameters as those in Sec.~\ref{subsec:corr_no_damp}.
In the presence of noise,
the calculated singular values $\sigma_i$ of the matrix $X_0$
drops rapidly
for a small index $i\lesssim 20$
and exhibit wide plateaulike behavior
for a large index $i\gtrsim 20$
[see Fig.~\ref{fig:ising_2d_corr_lmd_sgm_noise}(a)].
The plateaulike behavior is caused by the noise,
and the typical ratio $\sigma_i/\sigma_0$
at the plateau is found to be proportional to
$\epsilon_{\rm noise}$
(see Appendix~\ref{sec:estimate_noise_level}).
Therefore, even if we do not know the noise level {\it a priori},
we can estimate the strength of noise
in the input data by finding the position of the plateau
in the plot of singular values
$\sigma_i/\sigma_0$ as a function of index $i$.

The cutoff is chosen to be larger than the value at the plateau
so that the irrelevant modes below the noise level are excluded.
We choose the cutoff $\epsilon = 0.01$,
which is the same as
the parameter in Sec.~\ref{subsec:corr_no_damp}.
With this choice,
the rank $R$ of the truncated SVD becomes $R=15$.
The predicted long-time dynamics would be stable
because the absolute values of the eigenvalues $|\lambda_i|$
of the matrix $\tilde{A}$ are smaller than
or equal to unity [see
Fig.~\ref{fig:ising_2d_corr_lmd_sgm_noise}(b)].

We show the selected time evolution of the
predicted correlation function
for $t \in [0,110]$,
$t \in [400,510]$,
and
$t \in [800,910]$
in Figs.~\ref{fig:ising_2d_corr_dmd_noise}(a)--\ref{fig:ising_2d_corr_dmd_noise}(c).
For $t \lesssim 500$,
the DMD prediction 
and the exact result 
are in good agreement,
just as in the case without noise
in Fig.~\ref{fig:ising_2d_corr_dmd}.
For $t \in [800,910]$,
the amplitude of the oscillation
in the DMD prediction gets larger than that in the exact result.
The extent of the deviation from the exact result
is also larger than that without noise.
On the other hand,
the period of the oscillation in the DMD prediction
remains the same as that in the exact result.
Therefore, we conclude that
the DMD prediction reproduces the exact result
for a long time,
which is still
longer than five times the duration of the noisy input data.

We then calculate the Fourier transform of the correlation function
in the presence of noise.
We examine how the DMD prediction reproduces the exact result
when using the input data with noise
in Fig.~\ref{fig:ising_2d_corr_fourier_noise}(a).

The DMD prediction reproduces the exact result
fairly well even in the presence of noise
[see Figs.~\ref{fig:ising_2d_corr_fourier_noise}(b)].
The dominant peaks in the Fourier spectrum
are located at the same positions as those in the exact result.
The relative difference is smaller than
$10\%$ for most of the frequencies $\omega$.
On the other hand,
the relative difference between the exact result and the DMD prediction
is larger than that without noise
[see Fig.~\ref{fig:ising_2d_corr_fourier_noise}(c)].
As the intensity of the peaks in the Fourier spectrum increases,
the relative difference between the exact result and the DMD prediction
becomes smaller
[see Fig.~\ref{fig:ising_2d_corr_fourier_noise}(d)].

\subsection{Error analysis}
\label{subsec:error}

In general,
the prediction of short-time dynamics
is more accurate than that of long-time dynamics
because the error is accumulated
as the time increases.
It would be helpful to know
the reliability of the DMD prediction
as a function of time.
In this section, we
estimate the error of the DMD prediction
by performing the statistical analysis of the predicted data.

For simplicity,
we consider a noiseless time-series data and focus on
the case without damping,
i.e.,
the equal-time longitudinal spin-spin correlation function
after a sudden quench in the transverse-field Ising model
on a finite-size square lattice.
To reduce the computational cost,
we choose a larger time step $\Delta t = 0.2$
and a smaller number of points in each snapshot $M = 100$.
On the other hand,
we choose a larger number of snapshots $N = 5M$
so that the time interval of the input data
$t_{\rm input} = N\cdot\Delta t = 100$
is the same as that in Sec.~\ref{subsec:corr_no_damp}.
We numerically find that
the DMD prediction is rather stable
without introducing the cutoff $\epsilon$.

The DMD prediction contains two types of errors:
One is the systematic error
and the other is the statistical error.
The systematic error is
evaluated by the average difference
between the DMD prediction and the exact result.
On the other hand,
the statistical error
is characterized by
the standard deviation
of the DMD prediction
when the input data is
selected randomly.
In the present setup,
we can calculate the exact input data
of small systems for arbitrarily long times.
Therefore, we will estimate both systematic and statistical errors
by comparing the DMD prediction with the exact result.

\begin{figure}[t!]
\centering
\includegraphics[width=1.00\columnwidth]{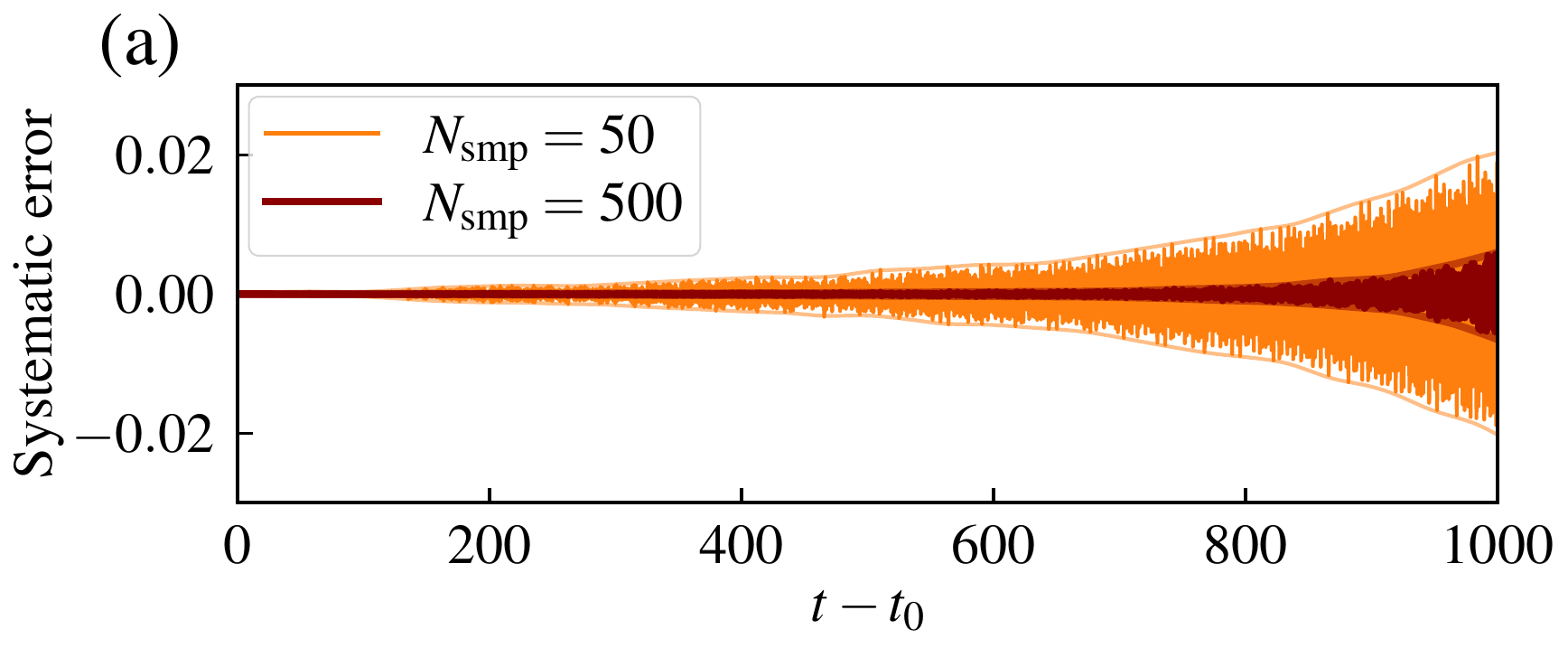}\\
\includegraphics[width=1.00\columnwidth]{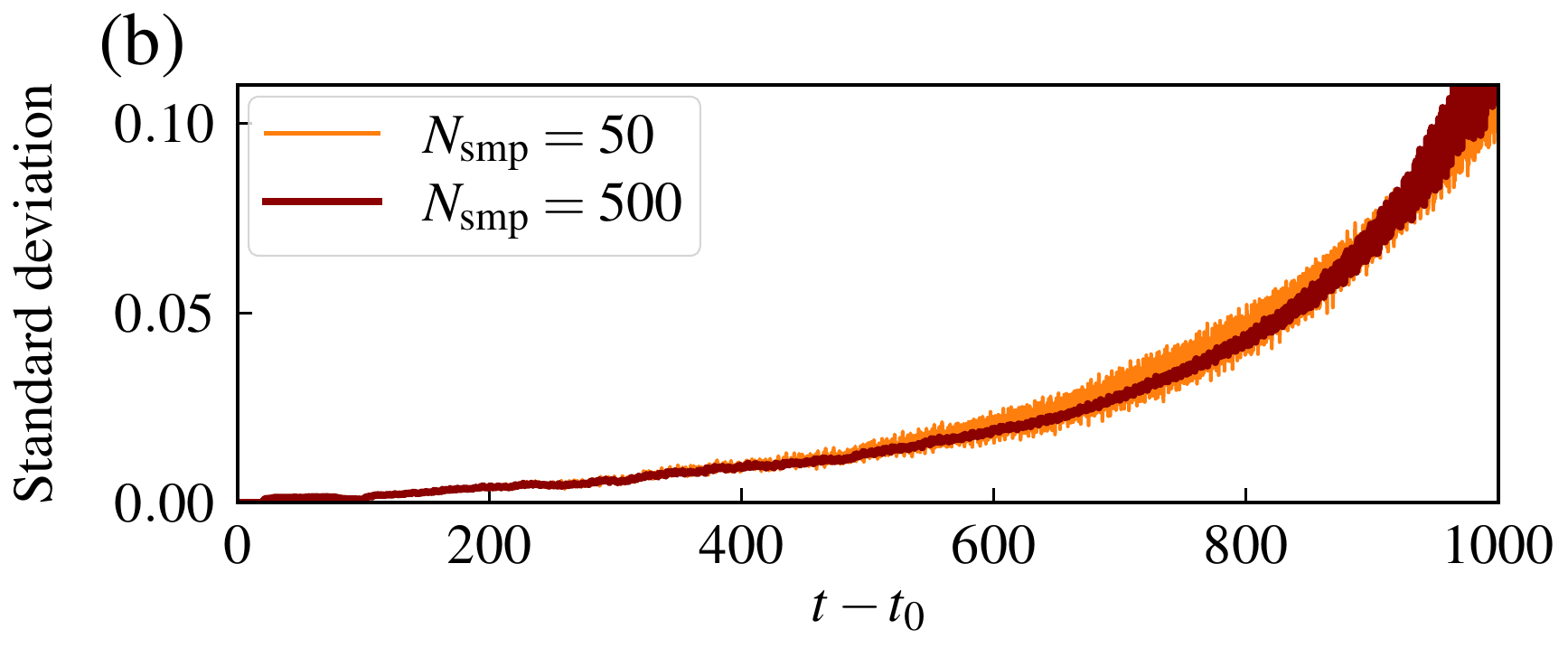}
\caption{%
Error analysis of the DMD prediction of the equal-time
longitudinal spin-spin correlation function
after a sudden quench in the
2D
transverse-field Ising model
on a square lattice.
We show (a) the systematic error
and (b) the statistical error
of the DMD prediction.
The former is obtained by the average of the differences
between the DMD predictions and the exact results
and the latter is obtained by the standard deviation
of the differences
for $N_{\rm smp} = 50$ and $N_{\rm smp} = 500$ samples.
The thin lines are the envelope of the systematic error.
}
\label{fig:ising_2d_corr_error}
\end{figure}

To estimate both systematic and statistical errors,
we prepare the DMD prediction data
by changing the initial time $t_0$ of the input data
and perform the statistical analysis.
We obtain the exact result
for the time interval $t \in [0, 500000)$
in this section and
arrange independent $N_{\rm smp} = 500$ samples
of the predicted data with different initial times
$t_0 = 0, 1000, 2000, \ldots, 499000$
and the maximum time
$t_{\rm max} = N_{\rm max}\cdot\Delta t = 1000$.

We obtain the systematic error
by taking the average of the differences between the DMD predictions
and the exact results for $N_{\rm smp}$ samples
in the time interval $t-t_0 \in [0, 1000)$
having different initial times $t_0$.
We also estimate the statistical error
from the standard deviation of the differences.
Since the initial times $t_0$ are sufficiently far apart,
the systematic error for different initial times should behave similarly.
At a given time-difference point $t-t_0$,
the statistical error of each sample
is expected to follow the same probability distribution.

We show the systematic error of the DMD prediction
in Fig.~\ref{fig:ising_2d_corr_error}(a).
The systematic error oscillates around zero,
which suggests that the DMD prediction
does not have a systematic bias
that leads to a deviation of the center of the oscillatory behavior.
Indeed, the systematic error is smaller than
the statistical error, as we will see below.
The systematic error approaches zero as $N_{\rm smp}$ increases.
Thus, it is less likely that the DMD prediction
has a systematic error.
Though the amplitude of the 
oscillatory systematic error
increases
as the time difference increases,
even at the latest time $t-t_0 = 1000$,
it is still smaller than $0.01$ for $N_{\rm smp} = 500$,
which is less than the amplitude of the spin-spin correlation function.

We also show the statistical error of the DMD prediction
in Fig.~\ref{fig:ising_2d_corr_error}(b).
The statistical error monotonically increases
as the time difference increases.
At $t-t_0 \approx 800$,
the statistical error reaches $0.05$,
which is comparable to the amplitude of the spin-spin correlation function.
Therefore,
the DMD prediction is reliable up to the timescale
$t-t_0 \approx 800$,
which is more than five times but slightly less than an order of magnitude
as compared to the duration of the input time.
As for the present data,
the statistical error has more influence on the accuracy of the DMD prediction
than the systematic error.

\section{Summary and outlook}
\label{sec:sum}

In conclusion,
we have applied the DMD to the dynamics of quantum many-body systems
in which the long-ranged quantum entanglement plays a crucial role
and discussed the accuracy and applicability of the DMD.
We have
studied the following cases:
(i)
Correlation functions that exhibit
multiple oscillatory components
(caused by the evolution of entanglement entropy, which stem from the quantum many-body effects)
and
(ii)
correlation functions that exhibit oscillatory behavior
and have a power-law decay.
The former case is observed in the equal-time
longitudinal spin-spin correlation function
after a sudden quench
in a finite-size system of
the
2D
transverse-field Ising model,
whereas the latter case is realized in
the unequal-time transverse spin-spin correlation function
in the
1D
transverse-field Ising model at the critical point.
We have found that
the DMD prediction is very accurate
when the eigenmodes are sufficiently included in the DMD procedure.
The DMD prediction is reliable up to the timescale 
typically more than five times to
nearly
an order of magnitude longer than 
the time interval of the initial input data.
We have also found that the DMD is superior to the GPR method,
which is a conventional machine learning method.

To deal with more realistic situations,
we have applied the DMD to noisy input data.
We have found that
the DMD prediction is still accurate
when the noise level is within a few percent of the noiseless part.
Moreover,
we have empirically found that
the singular values of the matrix
$X_0$ in Eq.~\eqref{eq:dmd_x0}
generated by the input data
exhibit a plateau region
when the input data are noisy.
The strength of noise is found to be
proportional to the plateau value.
This observation would be useful for
estimating the noise level
in experimental or numerical data
when the noise level is not known.

We have also estimated the
systematic and statistical errors of the DMD prediction.
We prepare independent samples of the predicted data
with different initial times.
Then,
the systematic error is obtained by the average of the differences
between the DMD predictions and the exact results,
whereas the statistical error is obtained by the standard deviation.
The statistical error is found to be
more influential on the accuracy of the DMD prediction
than the systematic error.
The statistical error reaches the amplitude of the spin-spin correlation function
when the time difference is nearly an order of magnitude longer
than the duration of the input time.
Extending the timescale of the reliable DMD prediction
by more than an order of magnitude longer
is important future work.

In this study, we restrict ourselves to the quantum dynamics generated by the time-independent Hamiltonian.
Simulating the dynamics generated by the time-dependent Hamiltonian would be challenging and is left for future studies.
If one applies the DMD to time-series data generated in time-dependent systems, then the assumption that the linear operator
$A$ defined in Eq.~\eqref{eq:dmd_a} is time-invariant is violated.
The difficulty in such a case already exists in classical time-series data.
However, periodically driven systems such as the Floquet systems would be tractable by the DMD
if one prepares input data sufficiently longer than the periodicity, which is an interesting future subject.

Our findings open up the possibility of
applying the DMD to the dynamics of
other quantum many-body systems,
which are difficult to simulate by conventional
brute-force numerical methods that require
calculating time-evolved wave functions.
We have shown that the DMD is a powerful and versatile tool even at 
quantum critical points with temporary long-ranged quantum fluctuations characteristic of the growing quantum entanglement.
Even without using complex and complicated methods, we were able to
provide high-precision solutions to difficult problems
of dynamics in quantum many-body systems by using the
simple approach that has been used
in classical systems.
Thanks to its simplicity and high accuracy,
the DMD may become a universal method that will be widely used in the next generation.
It is also interesting to apply the DMD to
the time series of experimental data
in quantum many-body systems.
Such an application is expected to be useful
and will be discussed in future work.

\begin{acknowledgments}
The authors acknowledge fruitful discussions with
Kota Ido,
Shimpei Goto, and Daichi Kagamihara.
This work was financially supported by 
MEXT KAKENHI, Grant-in-Aid for Transformative Research Area
(Grants No.\ JP22H05111 and No.\ JP22H05114).
R.K.\ was supported by
JSPS KAKENHI
(Grant No.\ JP21K13855).
T.O.\ was supported by
JSPS KAKENHI
(Grants No.\ JP19H00658 and No.\ JP22H05114),
and CREST (Grant No.\ JPMJCR18T4).
M.I.\ was supported by MEXT as 
``Program for Promoting Researches on the Supercomputer Fugaku''
(Simulation for basic science: approaching the new quantum era, 
Grant No.\ JPMXP1020230411).
Y.K.\ was supported by MEXT KAKENHI, Grant-in-Aid for Transformative Research Area (Grants No.\ JP22H05111 and No.\ JP22H05117), and CREST (Grant No.\ JPMJCR1912).
The numerical computations were performed on computers at
the Supercomputer Center, the Institute for Solid State Physics,
the University of Tokyo.
\end{acknowledgments}

\appendix

\section{%
Entanglement entropy dynamics
}
\label{sec:ee_dynamics}

\begin{figure}[t!]
\centering
\includegraphics[width=1.00\columnwidth]{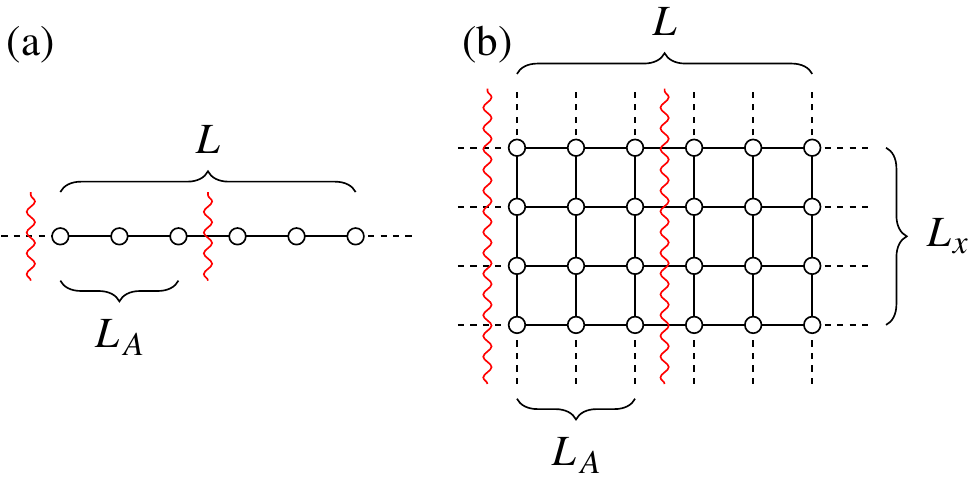}
\caption{%
Spatial partitions of (a) 1D and (b) 2D systems under periodic
boundary conditions that are used to calculate the von Neumann
entanglement entropy. Circles denote the lattice sites and bonds
correspond to the interaction between two sites.
In 1D, the total system size is $L$ and the subsystem size is $L_A$.
In 2D, the total system size is $N_{\mathrm{s}} = L_x L$
and the subsystem size is $N_A = L_x L_A$. The system is
divided into two parts at the boundaries drawn with wavy lines.
}
\label{fig_supp:partition_ee}
\end{figure}

\begin{figure}[t!]
\centering
\includegraphics[width=1.00\columnwidth]{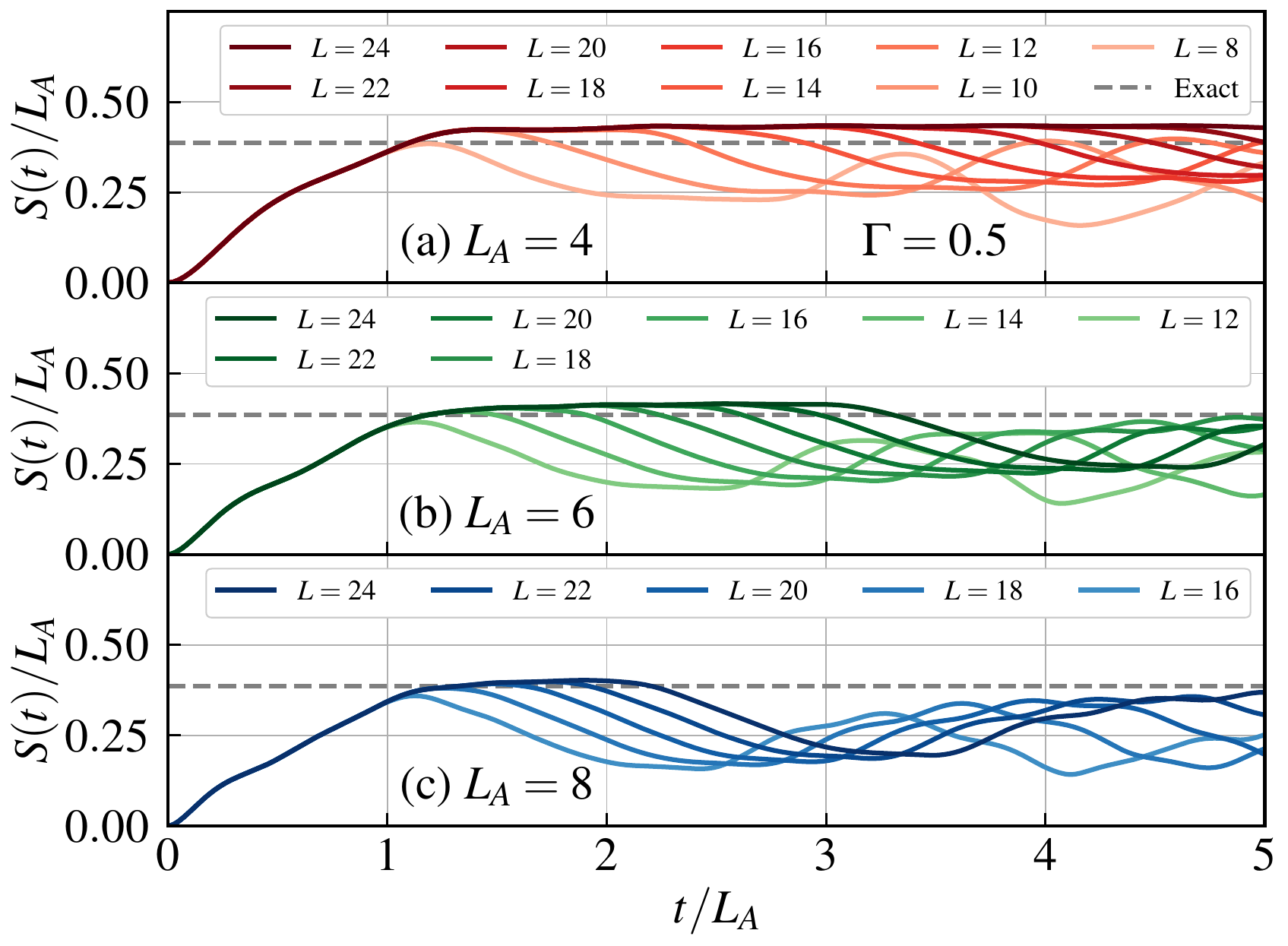}
\caption{%
Time evolution of the entanglement entropy
after a sudden quench to the critical point
($\Gamma/J = 0.5$)
in the 1D transverse-field Ising model.
We choose the subsystem sizes
(a) $L_{A} = 4$,
(b) $L_{A} = 6$,
and (c) $L_{A} = 8$
and the total system size $L \in [2L_{A}, 24]$.
The exact entanglement entropy per subsystem size
for infinite-size systems ($L \to \infty$, $L_{A} \to \infty$)
in the infinite-time limit ($t \to \infty$)
is given by
$S/L_{A} \to 2\ln 2 - 1 \approx 0.3863$~\cite{calabrese2005}
and is shown by the dashed line.
}
\label{fig_supp:1dtfising_05_ee_t}
\end{figure}

\begin{figure}[t!]
\centering
\includegraphics[width=1.00\columnwidth]{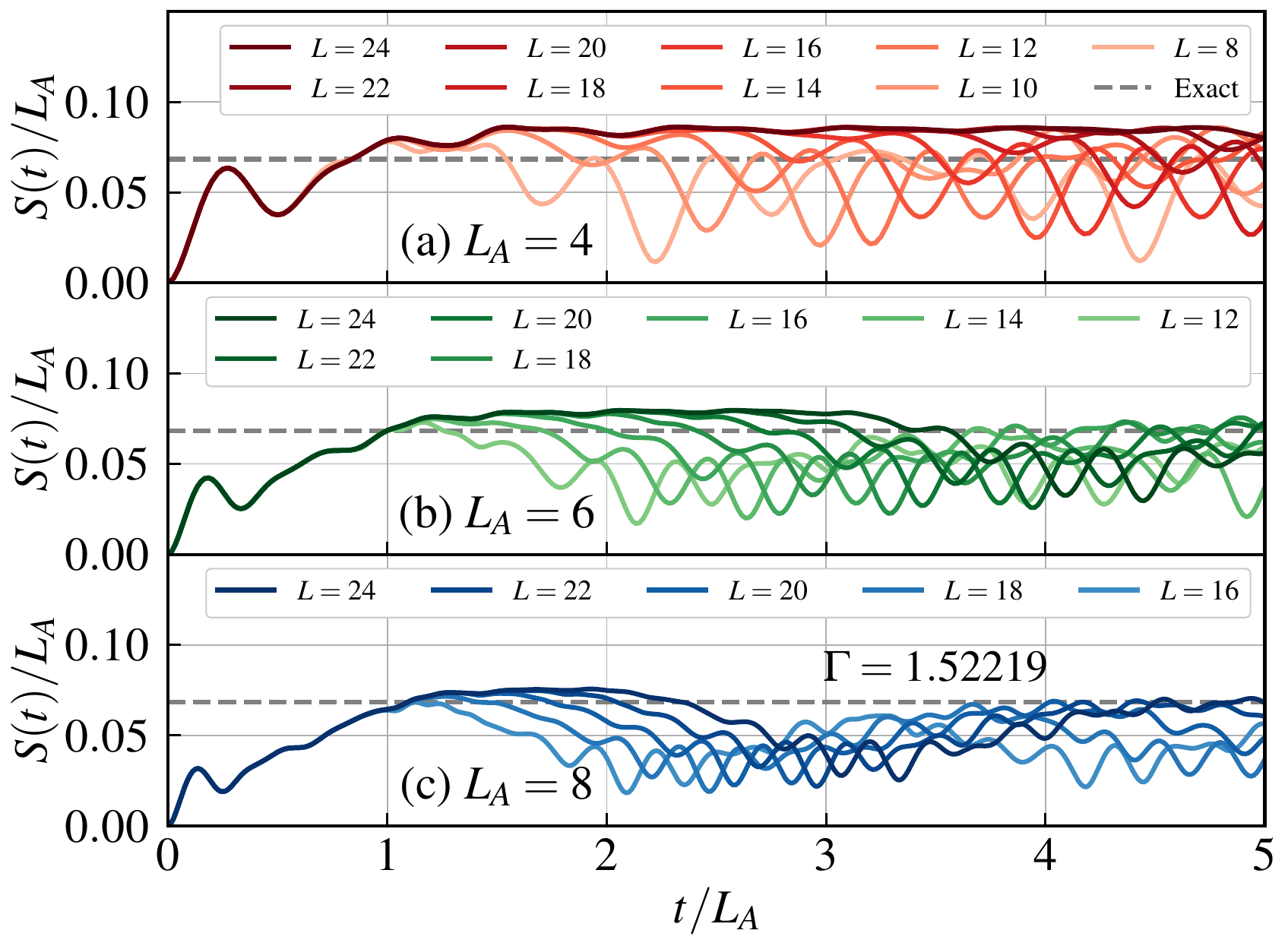}
\caption{%
Time evolution of the entanglement entropy
after a sudden quench to the transverse field
$\Gamma/J = 1.52219$
in the 1D transverse-field Ising model.
As in Fig.~\ref{fig_supp:1dtfising_05_ee_t},
we choose the subsystem sizes
(a) $L_{A} = 4$,
(b) $L_{A} = 6$,
and (c) $L_{A} = 8$
and the total system size $L \in [2L_{A}, 24]$.
The exact entanglement entropy per subsystem size
for infinite-size systems ($L \to \infty$, $L_{A} \to \infty$)
in the infinite-time limit ($t \to \infty$)
is given by
$S/L_{A} \approx 0.06837$~\cite{calabrese2005}
and is shown by the dashed line.
}
\label{fig_supp:1dtfising_152219_ee_t}
\end{figure}

\begin{figure}[t!]
\centering
\includegraphics[width=1.00\columnwidth]{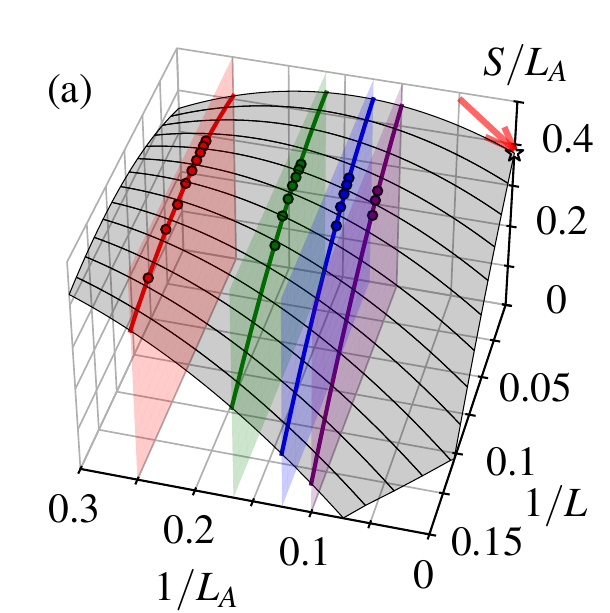}\\
\includegraphics[width=1.00\columnwidth]{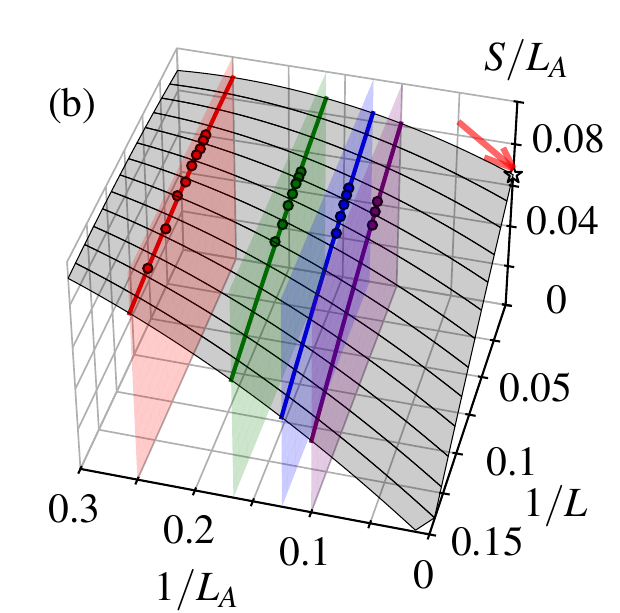}
\caption{%
Size extrapolation of the entanglement entropy
after a sudden quench to the transverse field
(a) $\Gamma/J = 0.5$
and (b) $\Gamma/J = 1.52219$
in the 1D transverse-field Ising model.
The fitting function is given by
$f_{\mathrm{1D}}(L,L_{A})
=S/L_{A} = s_{\mathrm{1D,\infty}} + c_{1}/L + c_{2}/L_{A}
+ c_{11}/L^{2} + c_{12}/(LL_{A}) + c_{22}/L_{A}^{2}$
with $s_{\mathrm{1D,\infty}}$, $c_{1}$, $c_{2}$, $c_{11}$, $c_{12}$,
and $c_{22}$ being fitting parameters.
The gray surface represents the fitting function
$f_{\mathrm{1D}}(L,L_{A})$
with the fitting parameters obtained by the least-squares method.
The extrapolated value for infinite-size systems
($L \to \infty$, $L_{A} \to \infty$)
in the infinite-time limit ($t \to \infty$)
is shown by stars.
They well agree with the exact results
shown by arrows.
}
\label{fig_supp:1dtfising_ee_fit}
\end{figure}

\begin{figure}[t!]
\centering
\includegraphics[width=1.00\columnwidth]{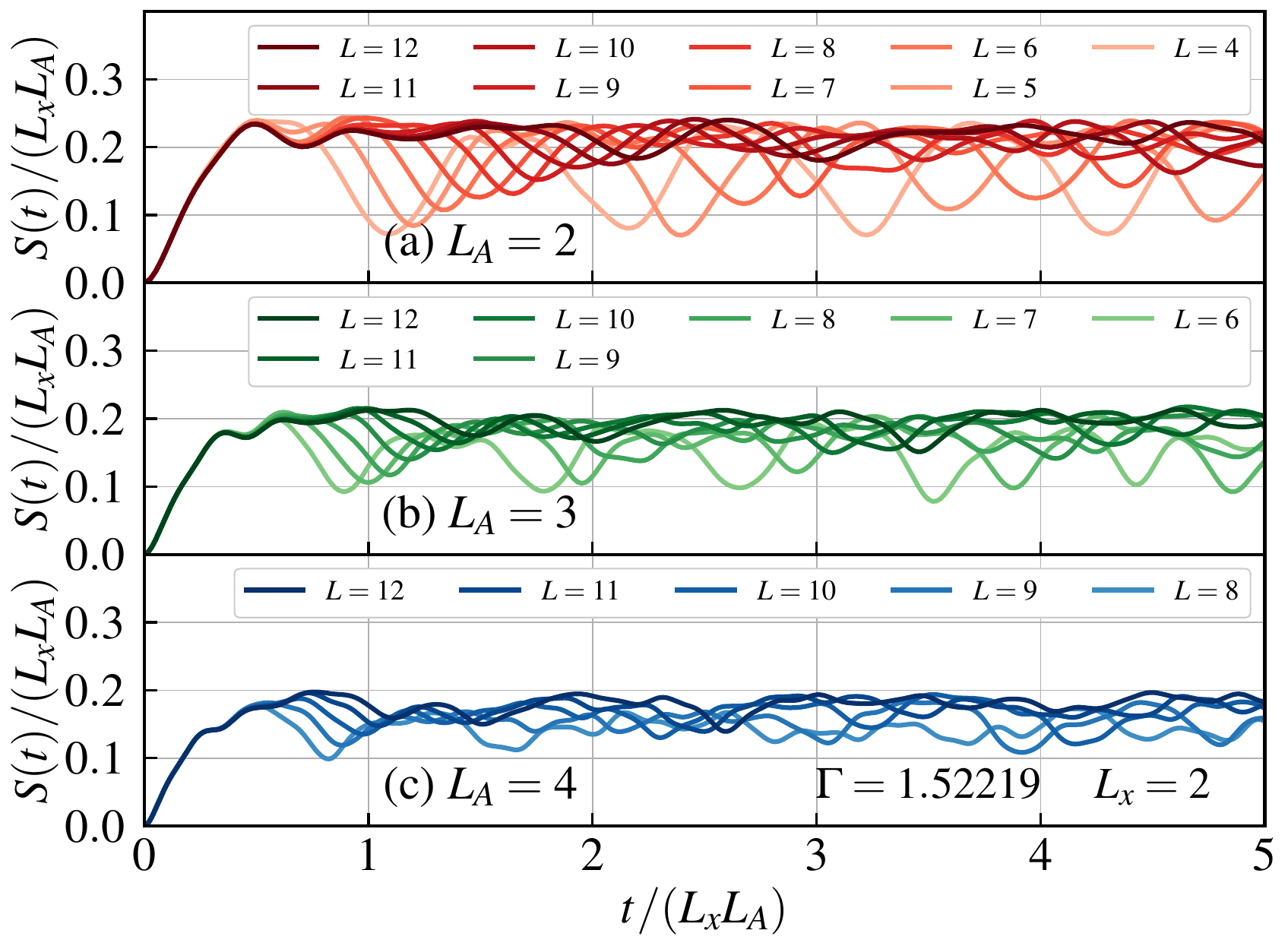}
\caption{%
Time evolution of the entanglement entropy
after a sudden quench to the critical point
($\Gamma/J = 1.52219$)
in the 2D transverse-field Ising model
with the system size $N_{\mathrm{s}} = L_{x} \times L$.
We choose the length along the $x$ direction
$L_{x} = 2$,
the length of the subsystem sizes along the $y$ direction
(a) $L_{A} = 2$,
(b) $L_{A} = 3$,
and (c) $L_{A} = 4$,
and the length of the total system size along the $y$ direction
$L \in [2L_{A}, 12]$.
}
\label{fig_supp:2dtfising_152219_ee_t}
\end{figure}

\begin{figure}[t!]
\centering
\includegraphics[width=1.00\columnwidth]{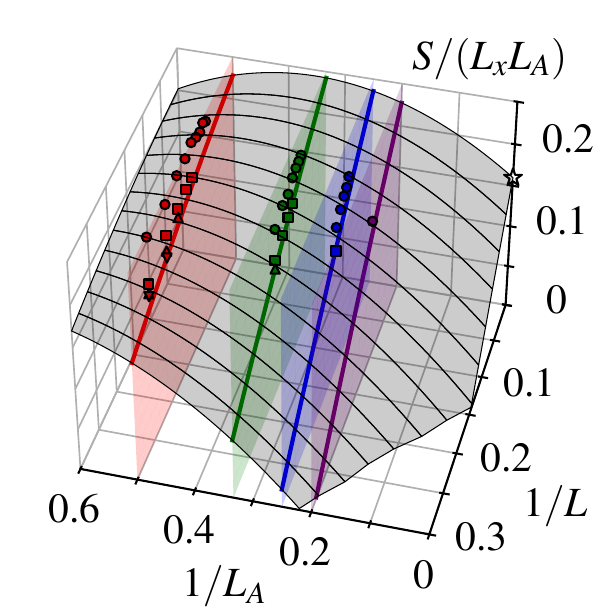}
\caption{%
Size extrapolation of the entanglement entropy
after a sudden quench to the critical point
($\Gamma/J = 1.52219$)
in the 2D transverse-field Ising model.
The fitting function is given by
$f_{\mathrm{2D}}(L,L_{A},L_{x})
= S/(L_{x}L_{A}) 
= s_{\mathrm{2D,\infty}} + d_{1}/L + d_{2}/L_{A} + d_{3}/L_{x}
+ d_{11}/L^{2} + d_{12}/(LL_{A}) + d_{13}/(LL_{x}) + d_{22}/L_{A}^{2}
+ d_{23}/(L_{A}L_{x}) + d_{33}/L_{x}^{2}$
with $s_{\mathrm{2D,\infty}}$, $d_{1}$, $d_{2}$, $d_{3}$, $d_{11}$, $d_{12}$,
$d_{13}$, $d_{22}$, $d_{23}$, and $d_{33}$ being fitting parameters.
The data points for $L_{x} = 2$, $3$, $4$, and $5$
are shown by circles, squares, up-pointing triangles, and down-pointing
triangles, respectively.
The gray surface represents the fitting function
$f_{\mathrm{2D}}(L,L_{A},L_{x})$
with the fitting parameters obtained by the least-squares method
and $L_{x} \to \infty$.
The extrapolated value for an infinite-size system
($L \to \infty$, $L_{A} \to \infty$, $L_{x} \to \infty$)
in the infinite-time limit ($t \to \infty$)
is shown by a star.
}
\label{fig_supp:2dtfising_ee_fit}
\end{figure}

\begin{figure}[t!]
\centering
\includegraphics[width=1.00\columnwidth]{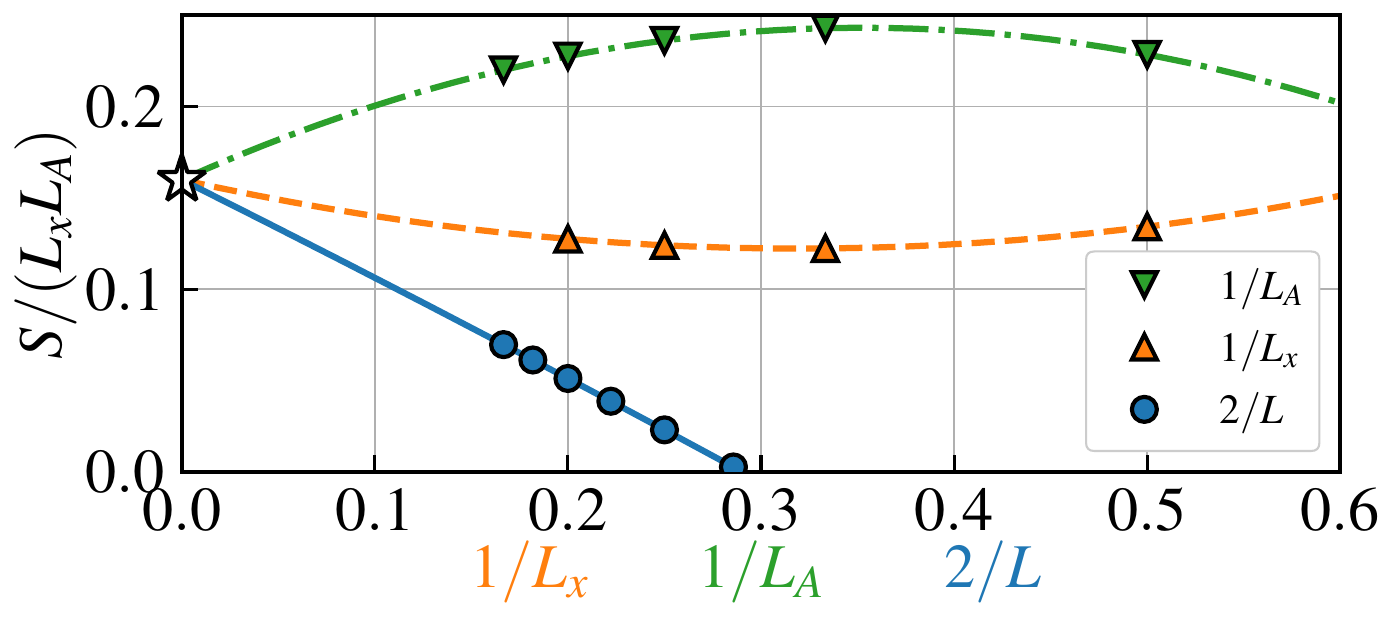}
\caption{%
Size extrapolation of the entanglement entropy
after a sudden quench to the critical point
($\Gamma/J = 1.52219$)
in the 2D transverse-field Ising model
in the projected spaces.
The data are equivalent to those in
Fig.~\ref{fig_supp:2dtfising_ee_fit}.
The fitting function for $L, L_{x} \to \infty$
is given by a dash-dotted line with down-pointing triangles,
that for $L, L_{A} \to \infty$
is given by a dashed line with up-pointing triangles,
and that for $L_{x}, L_{A} \to \infty$
is given by a solid line with circles.
In the thermodynamic limit ($L, L_{A}, L_{x} \to \infty$),
the entanglement entropy per subsystem size
converges to a nonzero value,
suggesting the presence of the volume-law entanglement entropy.
}
\label{fig_supp:2dtfising_ee_fit_invL}
\end{figure}

\begin{figure}[t!]
\centering
\includegraphics[width=1.00\columnwidth]{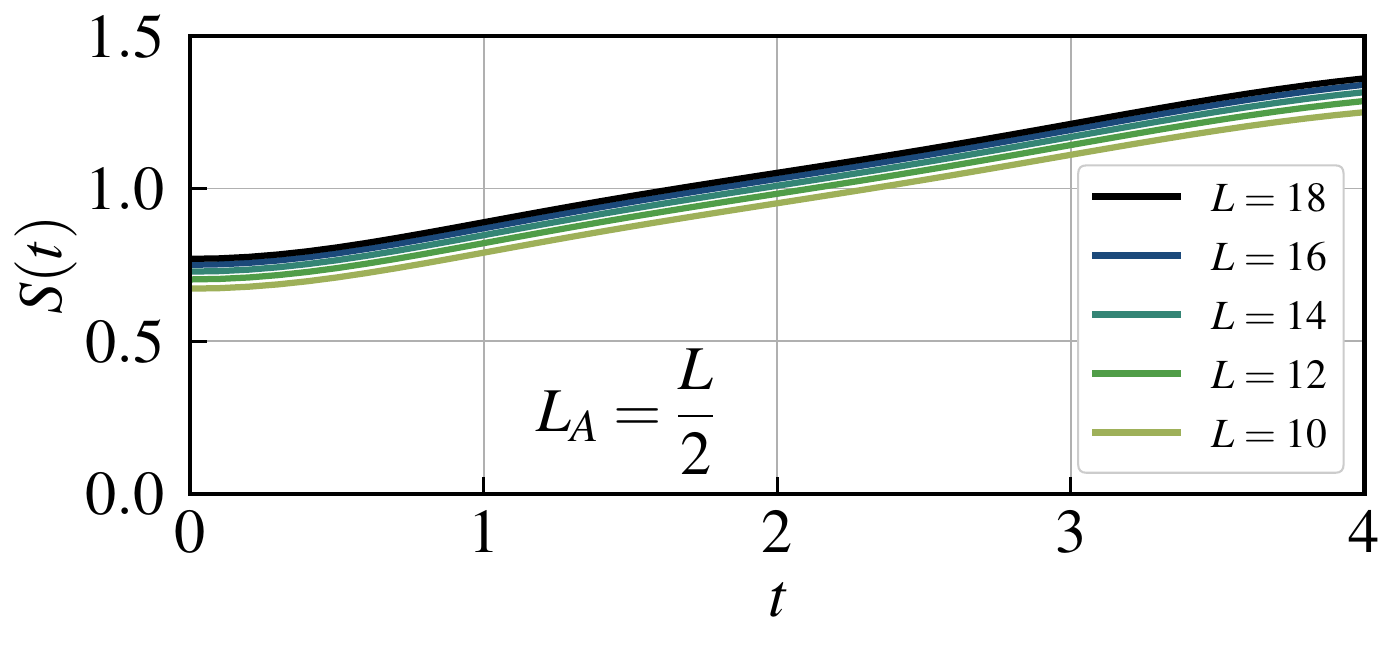}
\caption{%
Time evolution of the entanglement entropy
corresponding to the time-evolved state
at the critical point
($\Gamma/J = 0.5$)
for the 1D transverse-field Ising model.
Because the ground state is located at the critical point,
the initial state at $t=0$ has a nonzero entanglement entropy.
The entanglement entropy grows almost linearly with time,
just as in the case of the sudden quench
(see Figs.~\ref{fig_supp:1dtfising_05_ee_t}
and \ref{fig_supp:1dtfising_152219_ee_t}).
}
\label{fig_supp:1dtfising_ee_unequal_time}
\end{figure}

In Sec.~\ref{sec:application},
we apply the DMD to the dynamics of quantum many-body systems.
In this section,
we show that the data that we use in Sec.~\ref{sec:application}
are derived from the highly entangled quantum states
and they satisfy the volume-law entanglement entropy.
Our data are qualitatively different from the data
that are generated by the classical systems.

We first discuss the time evolution of the
von Neumann entanglement entropy
in the 1D transverse-field Ising model.
When we divide a pure state $|\psi\rangle$ in a system
into two subsystems $A$ and $B$,
the von Neumann entanglement entropy is defined as
$S = - \mathrm{Tr}_A \, \rho_A \ln \rho_A$,
where $\rho_A = \mathrm{Tr}_B \, \rho$ is the reduced density matrix
of $\rho = |\psi\rangle \langle\psi|$
and $\mathrm{Tr}_{A(B)}$
means the trace over the basis of subsystem $A$ ($B$). 
We use the spatial partitions defined in
Fig.~\ref{fig_supp:partition_ee}.
In 1D,
the system is integrable,
and the exact entanglement entropy
is known to satisfy the volume-law scaling
after a long-time evolution in the thermodynamic
limit~\cite{calabrese2005,fagotthi2008,alba2017,alba2018}.
The entanglement entropy per subsystem size
for infinite-size systems
in the infinite-time limit
is given by
\begin{align}
\label{eq:1dtfising_ee_exact}
 s_{\mathrm{1D,exact}}
 &= \frac{1}{\pi}
 \int_{0}^{\pi/2} d\phi
 \nonumber
 \\
 &~\phantom{=}~
 \times
 H\left(
 \frac{ 1 - (h+h_0)\cos\phi + hh_0 }
 {\sqrt{(1-2h\cos\phi+h^2)(1-2h_0\cos\phi+h_0^2)}}
 \right),
\end{align}
where 
$h_0=2\Gamma_0/J$ and $h=2\Gamma/J$
are the transverse fields 
before and after a sudden quench, respectively,
and
$H(x) = - \frac{1+x}{2}\ln\frac{1+x}{2}
- \frac{1-x}{2}\ln\frac{1-x}{2}$~\cite{calabrese2005}.

Although the exact entanglement entropy after a sudden quench
is known for the 1D transverse-field Ising model,
as a benchmark,
we apply the exact diagonalization method to small 1D systems on purpose
and examine the time evolution of the entanglement entropy.
By analyzing the time evolution of the entanglement entropy
for systems up to $24$ sites
using the \textsc{QuSpin} library~\cite{weinberg2017,weinberg2019},
we successfully reproduce the volume-law scaling
including the prefactor
expected in the analytical results.
These results would be useful for understanding the dynamics
of the entanglement entropy also in 2D systems
using the exact diagonalization method.

We calculate the time evolution of the entanglement entropy
after a sudden quench from the infinite transverse field
($\Gamma_0/J = \infty$)
to the critical point
($\Gamma/J = 0.5$)
in the 1D transverse-field Ising model,
as shown in Fig.~\ref{fig_supp:1dtfising_05_ee_t}.
We use the periodic boundary condition
and choose the system sizes up to $L=24$.
The subsystem size is chosen as $L_{A} \le L/2$
for each total system size $L$.
We specifically show the results for $L_{A} = 4$, $6$, and $8$
in Figs.~\ref{fig_supp:1dtfising_05_ee_t}(a)--\ref{fig_supp:1dtfising_05_ee_t}(c).
For $t/L_{A}\lesssim 1$
(using units in which the speed of the elementary excitations equals
unity),
the entanglement entropy grows almost linearly with
time~\cite{calabrese2005,fagotthi2008,alba2017,alba2018}.
For $t/L_{A}\gtrsim 1$,
when $L$ is sufficiently large,
we observe a nearly constant entanglement entropy
up to a certain timescale.
As the subsystem size $L_{A}$ increases,
the constant value of the entanglement entropy
approaches the exact result in the thermodynamic limit
(see a dashed line in Fig.~\ref{fig_supp:1dtfising_05_ee_t}).

To examine the entanglement entropy dynamics
away from the critical point,
we also calculate the time evolution of the entanglement entropy
after a sudden quench from the infinite transverse field
($\Gamma_0/J = \infty$)
to the different field
($\Gamma/J = 1.52219$, corresponding to the critical point
in the 2D transverse-field Ising model)
in the 1D transverse-field Ising model,
as shown in Fig.~\ref{fig_supp:1dtfising_152219_ee_t}.
For $t/L_{A}\lesssim 1$,
we observe a hump structure in the entanglement entropy
as a function of time.
This hump seems to disappear
as the subsystem size $L_{A}$ increases.
When $L_{A}=8$,
we see an almost linear growth of the entanglement entropy
with time~\cite{calabrese2005,fagotthi2008,alba2017,alba2018}.
For $t/L_{A}\gtrsim 1$,
the entanglement entropy approaches a nearly constant value
when $L$ is sufficiently large.
The constant value of the entanglement entropy
approaches the exact result in the thermodynamic limit
as the subsystem size $L_{A}$ increases
(see a dashed line in Fig.~\ref{fig_supp:1dtfising_152219_ee_t}).

To quantitatively examine the accuracy of the entanglement entropy
dynamics obtained by the exact diagonalization method,
we compare the entanglement entropy per subsystem size
estimated from data of small systems using the exact diagonalization method
with that obtained by the exact analytical calculation
in Eq.~\eqref{eq:1dtfising_ee_exact}.
In finite-size systems,
the entanglement entropy exhibits recurrent behavior
and is often smaller than
the nearly constant entanglement entropy
observed just after the time $t/L_{A}\approx 1$.
Therefore,
for each total system size $L$ and subsystem size $L_{A}$,
we regard the maximum value of the entanglement entropy
in a sufficiently long time interval
as the saturated constant value of the entanglement entropy.
We typically choose the maximum value of the entanglement entropy as
\begin{align}
\label{eq:1dtfising_ee_max}
 S_{\mathrm{1D,max}}(L,L_{A})
 = \max_{t\in [0,100]} S(t,L,L_{A}),
\end{align}
where $S(t,L,L_{A})$ is the entanglement entropy
at time $t$ for the total system size $L$ and subsystem size $L_{A}$.
We then perform the size extrapolation
of the entanglement entropy
per subsystem size $S_{\mathrm{1D,max}}(L,L_{A})/L_{A}$
by using the fitting function
\begin{align}
\label{eq:1dtfising_ee_fit}
 f_{\mathrm{1D}}(L,L_{A})
 = s_{\mathrm{1D,\infty}} + \frac{c_{1}}{L} + \frac{c_{2}}{L_{A}}
 + \frac{c_{11}}{L^{2}} + \frac{c_{12}}{LL_{A}}
 + \frac{c_{22}}{L_{A}^{2}},
\end{align}
where $s_{\mathrm{1D,\infty}}$, $c_{1}$, $c_{2}$, $c_{11}$, $c_{12}$,
and $c_{22}$ are fitting parameters.
After the least-squares fitting,
we obtain the extrapolated value $s_{\mathrm{1D,\infty}}$
corresponding to the entanglement entropy per subsystem size
for infinite-size systems
in the infinite-time limit
(see Fig.~\ref{fig_supp:1dtfising_ee_fit}).
In the case of $\Gamma/J = 0.5$,
the estimated value of the entanglement entropy per subsystem size is
$s_{\mathrm{1D,\infty}} \approx 0.379(17)$,
which is consistent with the exact result
$s_{\mathrm{1D,exact}} = 2\ln 2 - 1 \approx 0.3863$.
In the case of $\Gamma/J = 1.52219$,
the estimated value is
$s_{\mathrm{1D,\infty}} \approx 0.0654(38)$,
which also agrees with the exact result
$s_{\mathrm{1D,exact}} \approx 0.06837$.
Therefore, we have numerically confirmed that
the data for small systems up to $L=24$
are sufficient to estimate the entanglement entropy
per subsystem size for infinite-size systems.

After confirming the effectiveness of the finite-size scaling
for small systems in the 1D transverse-field Ising model,
we next examine the time evolution of the
entanglement entropy in the 2D transverse-field Ising model.
Because the system is nonintegrable,
the system is expected to thermalize
and
the entanglement entropy is expected to satisfy the volume-law scaling
after a long-time evolution.
However, to the best of our knowledge,
there are no detailed numerical analyses
of the entanglement entropy dynamics
extrapolated to the thermodynamic limit
for the nonintegrable 2D transverse-field Ising model.
Even for integrable systems, the study of entanglement entropy dynamics
in 2D systems is currently limited to the very recent research
on the free fermion systems~\cite{gibbins2024,yamashika2024}.
Therefore, we numerically calculate the time evolution
of the entanglement entropy using the exact diagonalization method
and explicitly show that the entanglement entropy
satisfies the volume-law scaling
in the 2D transverse-field Ising model.

Just as in the case of the analysis of the 1D transverse-field Ising model,
we first analyze the entanglement entropy dynamics
of finite-size systems
in the 2D transverse-field Ising model
under the periodic boundary condition.
We focus on a sudden quench from the infinite transverse field
($\Gamma_0/J = \infty$)
to the critical point
($\Gamma/J = 1.52219$),
as shown in Fig.~\ref{fig_supp:2dtfising_152219_ee_t}.
The corresponding spin-spin correlation function
for the $4 \times 4$ system
is shown in Fig.~\ref{fig:ising_2d_corr}
in the main text.
In 2D, as shown in Fig.~\ref{fig_supp:partition_ee}(b),
we have three length scales ($L_{x}$, $L$, and $L_{A}$);
the total system size $N_{\mathrm{s}} = L_x \times L$
contains the length along the $x$ direction ($L_x$)
and that along the $y$ direction ($L$),
and the subsystem size $N_{A} = L_{x} \times L_{A}$
contains the length along the $x$ direction ($L_{x}$)
and that along the $y$ direction ($L_{A}$).
For simplicity,
we use the same $L_{x}$ for the total system size $N_{\mathrm{s}}$
and the subsystem size $N_{A}$.
We need to extrapolate the entanglement entropy
per subsystem size $S/(L_{x}L_{A})$
to the thermodynamic limit
($L \to \infty$, $L_{A} \to \infty$, $L_{x} \to \infty$).
For this purpose,
we choose the system sizes up to $N_{\mathrm{s}} = L_x \times L = 27$
with $L_{x} = 2, 3, 4, 5$.
The length of the subsystem size along the $y$ direction
is chosen as $L_{A} \in [2,L/2]$.

We specifically show the results for $L_{x}=2$
in Fig.~\ref{fig_supp:2dtfising_152219_ee_t}.
For $t/(L_{x}L_{A})\lesssim 0.5$,
the entanglement entropy grows almost linearly with time.
For $t/(L_{x}L_{A})\gtrsim 0.5$,
when $L$ is sufficiently large,
we observe a nearly constant entanglement entropy
up to a certain timescale.
The constant value of the entanglement entropy per subsystem size
is nearly $0.2$ for $L_{A}=2,3,4$
and seems to converge to a certain nonzero value
in the infinite $L_{A}$ limit.
We repeat the same calculation for $L_{x}=3,4,5$ (not shown).

We also observe recurrent behavior
in the 2D entanglement entropy dynamics,
similarly to the 1D case.
For a fixed $L_{x}$ and $L_{A}$,
the period of the recurrence becomes longer
as the total system size
$N_{\mathrm{s}} = L_{x} \times L$ increases.
As in the case of the analysis in 1D systems,
we regard the maximum value of the entanglement entropy
for each total system size $N_{\mathrm{s}}$ and subsystem size
$N_{A}$ as the saturated constant value of the entanglement
entropy.
We typically choose the maximum value of the entanglement entropy as
\begin{align}
 S_{\mathrm{2D,max}}(N_{\mathrm{s}},N_{A})
 = \max_{t\in [0,100]} S(t,N_{\mathrm{s}},N_{A}),
\end{align}
where $S(t,N_{\mathrm{s}},N_{A})$ is the entanglement entropy
at time $t$ for the total system size $N_{\mathrm{s}}$ and subsystem
size $N_{A}$.
We then perform the size extrapolation
of the entanglement entropy
per subsystem size
$S_{\mathrm{2D,max}}(N_{\mathrm{s}},N_{A})/(L_{x}L_{A})$
by using the fitting function
\begin{align}
\label{eq:2dtfising_ee_fit}
 f_{\mathrm{2D}}(L,L_{A},L_{x})
 &= s_{\mathrm{2D,\infty}} + \frac{d_{1}}{L} + \frac{d_{2}}{L_{A}}
 + \frac{d_{3}}{L_{x}}
 + \frac{d_{11}}{L^{2}} + \frac{d_{12}}{LL_{A}}
\nonumber
\\
 &~\phantom{=}~
 + \frac{d_{13}}{LL_{x}}
 + \frac{d_{22}}{L_{A}^{2}} + \frac{d_{23}}{L_{A}L_{x}}
 + \frac{d_{33}}{L_{x}^{2}},
\end{align}
where $s_{\mathrm{2D,\infty}}$, $d_{1}$, $d_{2}$, $d_{3}$, $d_{11}$, $d_{12}$,
$d_{13}$, $d_{22}$, $d_{23}$, and $d_{33}$ are fitting parameters.
After the least-squares fitting,
we obtain the extrapolated value $s_{\mathrm{2D,\infty}}$
corresponding to the entanglement entropy per subsystem size
for infinite-size systems
in the infinite-time limit
(see Figs.~\ref{fig_supp:2dtfising_ee_fit}
and \ref{fig_supp:2dtfising_ee_fit_invL}
and the Supplemental Material).
The estimated value of the entanglement entropy per subsystem size is
$s_{\mathrm{2D,\infty}} \approx 0.16(3)$,
which is consistent with the expected volume-law scaling.

Last but not least,
we also examine the time evolution of the
bipartite von Neumann entanglement entropy
of the time-evolved state
[$S^x_r(t)|\psi'_0\rangle$ in Eq.~\eqref{eq:ising_1d_corr}]
giving the unequal-time spin-spin correlation
function in Fig.~\ref{fig:ising_1d_corr}
at the critical point
in the 1D transverse-field Ising model.
Note that even though
the system is not quenched and
the state $|\psi'_0\rangle$
is the eigenstate of the Hamiltonian,
the time-evolved state
$S^x_r(t)|\psi'_0\rangle$
is no longer the eigenstate of the Hamiltonian.
Therefore, the corresponding entanglement entropy is expected to grow
as the time $t$ increases~\cite{perales2008}.

We show the time evolution of the entanglement entropy
of the time-evolved state
[$S^x_r(t)|\psi'_0\rangle$ in Eq.~\eqref{eq:ising_1d_corr}]
obtained by the exact diagonalization method
in Fig.~\ref{fig_supp:1dtfising_ee_unequal_time}.
In contrast to the case of the sudden quench
from the infinite transverse field,
the initial state at $t=0$ has a sizable entanglement entropy
because the initial state is
the ground state of the Hamiltonian
at the critical point.
Up to this constant shift,
the entanglement entropy grows almost linearly with time,
just as in the case of the sudden
quench~\cite{calabrese2005,fagotthi2008,alba2017,alba2018}.
The behavior of the entanglement entropy is consistent
with that observed in the sudden quench
for the 1D transverse-field Ising model;
however, because of this initial entanglement entropy
caused by the critical phenomena,
the entanglement entropy of the time-evolved state is
expected to be larger than that in the sudden quench.

\section{%
Prediction by the GPR method
}
\label{sec:gpr}

\begin{figure}[t!]
\centering
\includegraphics[width=1.00\columnwidth]{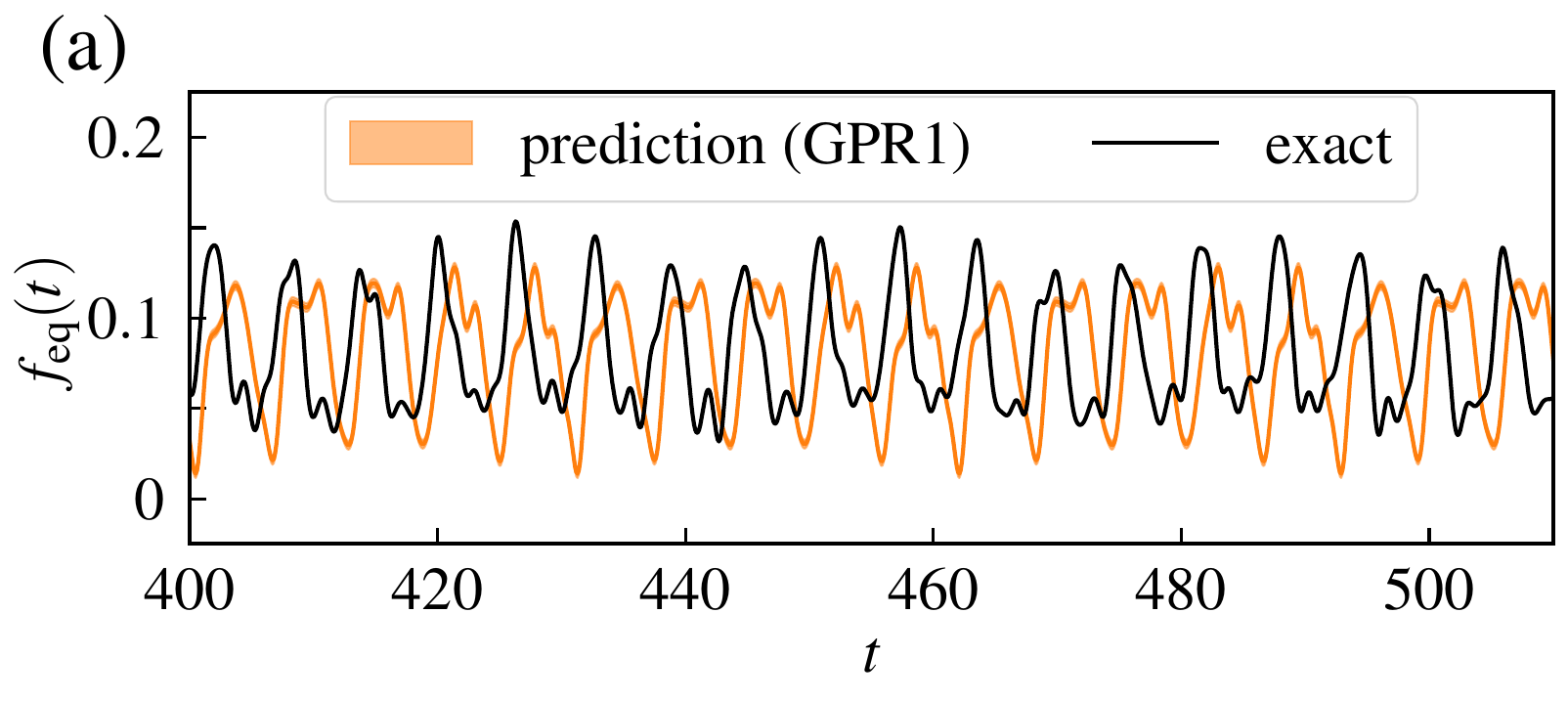}\\
\includegraphics[width=1.00\columnwidth]{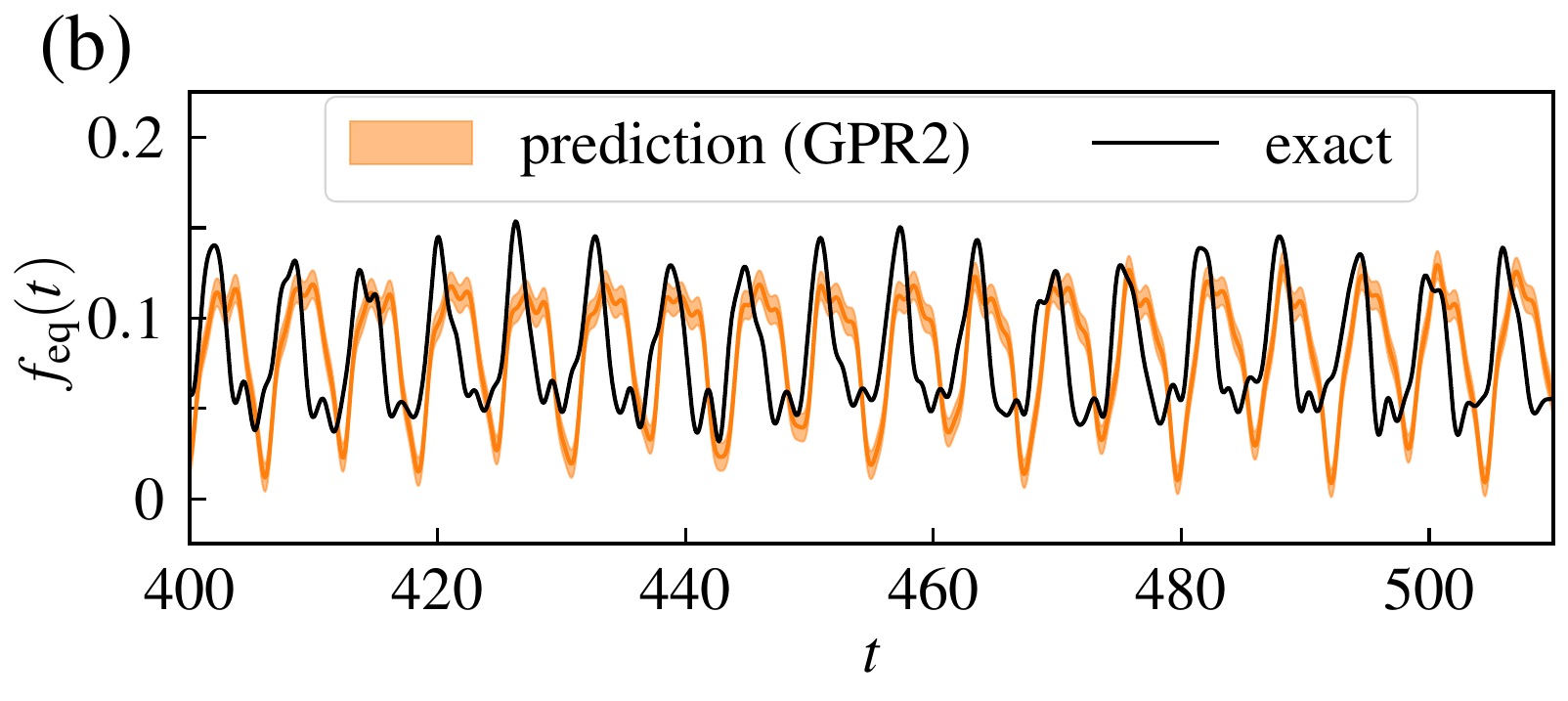}\\
\includegraphics[width=1.00\columnwidth]{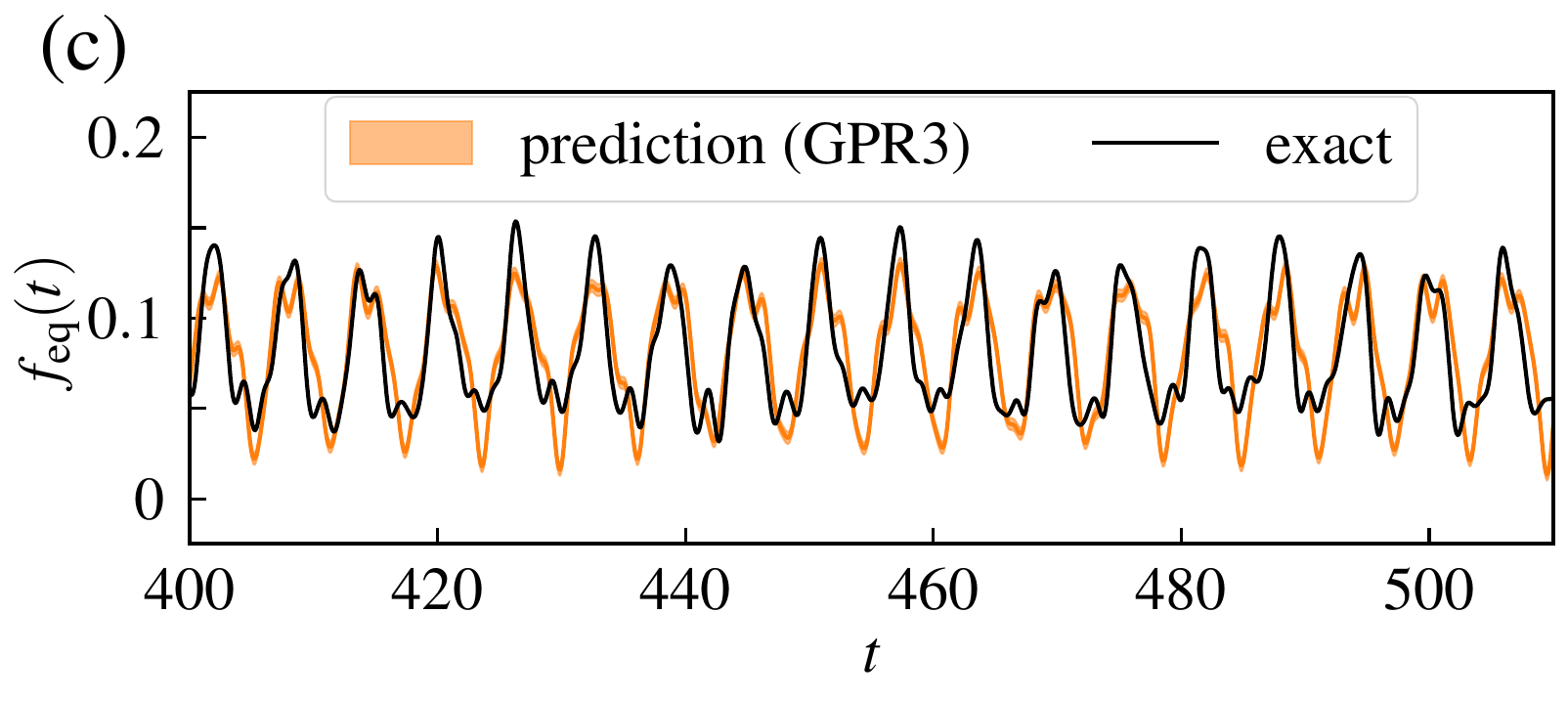}
\caption{%
GPR prediction of the equal-time
longitudinal spin-spin correlation function
in the 2D transverse-field Ising model.
We show the time evolution
for $t \in [400,510]$.
We show the results when the number of periodic kernels is
(a) $N_{\rm ker} = 1$,
(b) $N_{\rm ker} = 2$,
and (c) $N_{\rm ker} = 3$.
}
\label{fig_supp:gpr_corr}
\end{figure}

\begin{figure}[t!]
\centering
\includegraphics[width=1.00\columnwidth]{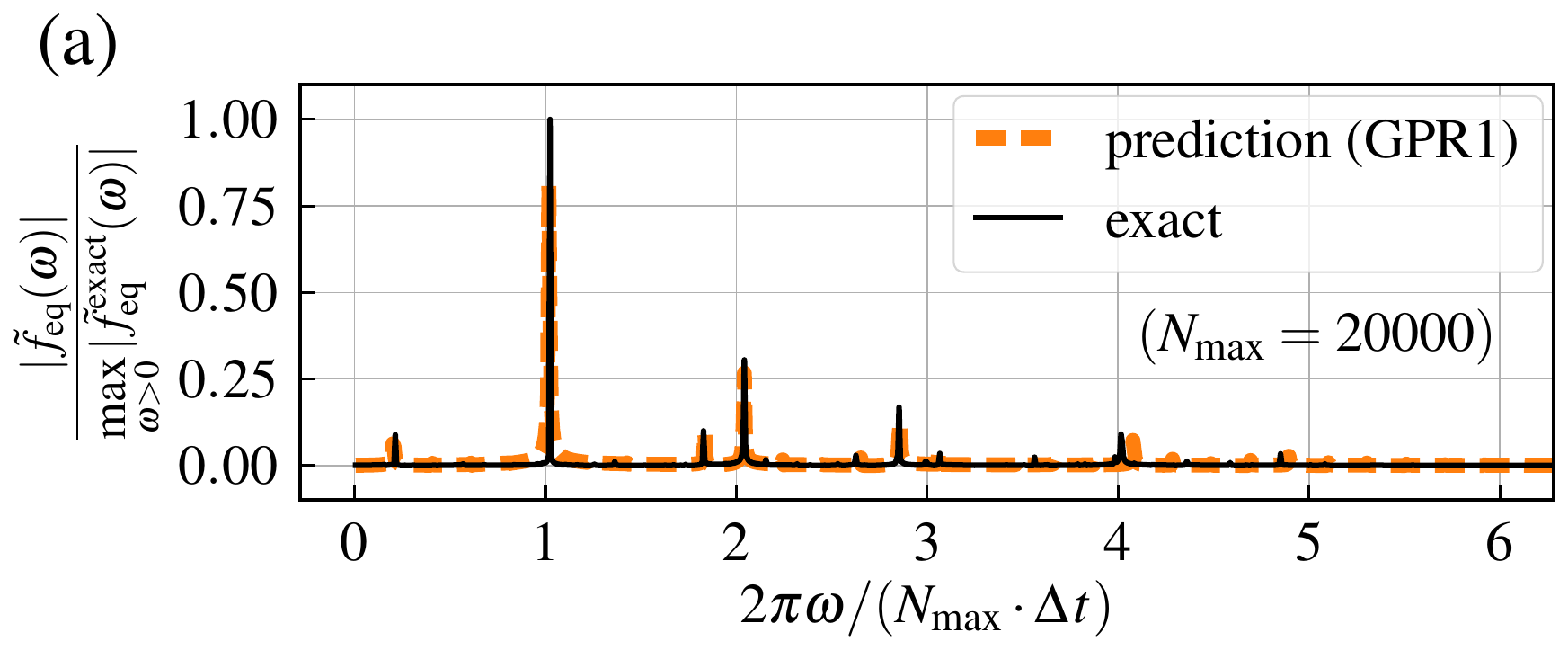}\\
\includegraphics[width=1.00\columnwidth]{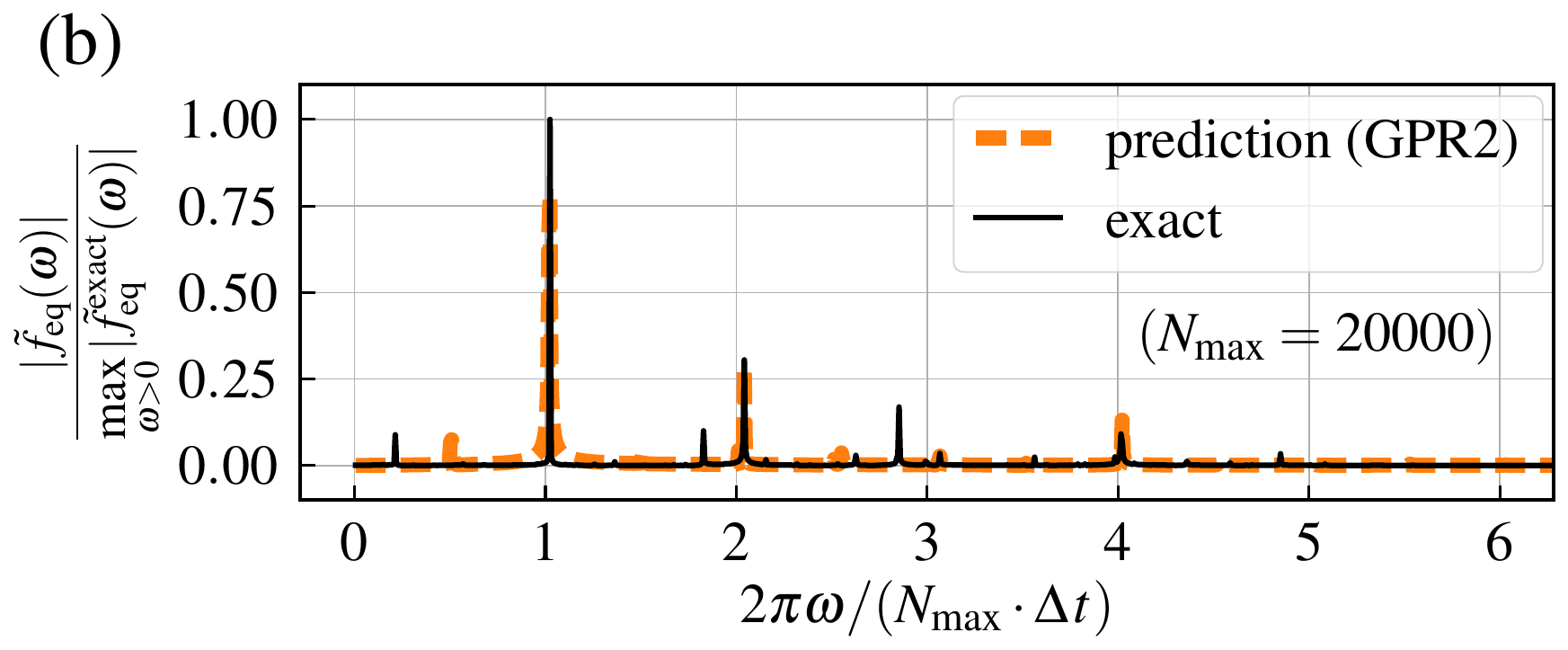}\\
\includegraphics[width=1.00\columnwidth]{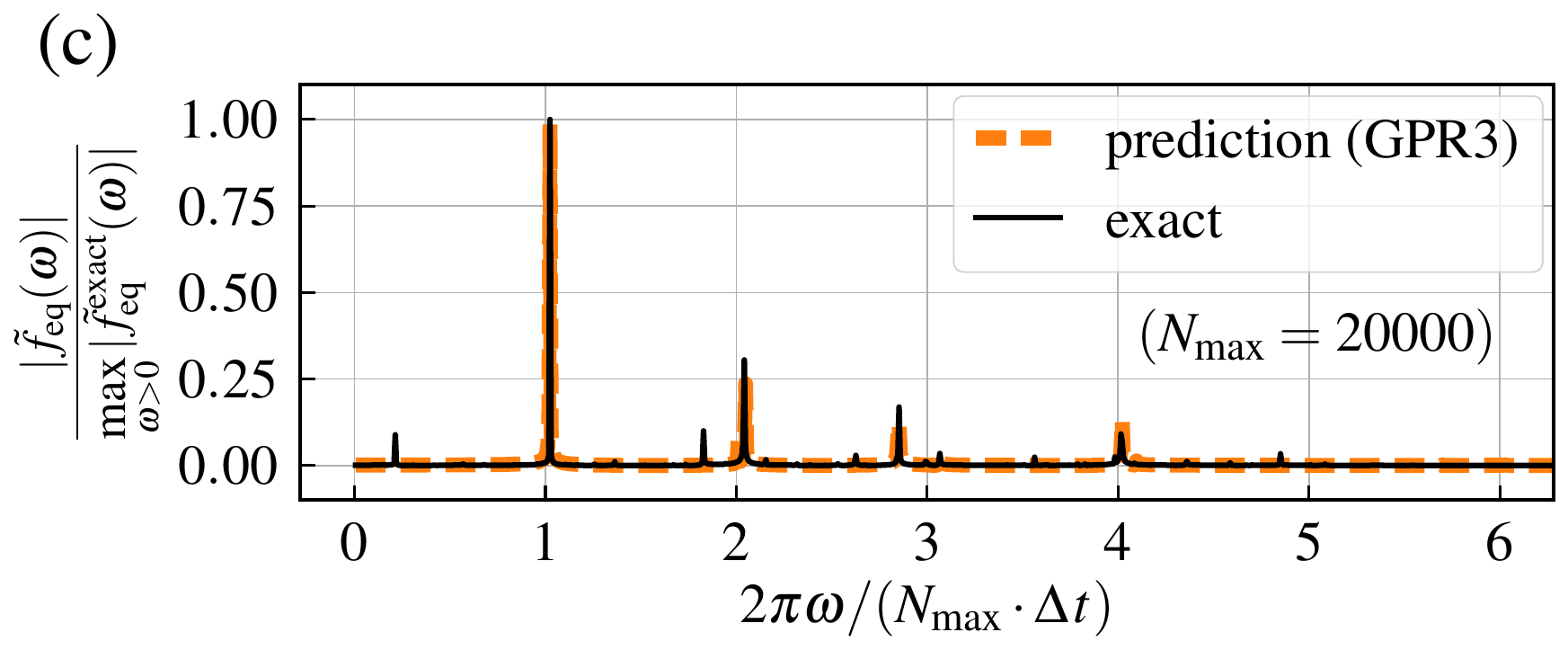}
\caption{%
Fourier transform of the GPR prediction of the equal-time
longitudinal spin-spin correlation function
in the 2D transverse-field Ising model.
We remove the value at $\omega=0$.
We show the results when the number of periodic kernels is
(a) $N_{\rm ker} = 1$,
(b) $N_{\rm ker} = 2$,
and (c) $N_{\rm ker} = 3$.
}
\label{fig_supp:gpr_fourier}
\end{figure}

\begin{figure}[t!]
\centering
\includegraphics[width=1.00\columnwidth]{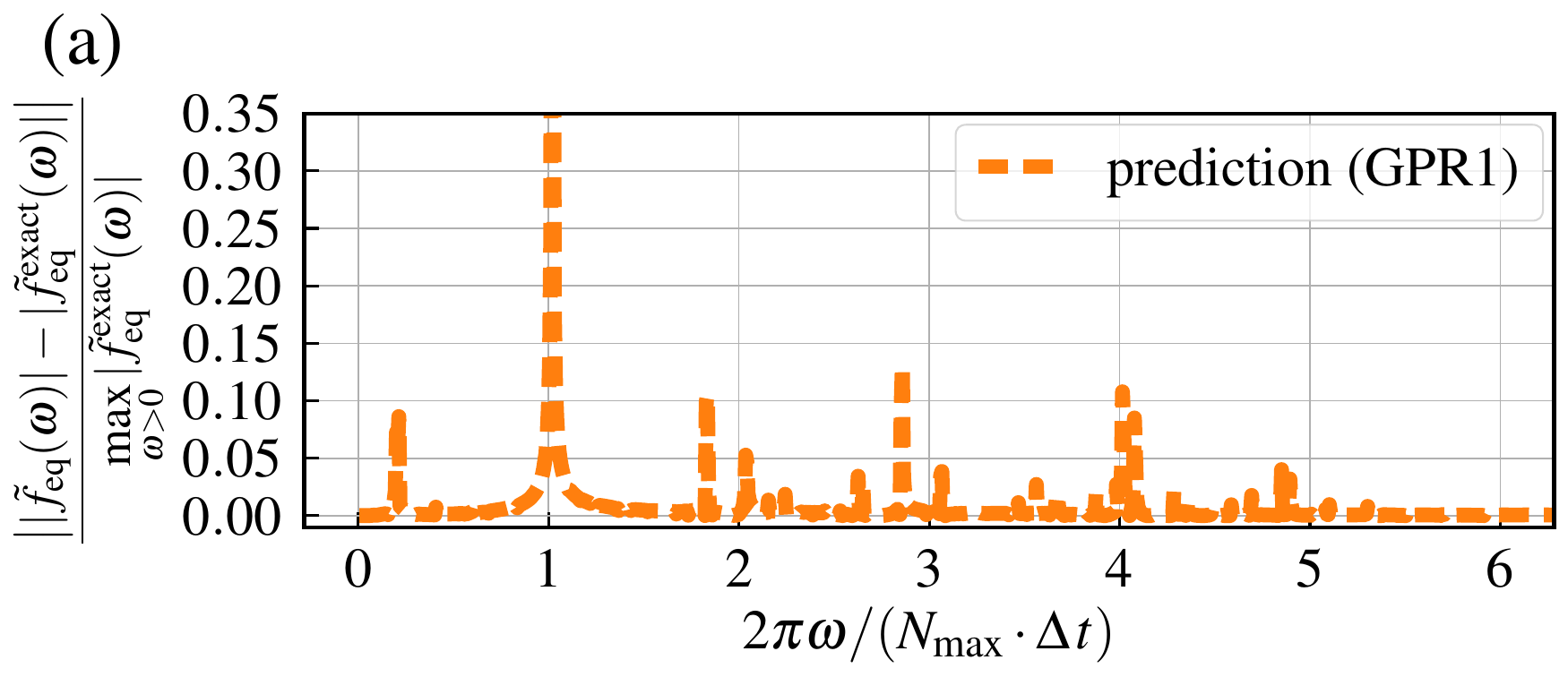}\\
\includegraphics[width=1.00\columnwidth]{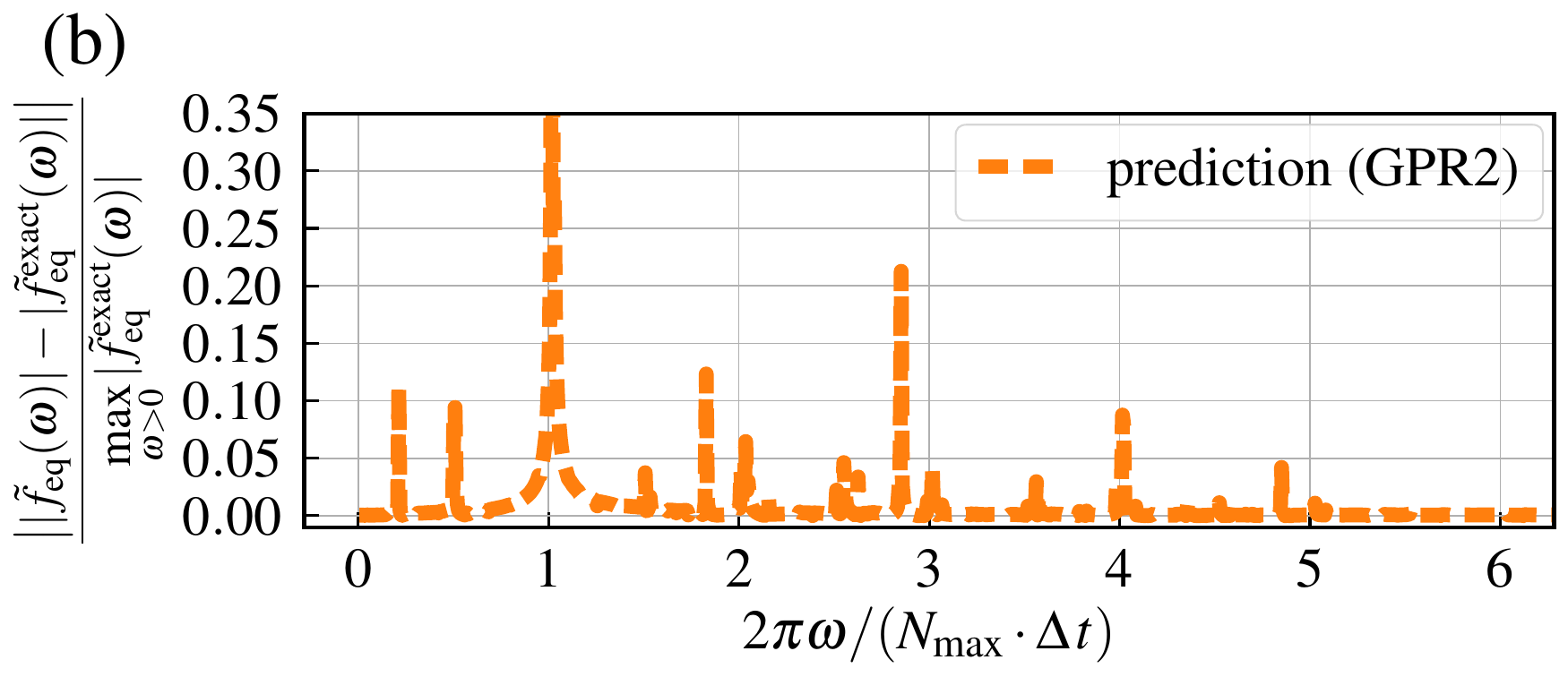}\\
\includegraphics[width=1.00\columnwidth]{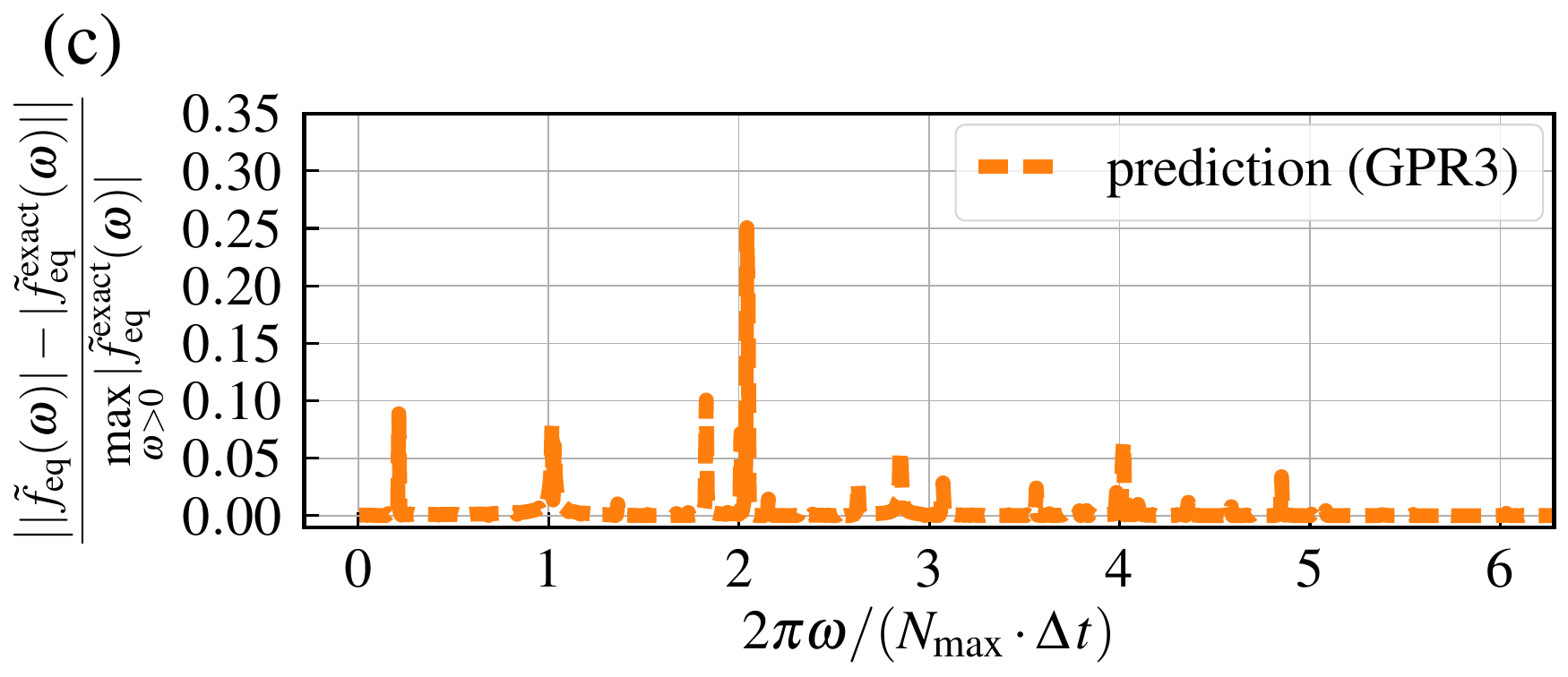}
\caption{%
Relative difference
between the Fourier transform of the GPR prediction
and the exact result
in the 2D transverse-field Ising model.
We show the results when the number of periodic kernels is
(a) $N_{\rm ker} = 1$,
(b) $N_{\rm ker} = 2$,
and (c) $N_{\rm ker} = 3$.
}
\label{fig_supp:gpr_fourier_diff}
\end{figure}

\begin{figure}[t!]
\centering
\includegraphics[width=1.00\columnwidth]{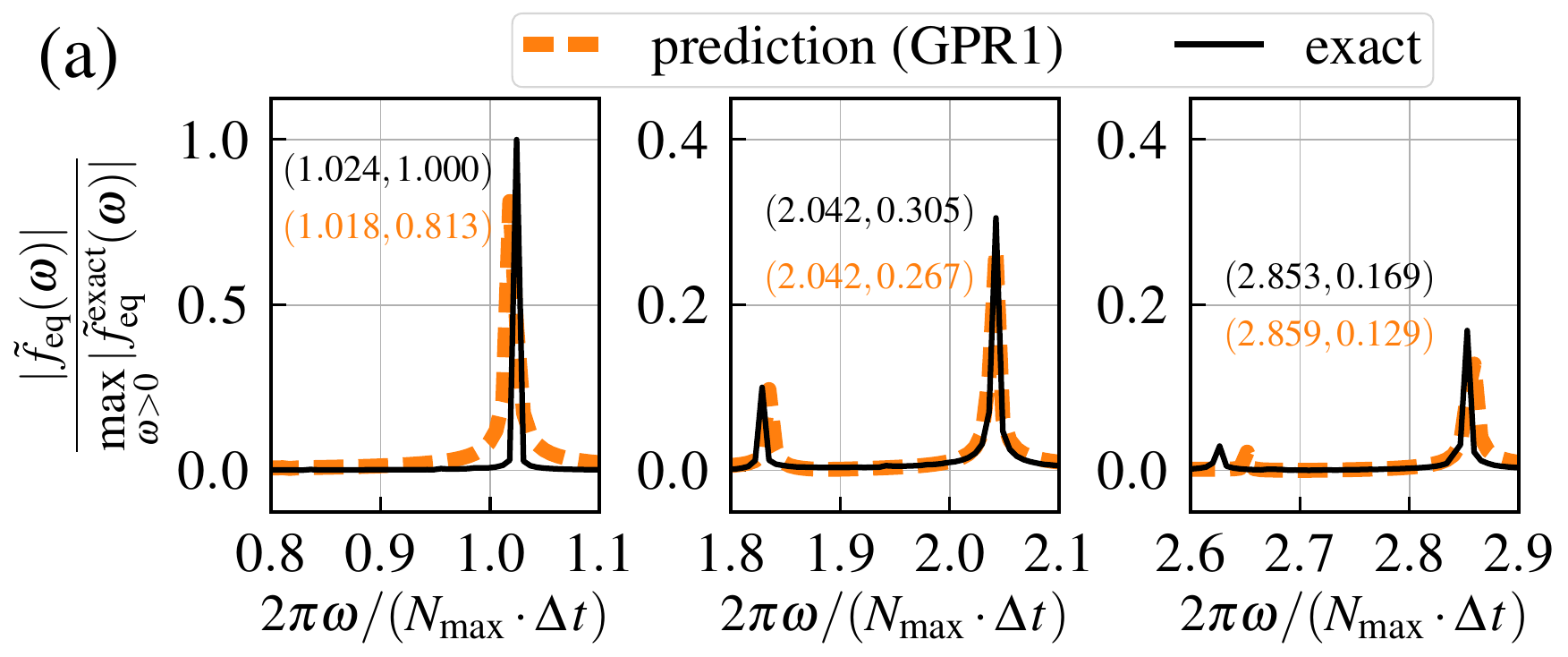}\\
\includegraphics[width=1.00\columnwidth]{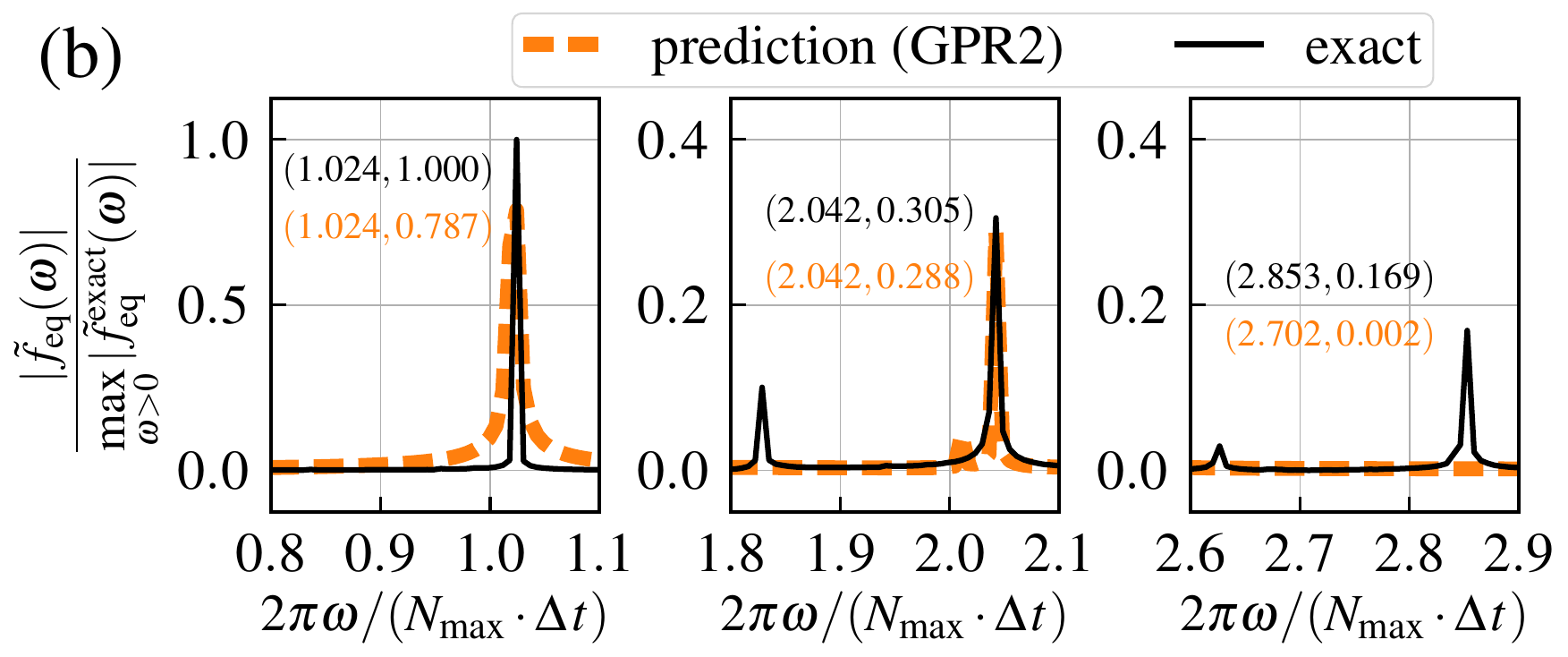}\\
\includegraphics[width=1.00\columnwidth]{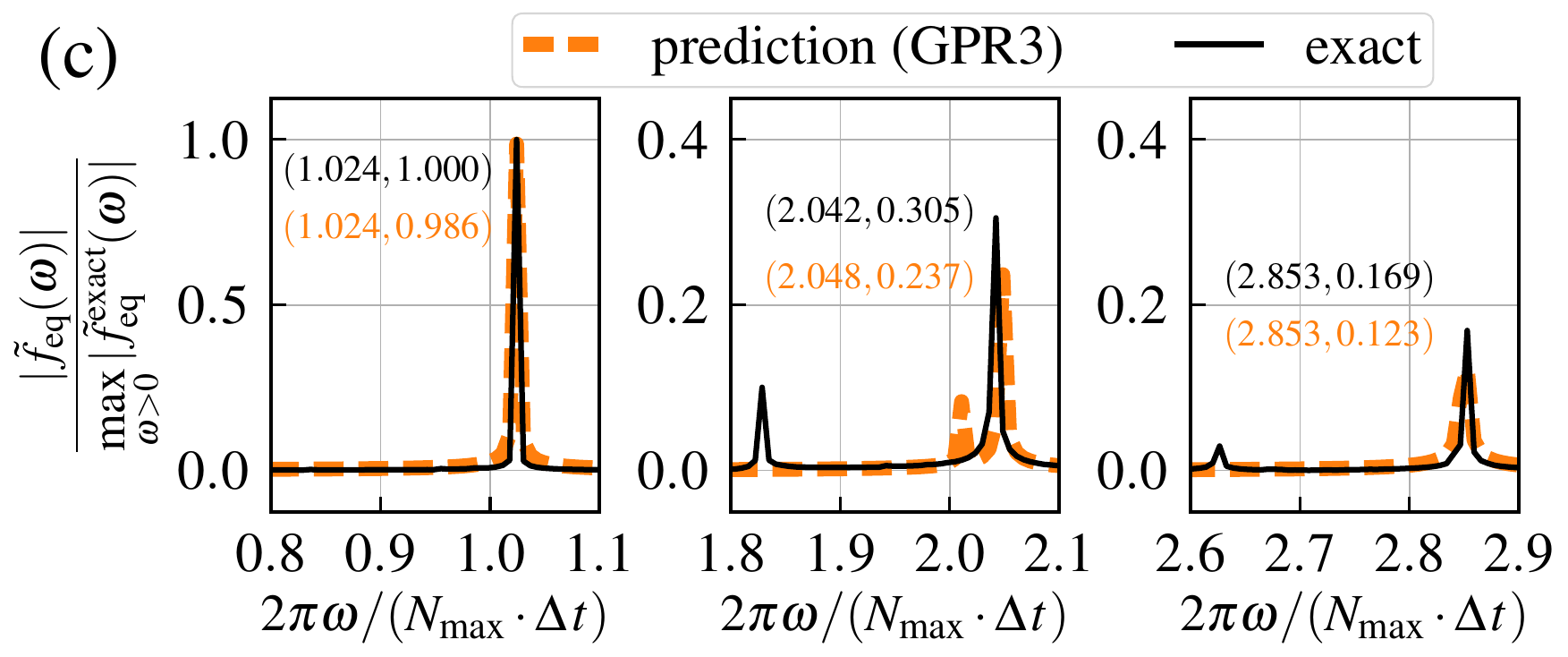}
\caption{%
Magnified views of Fig.~\ref{fig_supp:gpr_fourier}
in the 2D transverse-field Ising model.
We show the results when the number of periodic kernels is
(a) $N_{\rm ker} = 1$,
(b) $N_{\rm ker} = 2$,
and (c) $N_{\rm ker} = 3$.
The position and intensity of each peak
are also shown in the figure in the order (position, intensity).
}
\label{fig_supp:gpr_peak}
\end{figure}

\begin{figure}[t!]
\centering
\includegraphics[width=1.00\columnwidth]{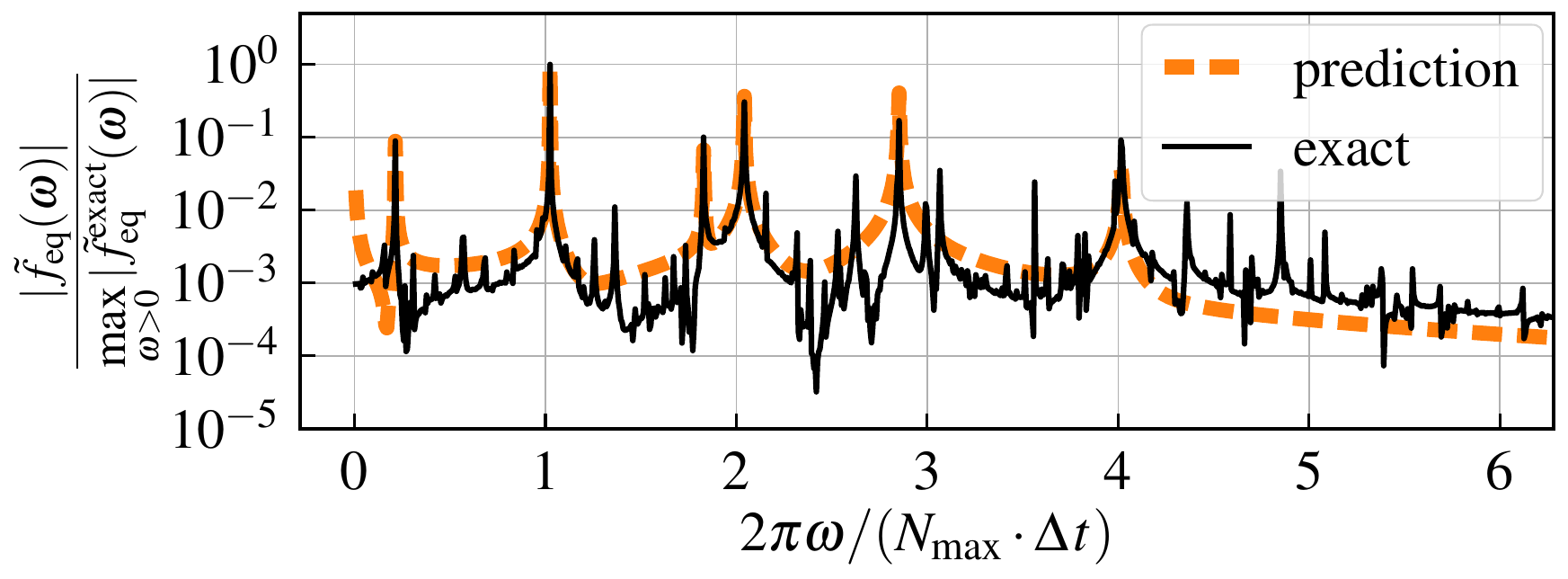}
\caption{%
Fourier transform of the equal-time longitudinal spin-spin correlation function
in the 2D transverse-field Ising model.
We compare the exact result (solid line)
and the DMD prediction (dashed line)
for the absolute value of the Fourier-transformed correlation function.
The vertical axis is in the logarithmic scale.
}
\label{fig_supp:ising_2d_corr_fourier_log}
\end{figure}

\begin{figure}[t!]
\centering
\includegraphics[width=1.00\columnwidth]{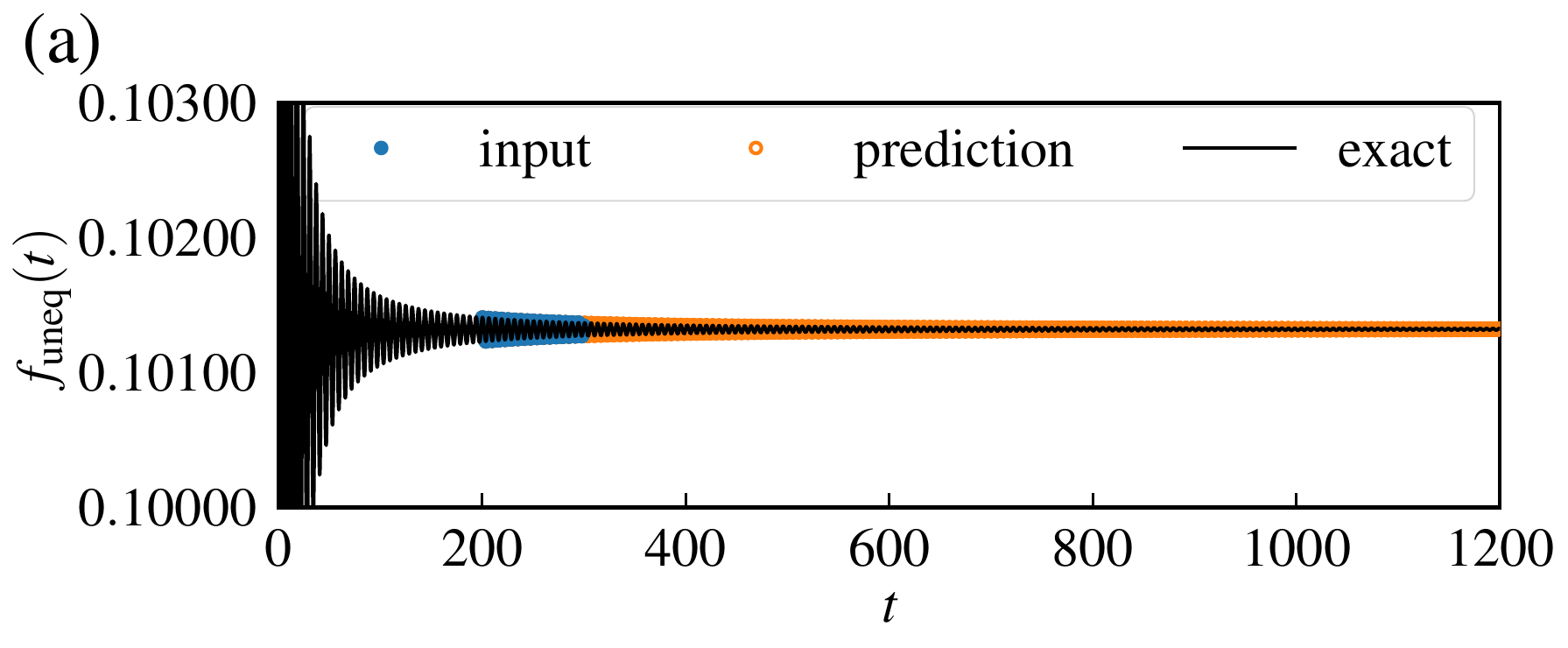}\\
\includegraphics[width=1.00\columnwidth]{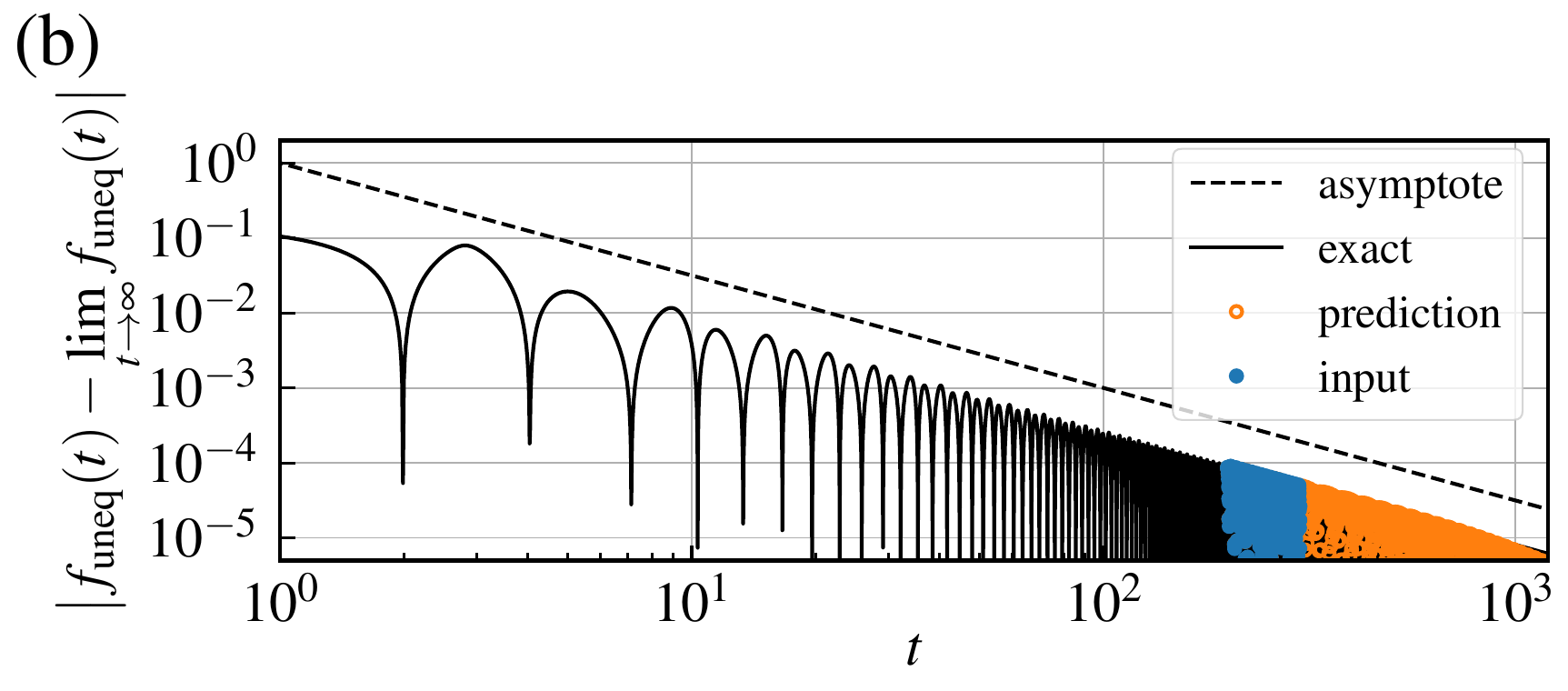}\\
\includegraphics[width=1.00\columnwidth]{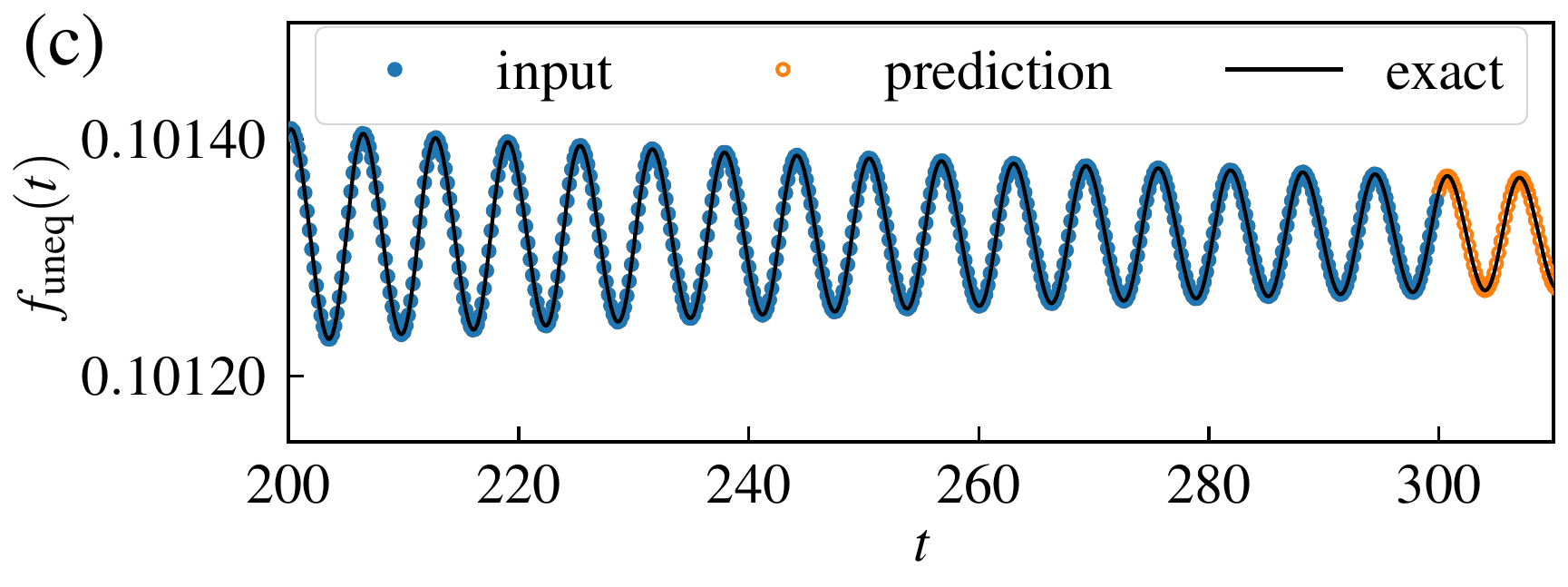}\\
\includegraphics[width=1.00\columnwidth]{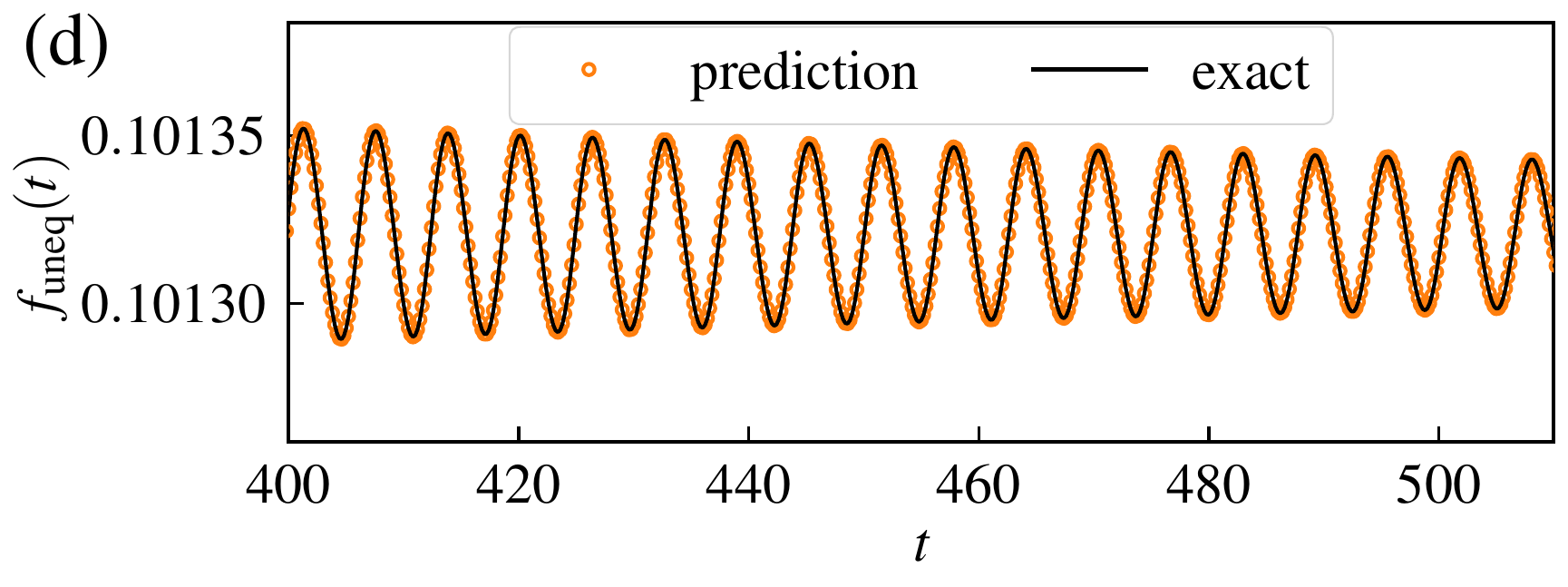}\\
\includegraphics[width=1.00\columnwidth]{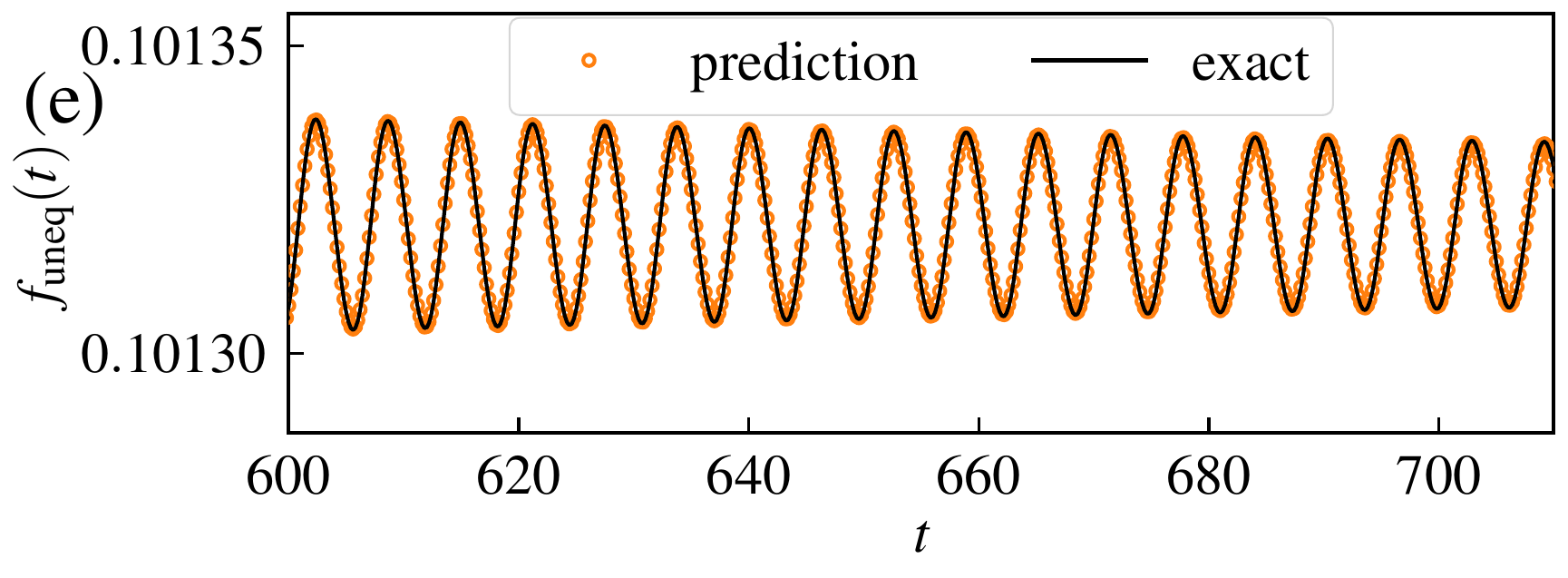}
\caption{%
DMD prediction of the unequal-time onsite transverse spin-spin correlation
function
in the 1D transverse-field Ising model
when the origin of the time series is shifted
by $t_{\rm 0} = 200$.
We show the time evolution of
the absolute value of the correlation function for
(a) $t \in [0, 1200)$
and
(b) that in the logarithmic scale.
The magnified views of the time evolution for
(c) $t \in [200, 310]$,
(d) $t \in [400, 510]$,
and
(e) $t \in [600, 710]$
are also shown.
The solid line is the exact result,
the blue filled circles are the input data,
and the orange open circles are the predicted data.
The data points $f_{{\rm uneq},n}=f_{\rm uneq}(n\cdot\Delta t)=f_{\rm uneq}(t)$ are plotted
only when $n$ is a multiple of $20$.
}
\label{fig_supp:ising_1d_corr_dmd}
\end{figure}

We examine the
accuracy of the GPR 
method~\cite{rasmussen2006,pedregosa2011}
compared to that of the DMD for the example of the equal-time longitudinal spin-spin correlation function after a sudden quench in a finite-size system of
the 2D transverse-field Ising model shown in Sec.~\ref{subsec:corr_no_damp}.
The GPR method is a technique used to interpolate function values
for unobserved data from input data.
It relies on the assumption that for similar input $t$,
the corresponding output $f(t)$ will also be similar.
Even when linear regression struggles to fit the data adequately,
the GPR method
is able to find a suitable fit in a number of cases.
The GPR method is also useful for estimating the error of the
prediction.
Moreover, by appropriately selecting the kernel function,
we can flexibly choose the model to fit the data.

We choose the kernel functions for the GPR method
in the following manner.
Because the equal-time longitudinal spin-spin correlation function
after a sudden quench
exhibits nearly periodic oscillations,
we choose the periodic kernel function
\begin{align}
 \label{eq:kernel_per}
 k_{\rm per}(t,t')
 =
 \sigma_{\rm per}^2
 \exp\left[
 -\frac{2\sin^2\left(\pi|t-t'|/T_{\rm per}\right)}{l_{\rm per}^2}
 \right]
\end{align}
with the variance
$\sigma_{\rm per}^2$,
the period
$T_{\rm per}$,
and the length scale
$l_{\rm per}$.
To cope with multiple oscillatory components
in the equal-time longitudinal spin-spin correlation function,
we study a sum of periodic kernel functions
with different 
variances
$\sigma_{\rm per}^2$,
periods
$T_{\rm per}$,
and length scales
$l_{\rm per}$.
Number of periodic kernel functions
($N_{\rm ker}$) is increased
to improve the accuracy of the prediction.
As for finite-size systems we considered,
the correlation function does not show damping.
Therefore, we do not
introduce additional kernel functions
such as the radial basis kernel function,
which is often used to model growing or decaying behavior.

For the optimization of the kernel parameters,
we take the following steps.
Here, we consider the time-series data
for $t \in [0,100]$ as the input data
and wish to predict the data for $t \in [0,1000]$.
\begin{enumerate}
 \item
We tentatively determine the kernel parameters
from the input data $t \in [0,50]$:
The initial values of the kernel parameters
are randomly chosen and are optimized by maximizing the log marginal likelihood.
 \item
Using the optimized kernel parameters,
we predict the correlation function
for $t \in [0,100]$.
Then,
the error of the prediction is estimated from the difference
between the exact result and the predicted data
by calculating the $L^2$ norm of the difference
and normalizing it by the number of data points.
 \item
The above steps 1 and 2
are repeated until the error of the prediction
becomes sufficiently small.
To accelerate the convergence,
we perform Bayesian
optimization~\cite{brochu2010_arxiv,snoek2012,nogueira2014}
to find the best kernel parameters.
The 
error of the prediction
calculated in step 2
is interpreted as the cost function to be
minimized
for Bayesian optimization.
We typically choose $20$ random initial points
and then perform $50$ iterations
of Bayesian optimization.
 \item
Finally, we predict the correlation function
for $t \in [0,1000)$
using the optimized kernel parameters.
\end{enumerate}

In practice,
we use the \textsc{Python} libraries
\textsc{Scikit-learn}~\cite{pedregosa2011}
and
\textsc{BayesianOptimization}~\cite{nogueira2014}
to perform the GPR and Bayesian optimization, respectively.
Choosing appropriate initial values
(in particular, the periods $T_{\rm per}$)
of the kernel parameters
with the help of Bayesian optimization is important
to avoid getting trapped to local minima
in the parameter space.

We show the prediction of the equal-time longitudinal spin-spin correlation
function by the GPR method
in Fig.~\ref{fig_supp:gpr_corr}.
With increase of the number of kernel functions,
the deviation of the prediction from the exact result
appears to decrease.
However, the accuracy of the prediction
is worse than that of the DMD prediction
[compare Fig.~\ref{fig_supp:gpr_corr} with
Fig.~\ref{fig:ising_2d_corr_dmd}(b)].
We have increased the number of kernel functions
up to $N_{\rm ker}=10$, but the deviation
from the exact result does not decrease further (not shown).
This might be because the GPR method
requires a large number of kernel parameters
to fit the data accurately,
and it is difficult to optimize
such a large number of parameters.
On the other hand,
the DMD method seems to be able to
extract the physically relevant information
from the short-time correlation function
by taking advantage of
the low-rank structure
of the correlation function data
via the SVD.

We also show the Fourier-transformed correlation function
obtained by the GPR prediction
in Fig.~\ref{fig_supp:gpr_fourier}.
The GPR prediction partially reproduces the peak positions
of the exact result.
The number of peaks that are reproduced
increases with increasing the number of kernel functions.
As shown in Fig.~\ref{fig_supp:gpr_fourier_diff},
the difference between the exact result and the GPR prediction
gets smaller with increasing the number of kernel functions.

On the other hand,
when we look at the Fourier-transformed correlation function
more carefully,
we find that the GPR prediction
fails to reproduce the exact peak positions.
In Fig.~\ref{fig_supp:gpr_peak},
we plot the magnified views of Fig.~\ref{fig_supp:gpr_fourier}.
It is difficult for the GPR method
to predict all the positions
of the largest, second-largest, and third-largest peaks
of the exact result simultaneously.
At least one of the peak positions
is slightly shifted from the exact result.
Again, the GPR method suffers from the difficulty
of optimizing a large number of kernel parameters,
whereas the DMD method seems to be able to
capture the physically relevant information
[see Sec.~\ref{fig:ising_2d_corr_fourier}(d)].

\begin{figure}[!t]
\centering
\includegraphics[width=.500\columnwidth]{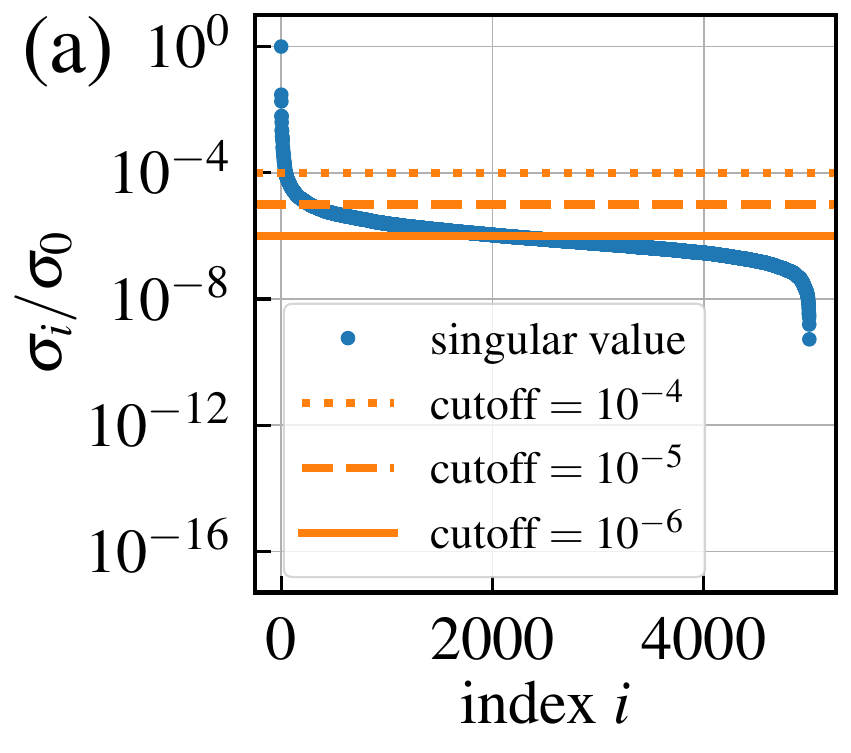}%
~%
\includegraphics[width=.455\columnwidth]{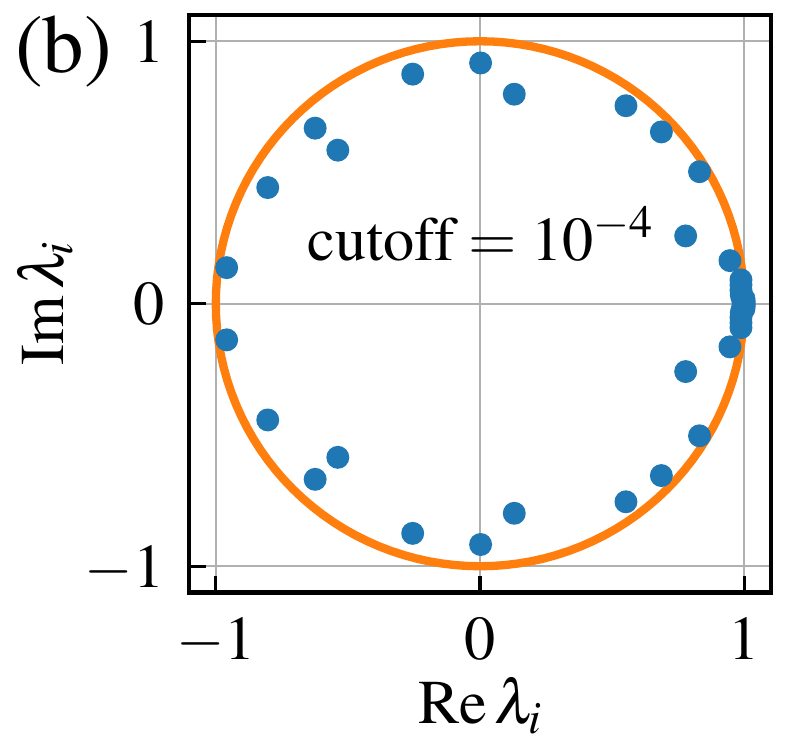}%
\\
~~~~~%
\includegraphics[width=.455\columnwidth]{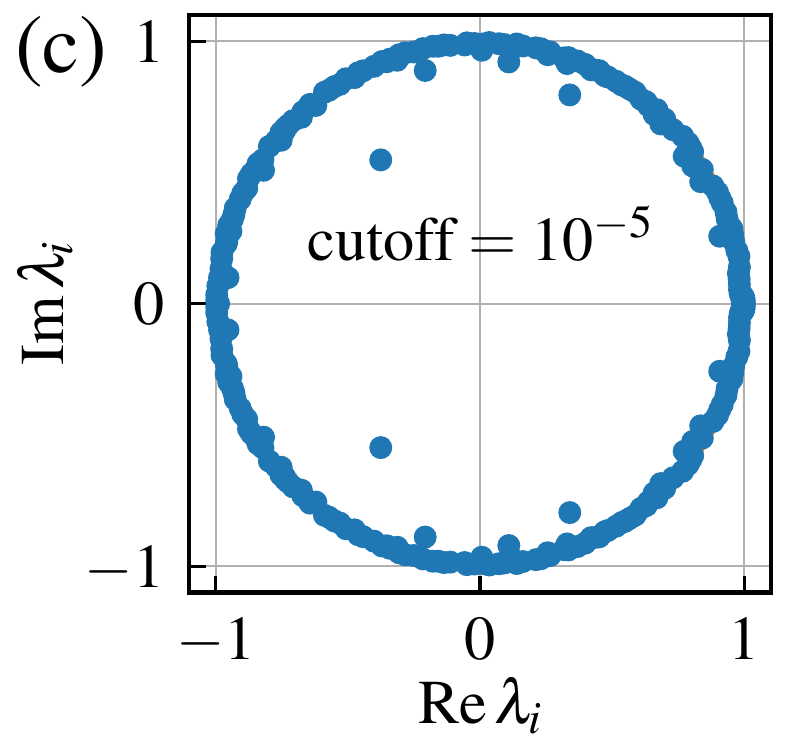}%
~%
\includegraphics[width=.455\columnwidth]{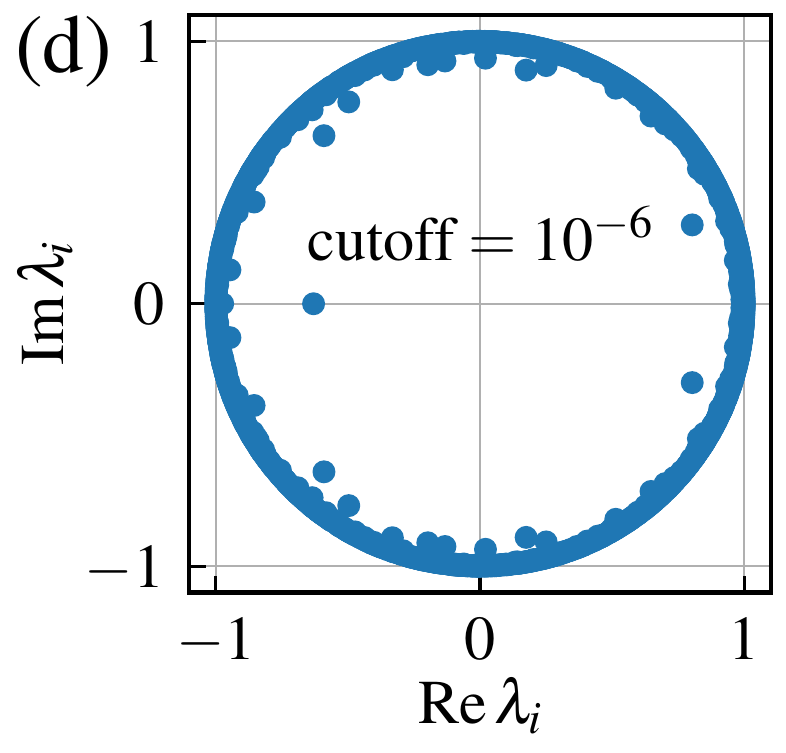}%
\caption{%
DMD parameters in the case of
the unequal-time onsite transverse spin-spin correlation function
in the 1D transverse-field Ising model
with noise.
(a) Singular values $\sigma_i$ of the truncated SVD of the matrix $X_0$.
Eigenvalues $\lambda_i$ of the matrix $\tilde{A}$
are plotted in the complex plane
for cutoff
(b) $\epsilon = 10^{-4}$,
(c) $\epsilon = 10^{-5}$,
and
(d) $\epsilon = 10^{-6}$.
}
\label{fig_supp:ising_1d_corr_lmd_sgm_noise}
\end{figure}

\begin{figure}[!t]
\centering
\includegraphics[width=1.00\columnwidth]{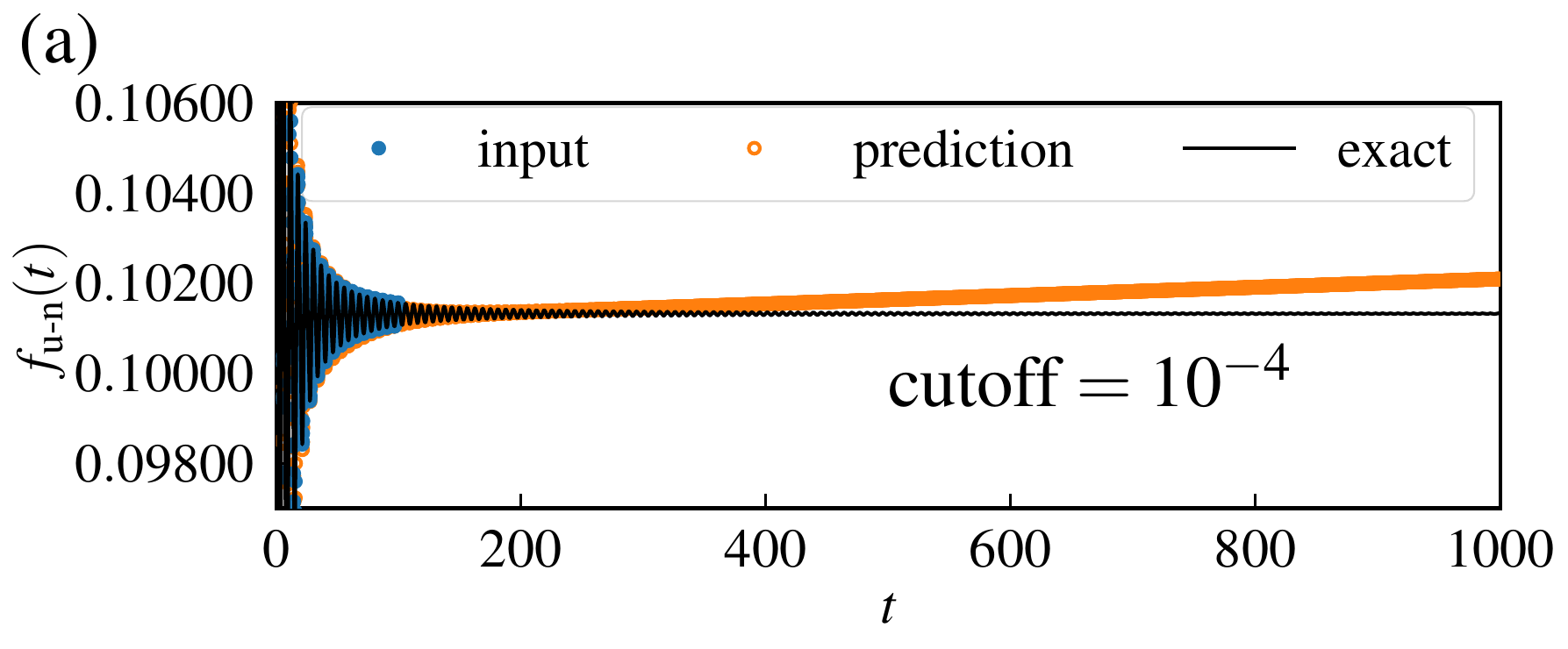}\\
\includegraphics[width=1.00\columnwidth]{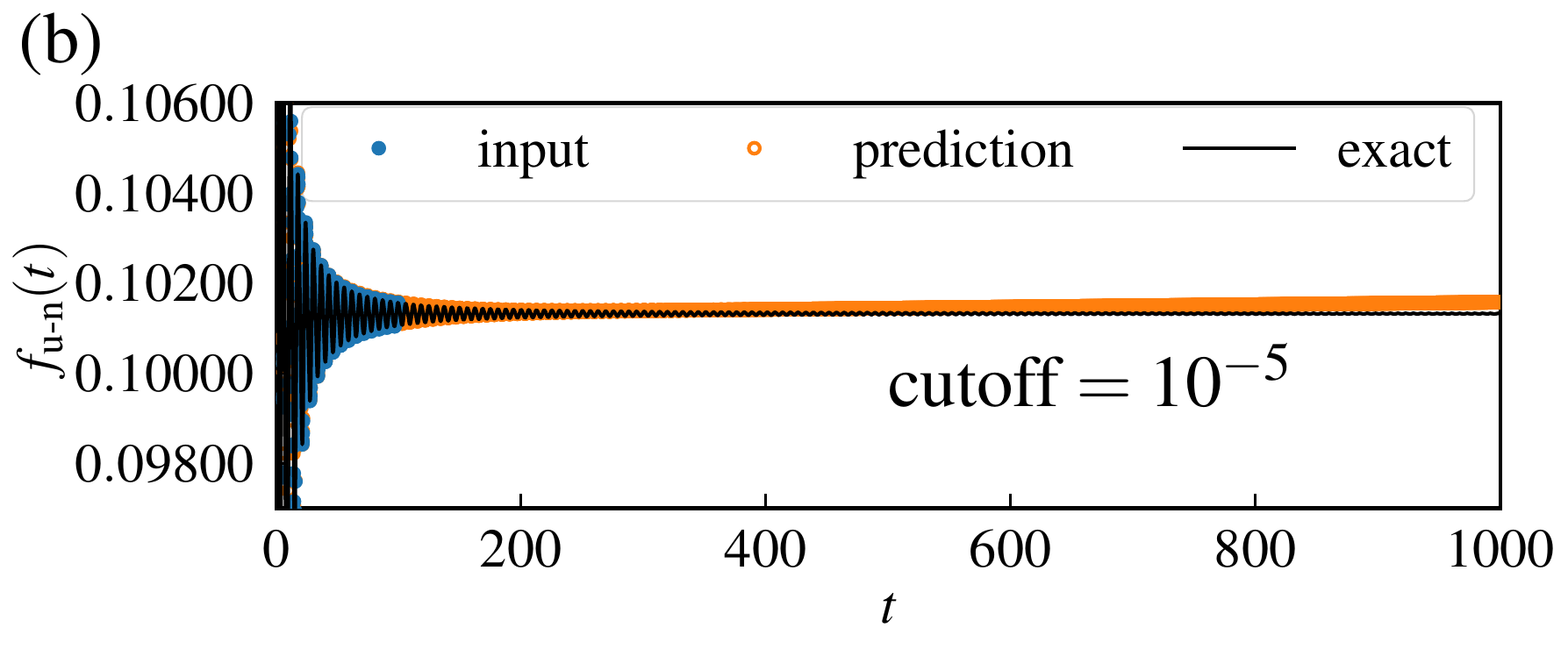}\\
\includegraphics[width=1.00\columnwidth]{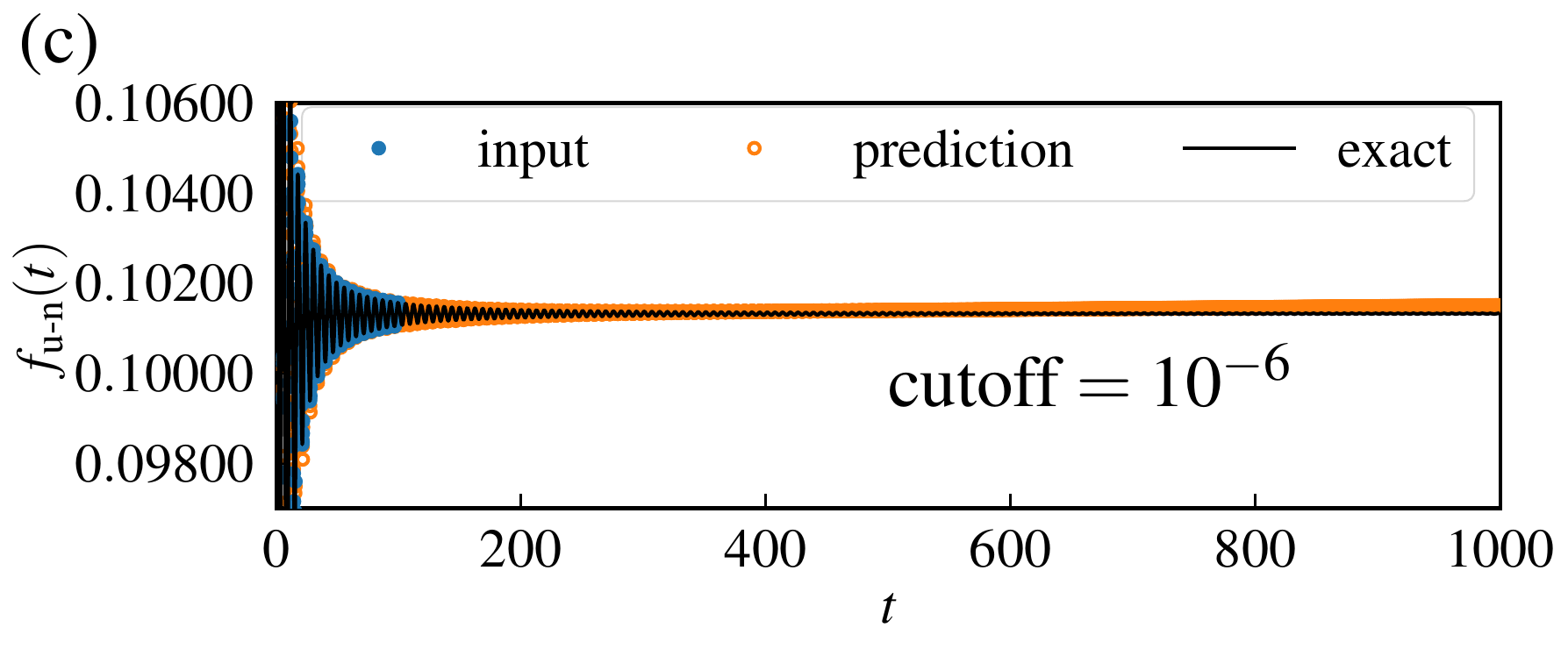}
\caption{%
DMD prediction of the unequal-time onsite transverse spin-spin correlation
function
in the 1D transverse-field Ising model
with noise.
The cutoff is set to
(a) $\epsilon = 10^{-4}$,
(b) $\epsilon = 10^{-5}$,
and
(c) $\epsilon = 10^{-6}$.
}
\label{fig_supp:ising_1d_corr_dmd_noise}
\end{figure}

\begin{figure}[!t]
\centering
\includegraphics[width=1.00\columnwidth]{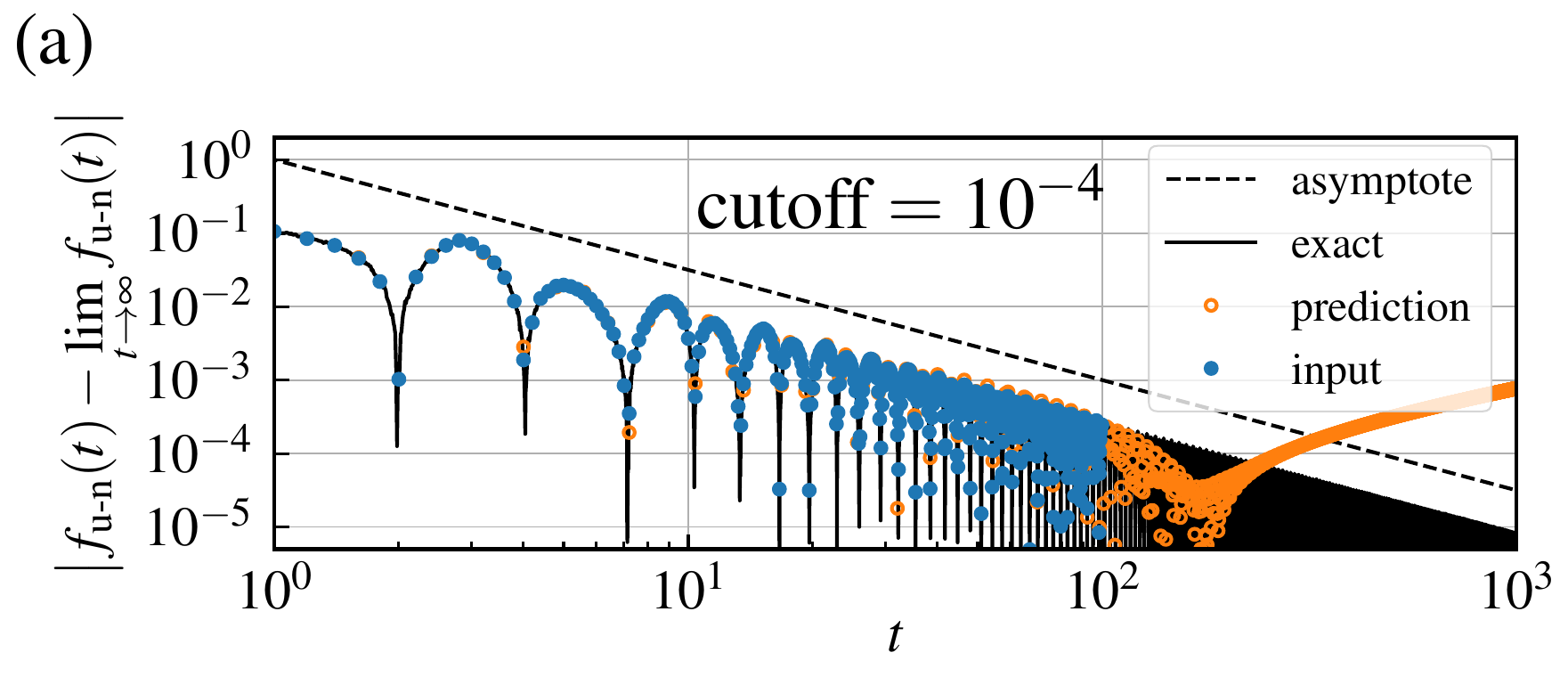}\\
\includegraphics[width=1.00\columnwidth]{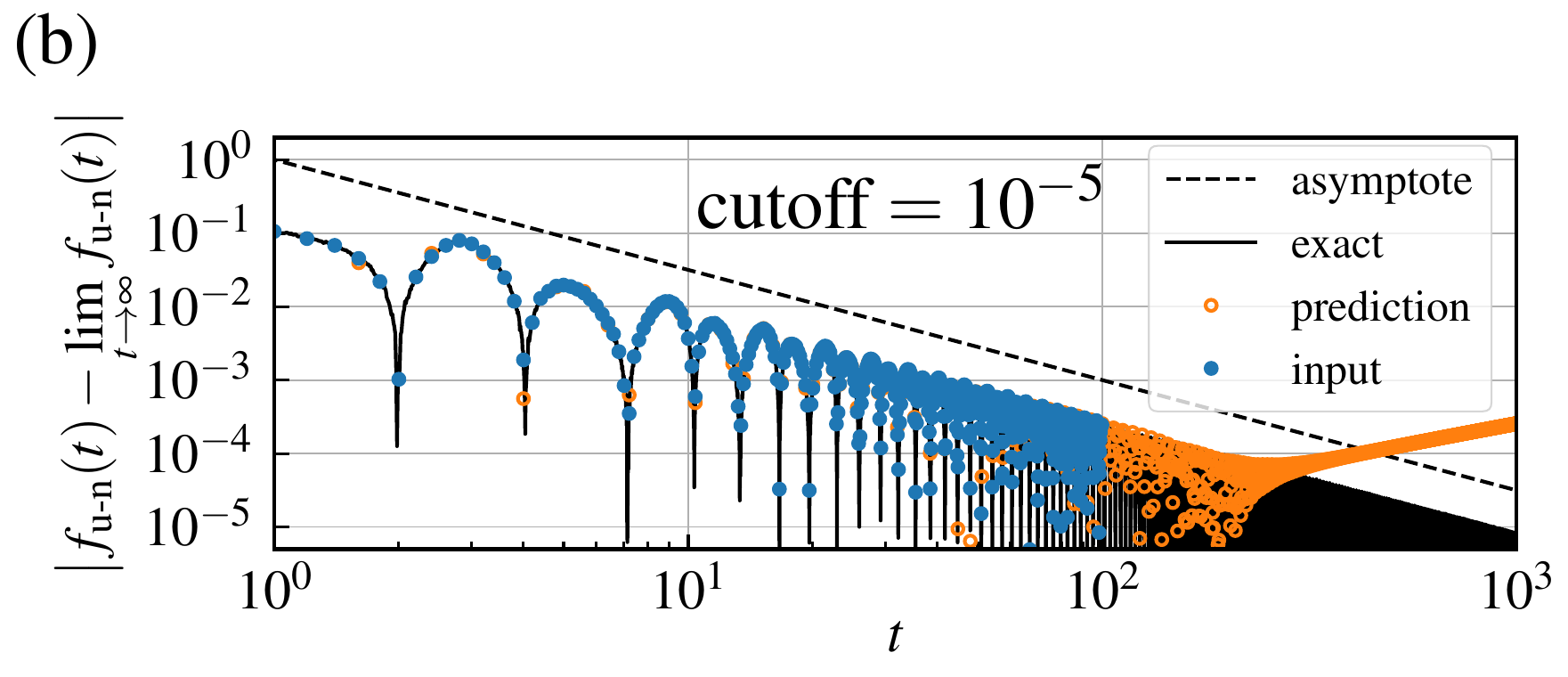}\\
\includegraphics[width=1.00\columnwidth]{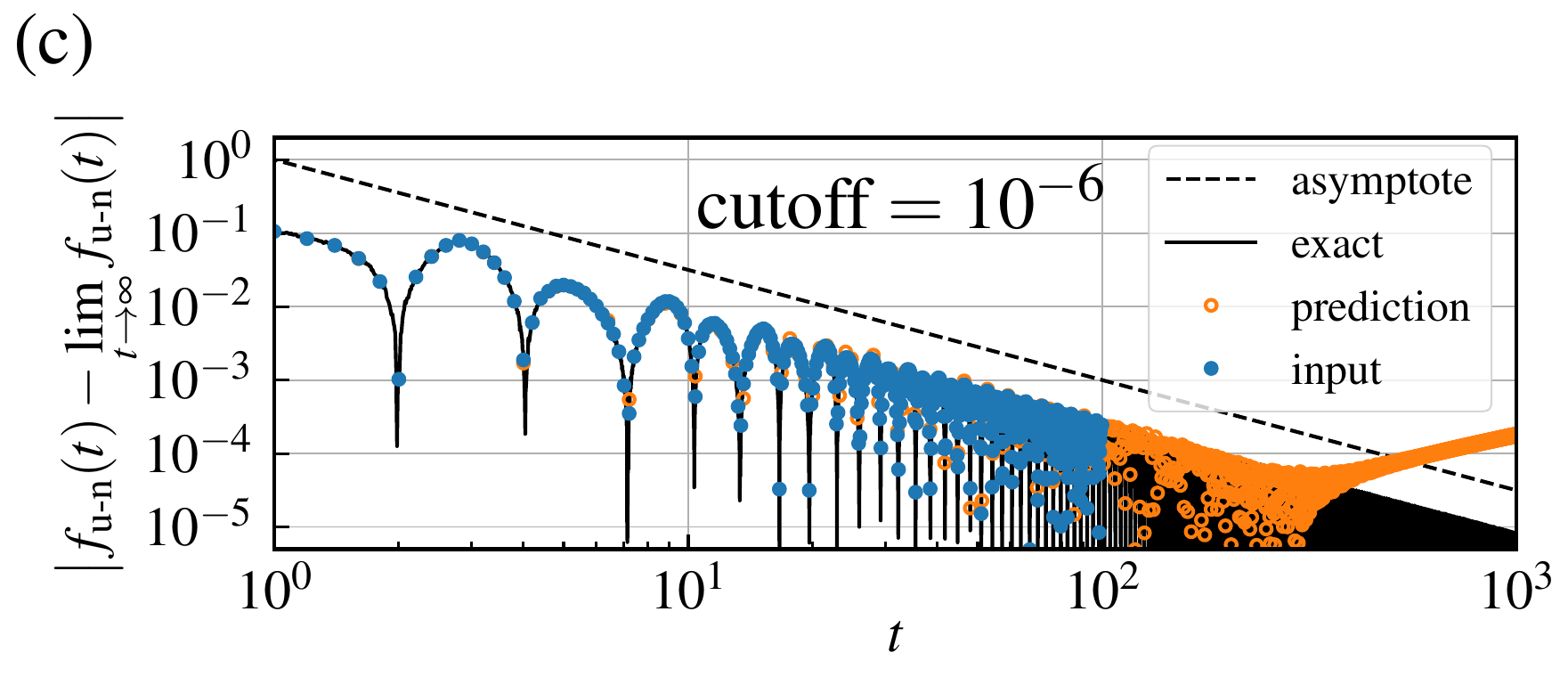}
\caption{%
DMD prediction of the unequal-time onsite transverse spin-spin correlation
function
in the 1D transverse-field Ising model
with noise
in the logarithmic scale.
The cut-off parameters are the same as those in
Fig.~\ref{fig_supp:ising_1d_corr_dmd_noise}.
}
\label{fig_supp:ising_1d_corr_dmd_noise_log}
\end{figure}

\section{Fourier-transformed correlation function
obtained by the DMD prediction in the logarithmic scale}
\label{sec:fourier_log}

In Sec.~\ref{subsec:corr_no_damp},
we predict the Fourier-transformed correlation function
in the 2D transverse-field Ising model by the DMD.
To carefully examine the difference between the exact result and the DMD prediction,
we plot it in the logarithmic scale in
Fig.~\ref{fig_supp:ising_2d_corr_fourier_log}.

The DMD prediction reproduces the exact peak positions
when the absolute value of the Fourier-transformed correlation function
is larger than $10\%$
of the maximum value of the exact result
$\max_{\omega>0} |\tilde{f}(\omega)|$.
This result indicates that
the DMD method can be used to extract
the physically relevant information
from the short-time correlation function.

On the other hand,
the DMD prediction fails to reproduce the exact peak positions
with small amplitudes.
These spiky structures with small intensities
originate from the finite-size effect.
When the system size is small,
the correlation function shows a recurrence
at a timescale proportional to the linear size of the system.
The dominant peaks in the Fourier spectrum
is mainly determined by
the short-time dynamics of the correlation function
until the recurrence time.
As the system size increases,
the number of peaks in the Fourier spectrum increases,
and
accumulates gradually to form a smooth curve.
In the thermodynamic limit then,
we would observe a continuous spectrum
aside from some dominant peaks or kinks
in the Fourier spectrum,
just as in the case of the 1D transverse-field Ising model
on an infinite chain [see Fig.~\ref{fig:ising_1d_corr_fourier}(a)].
We may be able to reproduce such a nearly smooth curve
in the Fourier spectrum
when the correlation function
in a sufficiently large system 
for a sufficiently long time
is available as the input data for the DMD,
although obtaining such input data
is practically not feasible
for the nonintegrable 2D transverse-field Ising model.

\section{Effects of shifting the origin of the time series}
\label{sec:shift_origin}

In Sec.~\ref{subsec:corr_damp},
the correlation function
in the 1D transverse-field Ising model is estimated by the DMD
by taking the time interval of $t_{\rm input}$ immediately after the sudden quench.
This choice can be quantitatively improved
by shifting the origin of the time series.
This is because data just after $t=0$ are strongly influenced by
the initial transient region.

We show the DMD prediction
when the origin of the time series is shifted
by $t_{\rm shift} = 200$
in Fig.~\ref{fig_supp:ising_1d_corr_dmd}.
In the logarithmic scale,
the DMD prediction with the shifted origin
looks reproducing the overall behavior of the exact result as much as
the case without shifting the origin
[compare Fig.~\ref{fig:ising_1d_corr_dmd}(b)
and Fig.~\ref{fig_supp:ising_1d_corr_dmd}(b)].
However, when one looks into more details, the DMD prediction better reproduces
the exact result
even up to $t = 700$.
By comparing the data for
$t-t_{\rm shift} \in [400, 510]$
with and without shifting the origin
[see Fig.~\ref{fig:ising_1d_corr_dmd}(e)
and Fig.~\ref{fig_supp:ising_1d_corr_dmd}(e)],
we can see a substantial improvement in the DMD prediction.

\begin{figure}[!t]
\centering
\includegraphics[width=0.495\columnwidth]{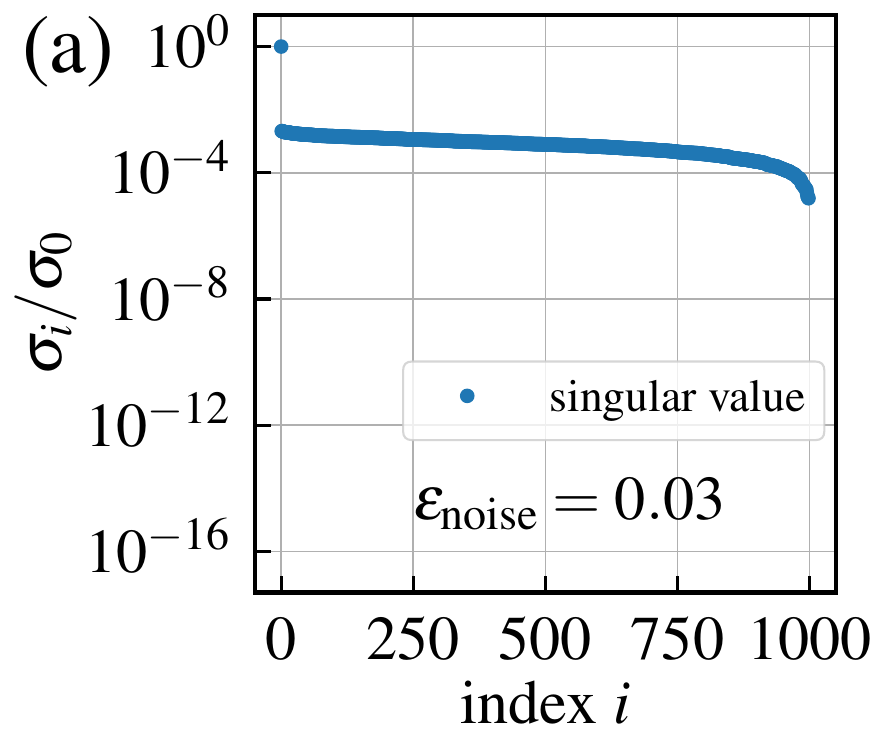}%
\hfil%
\includegraphics[width=0.495\columnwidth]{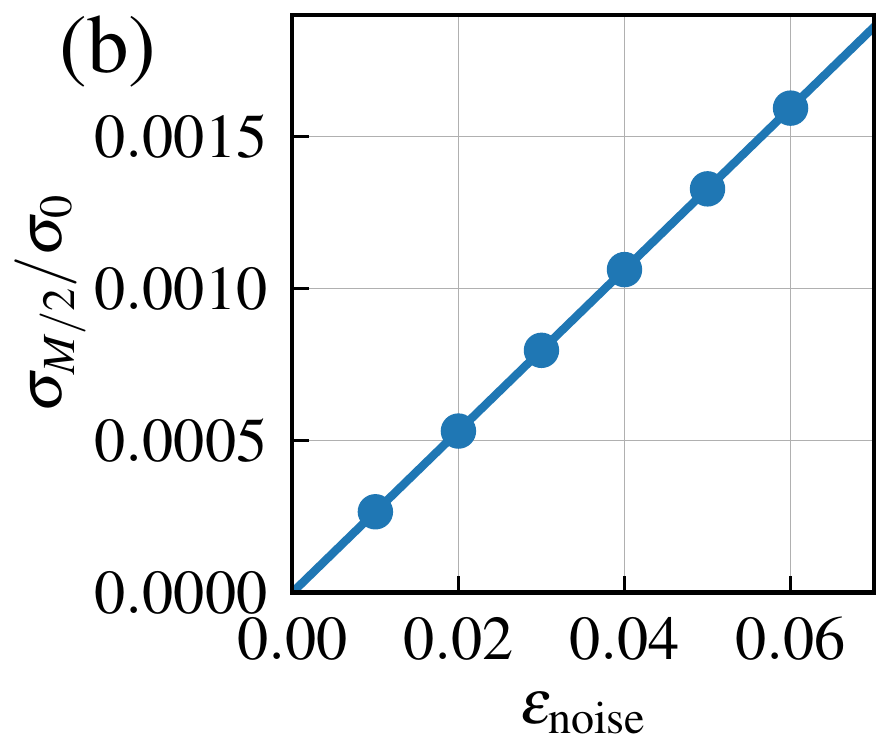}%
\caption{%
Singular values $\sigma_i$ of the truncated SVD of the matrix $X_0$
in the case of
a constant function
with noise $\epsilon_{\rm noise}$.
We choose the parameters $M = 1000$ and $N = 2M$
for the $M\times (N-M)$ matrix $X_0$.
(a) Normalized singular values $\sigma_{i}/\sigma_0$
for $\epsilon_{\rm noise} = 0.03$.
(b) Noise level $\epsilon_{\rm noise}$ dependence of
$(M/2)$th normalized singular values $\sigma_{M/2}/\sigma_0$.
}
\label{fig_supp:svd_vs_noise}
\end{figure}

\section{Effects of noise in input data with damping}
\label{sec:effect_noise_damping}

In Sec.~\ref{subsec:effect_noise},
we have discussed the effects of noise
in input data without damping
on the DMD prediction.
There, we have found that the DMD prediction
keeps up accuracy up to a certain level of
the noise level 
(typically less than a few percent of the maximum value
of input data).
Here, 
the effects is examined by adding the noise
that decays following a power law along with the main input
when the main dynamics is governed by damping.

For this purpose, we employ the 1D transverse-field Ising model
with damping
as in Sec.~\ref{subsec:corr_damp}.
When the noise level is larger than
the amplitude of the correlation function,
the DMD prediction clearly fails to reproduce
the exact result.
Therefore, in the case with damping,
we add a small amount of noise which is nearly
proportional to the amplitude of the correlation function.
We prepare time-series data
$f_{\rm u\text{-}n}(t)$ affected by such a relative noise $\eta(t)$,
which is chosen to be
Gaussian and white random variables with zero mean and
time-dependent variance $[\sigma_{\rm noise}(t)]^2$.
The standard deviation $\sigma_{\rm noise}(t)$
is chosen to be
\begin{align}
 \sigma_{\rm noise}(t)
 = \epsilon_{\rm noise} |f_{\rm orig}(0)| t^{-3/2},
\end{align}
with a small positive parameter $\epsilon_{\rm noise}$
that controls the noise level.
The value $|f_{\rm orig}(0)|$
corresponds to the amplitude
of the original correlation function
at $t=0$, which is $1/4$ in the present case.
Then, the time-series data $f_{\rm u\text{-}n}(t)$ affected by noise $\eta(t)$ is given by
\begin{align}
 f_{\rm u\text{-}n}(t) = |C^{xx}_{\rm uneq}(r=0, t)| + \eta(t),
\end{align}
where $C^{xx}_{\rm uneq}(r=0, t)$ is defined in
Eq.~\eqref{eq:ising_1d_corr_exact}.
We set $\epsilon_{\rm noise} = 0.01$ hereafter.

For the DMD, we choose the same parameters
as those in Sec.~\ref{subsec:corr_damp}.
Just as in the case without damping,
the singular values $\sigma_i$
of the matrix $\bm{X}_0$
constructed from the input data with damping
exhibit a plateaulike structure
[see Fig.~\ref{fig_supp:ising_1d_corr_lmd_sgm_noise}(a)].
We gradually decrease the cut-off parameter $\epsilon$
and examine the eigenvalues $\lambda_i$
of the matrix $\bm{A}$
[see Fig.~\ref{fig_supp:ising_1d_corr_lmd_sgm_noise}(b)--\ref{fig_supp:ising_1d_corr_lmd_sgm_noise}(d)].
As for $\epsilon \gtrsim 10^{-6}$,
$|\lambda_i|$ is found to be smaller than or equal to unity.
When $\epsilon$ is decreased further,
the divergence of time-series data
is observed for $t \approx 10^3$
because of some $|\lambda_i|$ slightly exceeding unity.
We thus choose $\epsilon \ge 10^{-6}$ as the cut-off parameter.
This cut-off parameter is comparable to the value of
$\sigma_i/\sigma_0$ at which the plateaulike structure appears.
The corresponding rank of the truncated SVD
for $\epsilon = 10^{-6}$
is $R = 2088$.

We show the DMD prediction
in Fig.~\ref{fig_supp:ising_1d_corr_dmd_noise}.
As the cut-off parameter $\epsilon$ is decreased,
the DMD prediction gradually reproduces
the long-time behavior of the exact result.
When $\epsilon = 10^{-6}$,
the predicted correlation function
nearly converges to a constant value,
which is consistent with the exact result.

To see tiny damped oscillatory behavior around the constant
$f_{\rm u\text{-}n}(t)\approx 0.101$, we have examined the DMD prediction in the logarithmic scale:
The predicted power-law decay to the constant
becomes extended to a longer time
as the cut-off parameter $\epsilon$ is decreased and progressively approaches the exact power-decay
(see Fig.~\ref{fig_supp:ising_1d_corr_dmd_noise_log}).
However,
even when $\epsilon = 10^{-6}$,
the DMD prediction for
$t \gtrsim 200$ starts deviating from the exact result and from the asymptotic $t^{-3/2}$ scaling.
If one is concerned with this tiny damped time dependence away from the constant,
then the DMD can predict the time evolution of
the correlation function
up to approximately twice the duration
of the input data.
In this regard, it is rather difficult to further improve the DMD prediction
when input data with damping are affected by noise.
Nevertheless, the essential and overall damping to the constant is well captured.

\section{Estimating the noise level from singular values}
\label{sec:estimate_noise_level}

In Sec.~\ref{subsec:effect_noise},
we examine the effects of noise
in input data
and find that singular values of the matrix $\bm{X}_0$
constructed from the input data
exhibit a plateaulike structure.
In this section,
by taking a simpler example of noisy input data,
we show that the position of the plateau
is proportional to the noise level
in input data.

To get insight into the plateaulike structure
observed in the singular values
for noisy input data,
we consider the case where
the original input data without noise
is a constant function, $f_{\rm orig}(t) = 1$.
We prepare time-series data $f_{\rm n}(t)$
affected by noise $\eta(t)$,
which is chosen to be
Gaussian white random variables with zero mean and
the standard deviation
$\sigma_{\rm noise} = \epsilon_{\rm noise}$.
Then, the time-series data are given by
$f_{\rm n}(t) = 1 + \eta(t)$.
We choose the parameters $M = 1000$ and $N = 2M$
as in Sec.~\ref{subsec:effect_noise}
and controls the noise level
$\epsilon_{\rm noise} = 0.01, 0.02, \dots, 0.06$.

We show the normalized singular values
$\sigma_i/\sigma_0$
for $\epsilon_{\rm noise} = 0.03$
in Fig.~\ref{fig_supp:svd_vs_noise}(a).
We observe a plateaulike structure
just as in the case of the correlation function
in the 2D transverse-field Ising model
with noise in Sec.~\ref{subsec:effect_noise}.
For the same noise level $\epsilon_{\rm noise}$,
the position of the plateau
is nearly the same
between the time series of the constant function
and those of the correlation function
[compare Fig.~\ref{fig_supp:svd_vs_noise}(a)
and Fig.~\ref{fig:ising_2d_corr_lmd_sgm_noise}(a)].
The only major difference is that,
for the time series of the constant function,
the largest singular value $\sigma_0$
is much larger than 
the second largest singular value $\sigma_1$.
This result can be understood
in the zero-noise limit,
where the matrix $\bm{X}_0$ becomes an all-ones matrix.
The singular values of the all-ones matrix
are zero except for the largest singular value $\sigma_0$.
The noise shifts the zero singular values
to nonzero but small values.

The position of the plateau in the singular values
is correlated with the noise level $\epsilon_{\rm noise}$.
As shown in Fig.~\ref{fig_supp:svd_vs_noise}(b),
the singular values $\sigma_i$ at the middle ($i=M/2$)
corresponding to the plateau region
is found to be proportional to $\epsilon_{\rm noise}$.
They satisfy
$\sigma_{M/2} \approx 0.025 \epsilon_{\rm noise}$.
Although the coefficient would be subject to change
depending on parameters $M$ and $N$,
this proportionality is expected to hold
even for more general time-series data
affected by noise.
Therefore,
we can estimate the noise level $\epsilon_{\rm noise}$
from the position of the plateau
in the singular values $\sigma_i$
even when 
we do not know to what extent
the original input data are affected by noise.
This would be useful when estimating the noise of the experimental data.

\onecolumngrid

\end{document}